\documentclass[11pt,letterpaper]{article}

\usepackage[margin=1in]{geometry}
\usepackage{lmodern}
\usepackage{microtype}

\usepackage{amsmath,amssymb,amsthm,mathtools}

\usepackage{booktabs}
\usepackage{array}
\usepackage{adjustbox}
\usepackage{enumitem}

\usepackage{xcolor}
\usepackage{tikz}
\usetikzlibrary{arrows.meta,positioning,calc,fit}
\usepackage{subcaption}

\usepackage[numbers,sort&compress]{natbib}
\usepackage[
    colorlinks=true,
    linkcolor=blue,
    citecolor=blue,
    urlcolor=blue
]{hyperref}
\usepackage[capitalize,nameinlink]{cleveref}

\usepackage{aliascnt}
\newtheorem{theorem}{Theorem}[section]
\newaliascnt{lemma}{theorem}
\newtheorem{lemma}[lemma]{Lemma}
\aliascntresetthe{lemma}
\newaliascnt{corollary}{theorem}

\aliascntresetthe{corollary}

\theoremstyle{definition}
\newaliascnt{definition}{theorem}

\aliascntresetthe{definition}

\theoremstyle{remark}
\newaliascnt{remark}{theorem}

\aliascntresetthe{remark}

\crefname{theorem}{Theorem}{Theorems}
\Crefname{theorem}{Theorem}{Theorems}
\crefname{lemma}{Lemma}{Lemmas}
\Crefname{lemma}{Lemma}{Lemmas}
\crefname{corollary}{Corollary}{Corollaries}
\Crefname{corollary}{Corollary}{Corollaries}
\crefname{definition}{Definition}{Definitions}
\Crefname{definition}{Definition}{Definitions}
\crefname{remark}{Remark}{Remarks}
\Crefname{remark}{Remark}{Remarks}
\crefname{algorithm}{Algorithm}{Algorithms}
\Crefname{algorithm}{Algorithm}{Algorithms}

\newcommand{\OPT}{\mathrm{OPT}}

\allowdisplaybreaks

\usepackage{authblk}

\usepackage{pgfplots,placeins,float}
\usepgfplotslibrary{groupplots}
\pgfplotsset{compat=1.18,every axis/.append style={font=\small,
 tick label style={font=\footnotesize},legend style={font=\footnotesize,draw=none},
 grid=major,grid style={gray!15},axis line style={gray!70},
 label style={font=\small},title style={font=\small}}}
\setlist{topsep=4pt,itemsep=2pt,parsep=0pt}
\newcommand{\E}{\mathbb E}
\newcommand{\Prb}{\mathbb P}
\newcommand{\R}{\mathbb R}
\newcommand{\Q}{\overline\Phi}
\newcommand{\U}{F_{\mathrm{upper}}}
\newcommand{\V}{F_{\mathrm{lower}}}
\newcolumntype{P}[1]{>{\raggedright\arraybackslash}p{#1}}

\hypersetup{pdftitle={Improved Asymptotic Approximation Guarantees for Continuous-Review Lost-Sales and Dual-Sourcing Inventory Models},pdfauthor={Weitian Tong and Yao Xu}}
\title{Improved Asymptotic Approximation Guarantees for
Continuous-Review Lost-Sales and Dual-Sourcing Inventory Models}

\author[1]{Weitian Tong\thanks{\textbf{Corresponding author}. Email: \texttt{wtong.research@gmail.com}}}
\author[1]{Yao Xu}
\affil[1]{%
\parbox[t]{0.6\textwidth}{%
\centering
School of Computing, Georgia Southern University,\newline
1332 Southern Dr, Statesboro, Georgia 30458, USA
}%
}
\date{}

\begin{document}
\raggedbottom
\maketitle
\begin{abstract}
We study continuous-review inventory models with unit Poisson demand and deterministic lead times: 
single sourcing with lost sales and dual sourcing with backlogging. 
%
%
Xin~\citep{Xin2022} establishes an asymptotic approximation factor of
$1.79$ for each model in its proportional-growth regime. Under the
same assumptions, we obtain a factor of $1.248977$ for the better of
\emph{base-stock} and \emph{constant-order} policies for lost sales and
a factor of $1.599528$ for the better of \emph{dual-index} and
\emph{tailored base-surge} policies for dual sourcing.
The benchmark is the optimal long-run average cost over all policies
using only current and past information. For lost sales, lead time and
the lost-sales penalty grow proportionally. For dual sourcing, the
lead-time difference grows proportionally to the \emph{express premium},
the extra unit cost of faster supply, while express lead time and
backorder penalty remain fixed. 
For sufficiently large fixed backorder penalties, the dual-sourcing factor improves to
1.248977. The improvements follow from stronger lower bounds on the
optimal cost. We retain intermediate constraints on mean inventory for
lost sales and use events determined only by future demand to bound
backlog costs for dual sourcing. 
\end{abstract}

\par


\section{Introduction}
\label{sec:introduction}

We study two continuous-review inventory models, in which orders may
be placed at any time. In the \emph{single-sourcing lost-sales model}, one
supplier delivers each order after a fixed \emph{lead time}. Customer requests
arrive according to a \emph{Poisson process}, independently over disjoint time
intervals, and each request requires one unit. A request is filled immediately
if stock is available; otherwise it is lost and incurs a one-time
\emph{lost-sales penalty}. The objective is to minimize the \emph{long-run average cost}, the expected
cost per unit time over an increasing operating horizon, including
holding costs for stock on hand and penalties for lost sales. A retailer facing
customers who leave when a product is unavailable is a natural example.

In the \emph{dual-sourcing model}, a firm can order from a slower, less
expensive \emph{regular supplier} and a faster, more expensive \emph{express supplier}.
Both lead times are fixed, and demand follows the same unit-request Poisson
model. Unfilled requests are \emph{backlogged}: they wait for later delivery,
with a \emph{backorder penalty} charged per waiting unit per unit of time.
The firm balances holding and backlog costs against the extra unit cost of
express supply, called the \emph{express premium}. A parts distributor,
for example, can use regular shipments to restock and express orders to
fill urgent backorders.
In both models, outstanding orders form the \emph{delivery pipeline}; their
quantities and arrival dates affect every subsequent ordering decision.

Simple policies summarize this pipeline by a few parameters. A
\emph{base-stock} (BS) policy replaces successful sales to maintain a target
\emph{inventory position}, the sum of on-hand stock and outstanding orders.
A \emph{constant-order} (CO) policy instead places unit orders at regular intervals. 
In the dual-sourcing model, \emph{net inventory} is the number of units
on hand minus the number of units of unmet demand awaiting delivery.
A positive value represents available stock; a negative value represents a backlog.
A \emph{dual-index} (DI) policy uses two targets: the \emph{regular
inventory position} adds all outstanding orders to net inventory, whereas
the \emph{express inventory position} counts only orders arriving within
the express lead time. A \emph{tailored base-surge}
(TBS) policy combines constant regular orders with an express base-stock rule.
We compare the better policy in each pair with the \emph{unrestricted optimum},
the infimum cost over all \emph{nonanticipating policies}, whose decisions
use current and past information. An \emph{asymptotic approximation factor}
bounds this cost ratio in the parameter limits specified below.

Xin~\citep{Xin2022} established an asymptotic factor of 1.79 for both
continuous-review models with Poisson demand and deterministic lead times. For
lost sales, the analysis compares the better of BS and CO with the optimal cost
when lead time grows proportionally to the lost-sales penalty. For dual sourcing,
it compares the better of DI and TBS with the optimal cost when the lead-time
difference grows proportionally to the express premium. \emph{Capped policies} add an order-frequency cap to
inventory targets. Their classes inherit these bounds because they contain the
two respective families as special or limiting cases. 
Hu~\citep{Hu2026} establishes a uniform $\sqrt{2}$ approximation
guarantee for ordinary base-stock policies in the same
continuous-review Poisson lost-sales model without setup costs.
The guarantee holds for every finite lead time and lost-sales
penalty. 
%

\subsection{Contributions and comparison with Xin}

We improve the lost-sales factor from 1.79 to 1.248977 and the dual-sourcing
factor from 1.79 to 1.599528. The dual-sourcing result covers the same parameter
region as Xin's guarantee. For sufficiently large fixed backorder penalties,
its factor also becomes 1.248977. The precise statements and order of limits
are given in \Cref{thm:main}. The improvements strengthen the performance
analysis of established policies; the capped policy classes inherit them by
the inclusions in Xin~\citep{Xin2022}.

The policy-cost upper bounds in the earlier analysis already permit a better
factor if the lower bound on the optimal cost can be raised. Xin's principal
lower comparison reduces the inventory problem to a one-period \emph{newsvendor}
calculation: choose stock before observing total demand over one lead-time
window, then charge for excess and shortage. The calculation retains aggregate
demand uncertainty but omits intermediate inventory requirements. Our lost-sales
analysis retains equal mean inventory at selected dates inside the window.
After \emph{time averaging}, these equalities hold for every admissible policy,
including nonstationary policies. Adding them restricts the comparison problem
and raises its optimal cost.

The main structural step determines how to meet an intermediate mean with
as little dispersion as possible. Starting from a random inventory level,
raise only its lowest outcomes to a common floor. This correction is least in
\emph{convex order} among all upward corrections with the prescribed mean: it minimizes
the expectation of every convex function of the corrected state. Keeping the path nonnegative with its smallest upward correction, called
\emph{minimal reflection}, and then applying the common floor gives an exact
\emph{feasibility test} for a finite-grid relaxation driven by \emph{Brownian motion},
a continuous Gaussian process with independent increments and increment
variance equal to elapsed time. It is the limit of centered Poisson demand. Intermediate dates
can strengthen this relaxation strictly. 
To establish the numerical guarantee, we divide a lead-time window into
two equal \emph{blocks}. The resulting relaxation gives convex necessary
conditions on normalized mean inventory and lost-demand rate. Tangents
to the convex functions in these conditions yield linear restrictions,
called \emph{supporting inequalities}. We select twelve such inequalities
for the numerical proof certificate. Together with nonnegativity, they
define a \emph{polyhedron} with thirteen vertices. Every feasible limiting
pair lies in this polyhedron, so minimizing the normalized cost over its
vertices gives a lower bound on the limiting normalized optimal cost.
We note that the choice of twelve is specific to this certificate, rather than a
requirement of the model; the smallest number sufficient to certify the
stated approximation factor is not determined here.

Dual sourcing requires a different treatment because inventory may be negative.
An unfavorable demand event can select a useful inventory-balance comparison,
but it can also be correlated with the backlog at the selected date. We charge
that backlog through a containing event determined only by subsequent demand.
Independent Poisson increments then justify the probability multiplying its
mean. Two-date and four-date Gaussian rules give the general factor. Finally,
a signed common-floor recursion shows how the lost-sales inequalities emerge
when a fixed backorder penalty becomes large. \Cref{tab:comparison}
summarizes the differences from the earlier analysis.

\begin{table}[htbp]
\centering\small
\caption{Asymptotic guarantees under Xin's proportional-growth assumptions
\citep{Xin2022}. The final column gives the lower-bounding method used for
each factor.}
\label{tab:comparison}
\begin{adjustbox}{max width=.9\linewidth}
\begin{tabular}{@{}P{.30\linewidth}lllP{.33\linewidth}@{}}
\toprule
Model and regime & Policies & \citep{Xin2022} & This paper & New lower-bound ingredient\\\midrule
Lost sales: \newline $L=\alpha p$, $p\to\infty$ & BS or CO & 1.79 & 1.248977 & Intermediate means and common-floor calibration\\
Dual sourcing: \newline $L=\alpha c$, $\alpha p\ge1$, $c\to\infty$ & DI or TBS & 1.79 & 1.599528 & Future-demand containing events for backlog charges\\
Dual sourcing: \newline sufficiently large fixed $p$, $\alpha p\ge1$ & DI or TBS & 1.79 & 1.248977 & Signed calibration and vanishing mean backlog\\
\bottomrule
\end{tabular}
\end{adjustbox}
\end{table}

\subsection{Related literature}

\paragraph{Lost-sales inventory.}
Bijvank and Vis~\citep{BijvankVis2011} review lost-sales models and
their replenishment policies. Goldberg, Reiman, and Wang~\citep{GoldbergReimanWang2021}
survey asymptotic methods for lost sales, dual sourcing, and assemble-to-order
systems. Their discussion distinguishes limits in which a large lead time
smooths replenishment from limits in which a large shortage penalty favors
responsive ordering. Proportional growth retains both effects.
Early continuous-review work connects inventory and lost sales to queueing
models \citep{Karush1957}. In \emph{periodic review}, decisions are made at
fixed review dates. Morton~\citep{Morton1971} derives ordering and cost
bounds with a newsvendor interpretation for a lagged lost-sales problem.
The dependence on outstanding replenishments has since motivated both
structural analysis and heuristic design. Zipkin~\citep{Zipkin2008Structure}
studies the structure of the periodic-review problem, and
Zipkin~\citep{Zipkin2008Methods} compares computational methods and simple policies.
Xin~\citep{Xin2022} compares base-stock and constant-order policies in
continuous review when lead time grows proportionally to the lost-sales penalty.
Hu~\citep{Hu2026} constructs a lower bound on the unrestricted
optimal cost using a potential function that tracks the ranks
and remaining delivery times of units on hand or in transit.
Concavity of this lower bound in the lost-sales penalty extends
base-stock cost comparisons at selected penalty values to all
penalties, yielding the uniform $\sqrt{2}$ guarantee.
With fixed lead times, Huh et al.~\citep{Huh2009} establish asymptotic
optimality of base stock as the lost-sales penalty grows, under their
periodic-review demand assumptions. Bu, Gong, and Chao~\citep{BuGongChao2024}
study the corresponding continuous-review regime.

With fixed cost parameters and growing lead time, Goldberg et
al.~\citep{Goldberg2016} prove constant-order asymptotic optimality in
periodic review. Xin and Goldberg~\citep{XinGoldberg2016} establish
exponential convergence using convexity comparisons that retain both
initial inventory and the replenishment rate. Xin~\citep{Xin2021}
introduces capped base-stock
policies and relates them to constant-order policies. For guarantees without an asymptotic limit, the marginal-cost accounting
and cost-balancing approach of Levi et al.~\citep{LeviPalRoundyShmoys2007}
gives constant-factor policies for backlogged inventory and stochastic lot sizing.
Levi, Janakiraman, and Nagarajan~\citep{Levi2008} develop a factor-two
dual-balancing policy for lost sales under their demand assumptions. A recent preprint by Xin~\citep{Xin2026} proves a $7/3\approx 2.33$
approximation for periodic-review capped base-stock policies. Its model and uniform finite-parameter guarantee differ from the
continuous-review proportional-growth guarantees studied here.

Projected inventory-level policies provide another way to use pipeline information.
Van Jaarsveld and Arts~\citep{vanJaarsveldArts2024} set a target for expected inventory
when an order arrives. They establish asymptotic optimality in the large-penalty
regime under demand-tail conditions and in the large-lead-time regime for exponential
demand. Moradi, Arts, and Drent~\citep{MoradiArtsDrent2025} report an iterated-limit
guarantee for large lead time and penalty in periodic review.
Yuan, Guan, and Jiang~\citep{YuanGuanJiang2026} report an exponential large-lead-time
guarantee for projected inventory-level policies in a recent periodic-review preprint.
Our lower bound averages over the history before the lead-time window
and then enforces mean inventory at selected dates inside it.

\paragraph{Dual sourcing.}
Fukuda~\citep{Fukuda1964} characterizes optimal policies under suitable
cost conditions when two periodic-review delivery lags differ by one period.
Moinzadeh and Schmidt~\citep{MoinzadehSchmidt1991} study an early
continuous-review emergency-replenishment model. Veeraraghavan and Scheller-Wolf~\citep{VeeraraghavanSchellerWolf2008} analyze
dual-index policies in capacitated systems. Sheopuri, Janakiraman, and
Seshadri~\citep{SheopuriJanakiramanSeshadri2010} extend this approach with
alternative order-up-to structures in periodic review. Allon and Van Mieghem~\citep{AllonVanMieghem2010}
develop tailored base-surge policies, which combine stable regular orders with
responsive express orders. Janakiraman, Seshadri, and Sheopuri~\citep{Janakiraman2015}
further analyze this policy class. Xin and Goldberg~\citep{XinGoldberg2018} prove
tailored base-surge asymptotic optimality as the regular lead time grows with other
parameters fixed. Chen, Li, and Xie~\citep{ChenLiXie2026} report an exponential
convergence bound for this periodic-review regime in a recent preprint.
Deng et al.~\citep{DengShaoSongYu2026} extend base-surge ideas to assembly
systems with several dual-sourced components and prove asymptotic optimality
as regular lead times grow. Our model has one item and jointly growing
lead-time and cost differences.
Moradi et al.~\citep{MoradiDrentArtsDrent2026} report asymptotic optimality of
projected expedited inventory-position policies in periodic review, both for
large regular lead time at fixed costs and for jointly large shortage and
express costs at fixed ratio and fixed lead times. These limits differ from
proportional growth of the lead-time
difference and express premium. Sun and Van Mieghem~\citep{SunVanMieghem2019} establish capped
dual-index optimality in a robust model. Jiang et al.~\citep{Jiang2026}
derive a lead-time-sensitive TBS approximation in a multiperiod distributionally
robust model with independent demand. These robust benchmarks differ from the
fixed-Poisson stochastic optimum considered here. Song and Zipkin~\citep{SongZipkin2009} provide the overflow
queueing framework underlying the specialized dual-index cost formula used below.
Song, Xiao, and Zhang~\citep{SongXiaoZhang2025} characterize optimal
policies under conditions in an order-tracking model with Erlang
processing times. Bai et al.~\citep{Bai2023} develop semi-open-loop
asymptotic control for Markov decision processes with long lead times,
under mixing conditions or linear dynamics and convexity. Bai et
al.~\citep{Bai2026} extend the analysis to stochastic lead times.

\paragraph{Learning and computational policies.}
Lyu, Zhang, and Xin~\citep{LyuZhangXin2024} study learning from censored
sales in periodic review and give regret guarantees relative to the best
base-stock or capped base-stock policy. Gijsbrechts et al.~\citep{Gijsbrechts2022}
evaluate deep reinforcement learning on lost-sales, dual-sourcing, and
multiechelon problems. We assume the Poisson demand rate is known and
compare the costs of specified policy families with the unrestricted
nonanticipating optimum.

\paragraph{Lower-bound methods.}
Janakiraman, Seshadri, and Shanthikumar~\citep{JanakiramanSeshadriShanthikumar2007}
compare optimal lost-sales and backlogged costs under periodic review and
relate their shortage-cost parameters. 
Xin~\citep{Xin2022} gives an asymptotic 1.79 guarantee for both
continuous-review models.
The policy upper bounds in that paper already
permit a smaller factor if the optimal-cost lower bound can be strengthened.
We retain the frozen-pipeline averaging idea related to the joint-convexity
construction of Xin and Goldberg~\citep{XinGoldberg2016}, but impose the mean
shortage budget at intermediate dates. Convex-order comparisons, for which
Shaked and Shanthikumar~\citep{ShakedShanthikumar2007} provide general background,
identify the least dispersed upward correction at each date. The contribution
is the inventory relaxation and its certified consequences, rather than a new
general characterization of convex order. Information-relaxation bounds such as
Brown, Smith, and Sun~\citep{BrownSmithSun2010} also use the information structure
of stochastic control. Chen, Li, and Xie~\citep{ChenLiXie2026} instead construct infinite-horizon
primal-dual bounds for closed-loop and semi-open-loop dual-sourcing systems.
Here we preserve adaptation and retain intermediate mean constraints after
conditional averaging of the pipeline. The supporting inequalities follow
directly from convexity; no external strong-duality result is invoked.

\paragraph{Organization.}
\Cref{sec:models} states the guarantees and gives the policy-cost formulas.
\Cref{sec:lost-sales-analysis,sec:dual-sourcing-analysis} establish the
lower bounds and the large-backorder-penalty result.
\Cref{sec:implementation} gives policy parameters and numerical comparisons.
\Cref{app:averaging} proves the averaging and diffusion reductions, and
\Cref{app:grid-extensions} extends the methods to finer grids.
\Cref{app:certificates} gives the inputs and error bounds for the theorem.
\Cref{app:finite-study} reports the finite-parameter comparisons.

\section{Models and main results}
\label{sec:models}

We follow the notation and policy terminology of Xin~\citep{Xin2022}.
Customer requests form a unit-demand Poisson process $D(t)$ of rate $\lambda>0$;
write $D(s,t]=D(t)-D(s)$ for demand in $(s,t]$. Holding cost is normalized to
$h=1$. Expectations are denoted by $\E$. 
In each model, $C(\pi)$ denotes the long-run average cost of policy $\pi$,
and $\OPT$ denotes its infimum over all admissible policies. We evaluate these costs from zero initial inventory and an empty pipeline.
The lower-bound arguments also allow initial states of finite expected size.
Subscripts identify the model when both are being compared. In particular, $p$ is the lost-sales
penalty in the first model and the backorder penalty rate in the second, as in Xin.
For a real number $w$, its \emph{positive part} is $w^+=\max\{w,0\}$ and
its \emph{negative part} is $w^-=\max\{-w,0\}$. We write $\mathbf1_E$ for
the \emph{indicator} of event $E$, and $\Prb(E)$ for its probability.

An admissible policy is \emph{nonanticipating}: it may use observed demand,
inventory, outstanding orders, and independent randomization, but not future
demand. Orders have deterministic lead times, no setup cost, and cannot be
cancelled. Disposal is excluded. Order and arrival processes are locally finite
with finite expectation on bounded intervals. Initial inventory and pipeline
have finite expected size and are independent of future demand. 
Future Poisson
increments remain independent of information available to the policy.
State variables are right-continuous. Deliveries already scheduled for a
request epoch are processed before the request. A zero-lead-time order
triggered by the request arrives after it at the same epoch. State-cost
integrals are locally integrable. These conditions permit
time-averaging arguments without assuming that an optimal stationary policy exists.

\subsection{Single-sourcing lost-sales model and policies}

Let $I(t)\ge0$ be inventory on hand. An order placed at time $t$ arrives at
$t+L$, where $L>0$ is the replenishment lead time. A request is filled if at least one
unit is available. Otherwise it is lost and incurs a penalty $p>0$. If $Y(T)$ counts lost requests by time $T$,
the objective is
\[
 C(\pi)=\limsup_{T\to\infty}\frac1T
 \E\!\left[\int_0^T I(t)\,dt+pY(T)\right].
\]
We fix $\alpha,\lambda>0$ and take $p\to\infty$ with $L=\alpha p$.

The BS policy $\pi_S$ maintains inventory position, defined as on-hand stock
plus outstanding replenishments, at an integer target $S\ge0$. Each successful
sale triggers one replacement order; lost demand triggers no order. If $S=5$
and two units are on hand while three are in transit, a sale reduces on-hand
stock to one and adds one unit to the pipeline. Inventory position returns to
five, although only one unit is available immediately.

The CO policy $\pi_r$ places a unit order every $r^{-1}$ time units, with
$0<r<\lambda$. It smooths replenishments rather than reacting to each sale.
A fixed clock phase is time-averaged, or its phase may be randomized.
The never-order policy is a separate endpoint. A \emph{capped base-stock policy}
combines a base-stock target with a minimum inter-order time; its best cost
is at most the smaller of the BS and CO infima
\citep{Xin2022}.

Let $H$ and $d$ be subsequential time averages of on-hand inventory and
lost-demand rate. The policy's long-run average cost is at least $H+pd$.

\subsection{Dual-sourcing model and policies}

The regular supplier has lead time $L_R$ and unit ordering cost $c_R$; the
express supplier has lead time $L_E<L_R$ and cost $c_E>c_R$. Following Xin~\citep{Xin2022},
set $L=L_R-L_E$, normalize $c_R=0$, and write $c=c_E-c_R>0$ for the
express premium. All unmet demand is backlogged. Here $I(t)$ is signed net
inventory, so $I(t)^+$ is stock on hand and $I(t)^-$ is backlog. A unit of
backlog costs $p>0$ per unit of time. If $Q_E(T)$ counts express units ordered
by $T$, then
\[
 C(\pi)=\limsup_{T\to\infty}\frac1T
 \E\!\left[\int_0^T\left(I(t)^++pI(t)^-\right)\,dt+cQ_E(T)\right].
\]
We take $L=\alpha c$ and let $c\to\infty$, keeping $\lambda$, $L_E\ge0$,
$\alpha>0$, and $p>0$ fixed. The main parameter region is $\alpha p\ge1$.
For example, regular and express lead times of sixteen and two days give
$L=14$ days and $L_E=2$ days.

The \emph{regular inventory position} is net inventory plus all outstanding
orders. The \emph{express inventory position} includes only orders arriving
within $L_E$. The DI policy $\pi_{S,U}$ maintains regular position at $S$ and
express position at least $S-U$, where $U\ge0$ is an integer target difference.
At a demand arrival it orders one unit from express if the number of regular
orders arriving later than $L_E$ equals $U$, and from regular otherwise.
The target $S$ is an integer and may satisfy $S\le U$.

The TBS policy $\pi_{r,S}$ places regular unit orders at rate $0<r<\lambda$
and uses express base-stock target $S\in\mathbb Z$. Just after a demand,
it places a unit express order when express inventory position falls below
$S$. A negative target is permitted because backlogging is allowed.
The regular stream supplies the predictable base quantity; express orders
respond to shortfalls. \emph{Capped dual-index policies} add a cap on regular order
frequency and contain DI and TBS as special or limiting cases
\citep{Xin2022}.

Write $H$ and $\bar B$ for time-average positive inventory and backlog,
and $d$ for the express-arrival rate. The regular-arrival rate is $\lambda-d$,
and $H+p\bar B+cd$ is a lower comparison for long-run cost. Express arrivals
are bounded by express orders plus the initial express pipeline, so this
comparison remains valid when premiums are charged on order placement.
\Cref{app:averaging} establishes these statements for nonstationary
policies. The simple policies have equal long-run order and arrival rates.
Delivery residual times remain part of the physical state; pipeline counts
alone need not be Markov.

For a policy family $\mathrm P$, define
$C_{\mathrm P}^*=\inf_{\pi\in\mathrm P}C(\pi)$.
This notation records an infimum without requiring an optimizer at a limiting
parameter value. All other model parameters are suppressed unless needed.

\subsection{Approximation guarantees}

\begin{samepage}
\begin{theorem}
\label{thm:main}
For fixed $\lambda>0$,
\begin{equation*}
 \sup_{\alpha>0}\limsup_{ p\to\infty}
 \frac{\min\{C_{\rm BS}^*(\alpha p, p),C_{\rm CO}^*( p)\}}
      {\OPT_{\rm LS}(\alpha p, p)}
 \le r_{\rm LS}:=\frac{1248977}{1000000}.
\end{equation*}
For fixed $\lambda>0$ and $L_E\ge0$,
\begin{equation*}
 \sup_{\substack{\alpha,p>0\\\alpha p\ge1}}\limsup_{c\to\infty}
 \frac{\min\{C_{\rm DI}^*(\alpha c,c,p,L_E),C_{\rm TBS}^*(c,p,L_E)\}}
      {\OPT_{\rm DS}(\alpha c,c,p,L_E)}
 \le r_{\rm DS}:=\frac{199941}{125000}.
\end{equation*}
There is a finite $p_0$ such that, for every fixed $p\ge p_0$, the supremum in
the dual-sourcing bound above over $\alpha p\ge1$ is at most $r_{\rm LS}$.
\end{theorem}
\end{samepage}

Each limit holds with $\alpha$ fixed, and the dual-sourcing limit also
holds with $p$ fixed. We then take the supremum over the indicated
parameters. The large-penalty result gives existence of a finite $p_0$.
An explicit threshold and finite-parameter convergence rates remain open.

\subsection{Policy costs and normal comparison functions}

We use the normal functions and cost-comparison notation of Xin~\citep{Xin2022}.
Let $\phi$ and $\Phi$ denote the standard normal density and \emph{cumulative
distribution function} (CDF), respectively, put $\Q(z)=1-\Phi(z)$, and define the \emph{normal hazard rate}
$\psi(z)=\phi(z)/\Q(z)$. Thus $\psi(-\beta)=\phi(\beta)/\Phi(\beta)$.
Let $Z\sim N(0,1)$. The \emph{normal loss function} is
$n(z)=\phi(z)+z\Phi(z)=\E\left[(z+Z)^+\right]$.

Let $\beta(\alpha)$ be the unique solution of
\begin{equation*}
 \psi(-\beta)\left(\psi(-\beta)+\beta\right)=\frac{\alpha}{1+\alpha}.
\end{equation*}
The trial target $\lfloor\lambda L+\beta(\alpha)\sqrt{\lambda L}\rfloor$
differs from mean lead-time demand by $\beta(\alpha)$ standard deviations,
up to integer rounding.
Define
\begin{align*}
 \U(\alpha)&=\sqrt\alpha\,\beta(\alpha)
       +(\sqrt\alpha+\alpha^{-1/2})\psi(-\beta(\alpha)),\\
 \V(\alpha)&=(\sqrt\alpha+\alpha^{-1/2})
       \phi\!\left(\Phi^{-1}\!\left(\frac1{1+\alpha}\right)\right),\\
 F_{\rm TBS}(p)&=\sqrt{2p\log(1+1/p)}.
\end{align*}
The first is the normalized BS cost comparison, also an upper comparison for
DI. The second is the newsvendor lower comparison, and the third is the TBS
upper comparison. For a fixed trial $\beta$, write
$F_{\mathrm{upper},\beta}(\alpha)=\sqrt\alpha\,\beta+
(\sqrt\alpha+\alpha^{-1/2})\psi(-\beta)$, so that
$\U(\alpha)=\inf_\beta F_{\mathrm{upper},\beta}(\alpha)$.
The BS and CO comparisons cross at the unique $\alpha^*\simeq0.69787464$
with $\U(\alpha^*)=\sqrt2$. The certificate uses the rational cutoff
$\alpha_0=69787/100000<\alpha^*$.

The four policy upper bounds in the next lemma are stated in
Xin~\citep{Xin2022}. We derive them from the exact policy costs and the
time-average CO inventory law in \eqref{eq:co-inventory-tail}.

\begin{lemma}
\label{lem:policy-facts}
For fixed parameters inside the limits,
\begin{align*}
 \limsup_{ p\to\infty}\frac{C_{\rm BS}^*}{\sqrt{\lambda p}}
 &\le\U(\alpha),&
 \limsup_{ p\to\infty}\frac{C_{\rm CO}^*}{\sqrt{\lambda p}}
 &\le\sqrt2,\\
 \limsup_{c\to\infty}\frac{C_{\rm DI}^*}{\sqrt{\lambda c}}
 &\le\U(\alpha),&
 \limsup_{c\to\infty}\frac{C_{\rm TBS}^*}{\sqrt{\lambda c}}
 &\le F_{\rm TBS}(p)<\sqrt2.
\end{align*}
Lost-sales normalized optimal cost has liminf at least $\V(\alpha)$. The same
dual-sourcing comparison holds when $\alpha p\ge1$.
Both actual optimal costs are nondecreasing in regular replenishment lead time
when all other model parameters are fixed.
\end{lemma}

The upper bounds follow by evaluating the following exact cost formulas
at explicit trial parameters.
For $q>0$ and an integer $S\ge0$, define the \emph{Erlang loss function}
$B(S,q)=(q^S/S!)/\sum_{j=0}^S q^j/j!$, with $B(0,q)=1$. Then
\begin{equation}
 C(\pi_S)=S-\lambda L+\lambda(L+p)B(S,\lambda L).
 \label{eq:bs-exact}
\end{equation}
For CO, put $\rho=r/\lambda$ and let $w>0$ solve
$\rho=(1-e^{-w})/w$. Mean inventory is $1/w$, and cost is
$1/w+p(\lambda-r)$; a deterministic lead time only shifts the order clock.

Let $I_r$ have the time-average CO inventory distribution, and put
$\sigma=e^{-w}$. Let $Q_n$ be inventory immediately after the $n$th unit
delivery and let $D_n$ be demand before the next delivery. Then
\[
 Q_{n+1}=1+(Q_n-D_n)^+,\qquad
 D_n\sim\operatorname{Pois}(\lambda/r),
\]
where $D_n$ is independent of $Q_n$. The geometric law
$\Prb(Q_n\ge j)=\sigma^{j-1}$, $j\ge1$, is invariant because
$\E\left[\sigma^{D_n}\right]
=e^{-(\lambda/r)(1-\sigma)}=\sigma$.
The delivery-epoch chain is irreducible and aperiodic, so it converges to
this law. Coupling the chain from the empty start with a stationary chain on
the same demand increments bounds its inventory tails by the stationary
tails and gives convergence of the inventory means.

In the invariant regime, the inventory tail at elapsed time
$t\in[0,1/r)$ after a delivery is
$\sigma^{j-1}e^{-\lambda t(1-\sigma)}$. Averaging over one delivery interval gives
\begin{equation}
 \label{eq:co-inventory-tail}
 \Prb(I_r\ge j)
 =r\int_0^{1/r}\sigma^{j-1}e^{-\lambda t(1-\sigma)}\,dt
 =\rho\sigma^{j-1},\qquad j\ge1.
\end{equation}
Consequently,
\[
 \Prb(I_r=0)=1-\rho,\qquad
 \Prb(I_r=j)=\rho(1-\sigma)\sigma^{j-1},\quad j\ge1.
\]
Summing the time-average tail gives $\E\left[I_r\right]=\rho/(1-\sigma)=1/w$.

For DI, let $N_1$ have law $(\operatorname{Pois}(\lambda L)\mid
\operatorname{Pois}(\lambda L)\le U)$, independently of
$N_2\sim\operatorname{Pois}(\lambda L_E)$. Xin~\citep{Xin2022} gives
\begin{equation}
 C(\pi_{S,U})=\lambda c B(U,\lambda L)
       +\E\left[(S-N_1-N_2)^++p(N_1+N_2-S)^+\right].
 \label{eq:di-exact}
\end{equation}
For TBS, Xin~\citep{Xin2022} gives
$c(\lambda-r)+\E\left[(I_r+S-N_2)^++p(N_2-S-I_r)^+\right]$,
where $I_r$ is independent of $N_2$.

We first derive the BS and DI upper bounds. Fix $\beta\in\R$, put
$q=\lambda L$, and take $S_q=\lfloor q+\beta\sqrt q\rfloor$.
The Poisson local limit theorem and central limit theorem give
$\sqrt q\,B(S_q,q)
 \longrightarrow \frac{\phi(\beta)}{\Phi(\beta)}
 =\psi(-\beta)$
as $q\to\infty$.
Substituting this limit into \eqref{eq:bs-exact}, with $L=\alpha p$, yields
$
 \frac{C(\pi_{S_q})}{\sqrt{\lambda p}}
 \longrightarrow F_{\mathrm{upper},\beta}(\alpha).
$

For DI, take $U=S_q$ and $S=U$ in \eqref{eq:di-exact}.
Since $N_1\le U$, the inventory-cost term differs from
$\E\left[U-N_1\right]$ by at most
$\max\{1,p\}\E\left[N_2\right]=\max\{1,p\}\lambda L_E$.
This difference vanishes after normalization by $\sqrt{\lambda c}$.
Using $\E\left[N_1\right]=q(1-B(U,q))$ and $L=\alpha c$ therefore gives
$
 \frac{C(\pi_{U,U})}{\sqrt{\lambda c}}
 \longrightarrow F_{\mathrm{upper},\beta}(\alpha).
$
Choosing $\beta=\beta(\alpha)$ establishes the BS and DI upper bounds.

For CO, the relation $\rho=(1-e^{-w})/w$ implies
$1-\rho=w/2+O(w^2)$ as $w\downarrow0$.
The trial $r=\lambda-\sqrt{\lambda/(2p)}$ is positive for sufficiently
large $p$ and satisfies $w\sqrt{\lambda p}\to\sqrt2$.
Hence,
\[
 \frac{C(\pi_r)}{\sqrt{\lambda p}}
 =\frac{1}{w\sqrt{\lambda p}}
  +\frac{p(\lambda-r)}{\sqrt{\lambda p}}
 \longrightarrow\sqrt2.
\]

For TBS, fix $p>0$ and put $b_p=\log(1+1/p)$ and
$a_p=\sqrt{2pb_p}$. For sufficiently large $c$, choose
\[
 r_c=\lambda-\frac{a_p}{2}\sqrt{\lambda/c},
 \qquad
 S_c=-\left\lfloor\frac{b_p}{a_p}\sqrt{\lambda c}\right\rfloor.
\]
Let $w_c>0$ satisfy $r_c/\lambda=(1-e^{-w_c})/w_c$.
The same expansion gives $w_c\sqrt{\lambda c}\to a_p$.
The time-average CO tail in \eqref{eq:co-inventory-tail} then yields
\[
 \frac{I_{r_c}}{\sqrt{\lambda c}}\Rightarrow X,
 \qquad
 \frac{\E\left[I_{r_c}\right]}{\sqrt{\lambda c}}
 \longrightarrow \E\left[X\right]=\frac1{a_p},
\]
where $X$ has an \emph{exponential distribution} with rate $a_p$.
Because $L_E$ is fixed,
$\E\left[N_2\right]/\sqrt{\lambda c}\to0$.
Substituting these limits into the exact TBS cost and using
$e^{-b_p}=p/(1+p)$ gives
\begin{equation*}
 \frac{C(\pi_{r_c,S_c})}{\sqrt{\lambda c}}
 \longrightarrow \frac{a_p}{2}
   +\E\left[(X-b_p/a_p)^++p(b_p/a_p-X)^+\right] =\frac{a_p}{2}+\frac{pb_p}{a_p}
 =F_{\rm TBS}(p).
\end{equation*}
These trials coincide with those specified in
\Cref{sec:implementation}. Since the optimal cost within each
policy family is at most its trial cost, the four upper bounds follow.

The lost-sales lower comparison is due to Xin~\citep{Xin2022}:
\[
 \OPT_{\rm LS}(L,p)\ge
 \inf_s\E\left[(s-D_L)^++(p/L)(D_L-s)^+\right],
 \qquad D_L\sim\operatorname{Pois}(\lambda L).
\]
Its normal limit is $\V(\alpha)$. \Cref{lem:one-date} proves the
dual-sourcing lower bound. For lead-time monotonicity, start both systems
empty and use the same demand path. Simulate a policy for the longer lead
time. In the shorter system, place each affected order at its simulated
order time plus the lead-time difference. Keep the simulated express order
times in dual sourcing. The two systems then have the same delivery and
inventory paths, and the same express costs in dual sourcing. Taking the
infimum gives the monotonicity claim.

\section{The lost-sales approximation}
\label{sec:lost-sales-analysis}

We use the policy upper bound $\min\{\U(\alpha),\sqrt2\}$ from
Xin~\citep{Xin2022}. To lower-bound the optimum, we impose mean-inventory
constraints inside a lead-time window. We solve the resulting fixed-grid
relaxation and use supporting inequalities from a two-block grid to
certify the approximation factor.

\subsection{Causal mean calibration}
\label{sec:causal}

\begin{figure}[htbp]
\centering
\begin{adjustbox}{max width=.6\linewidth}
\begin{tikzpicture}[>=Latex,font=\small]
\draw[->,thick] (-.2,0)--(11.5,0);
\foreach \x/\lab in {0/0,5.5/{L/2},11/L}{
 \fill (\x,0) circle (1.6pt);
 \node[below=5pt] at (\x,0) {$\lab$};
}
\draw[blue!65!black,very thick] (0,1.25)--(11,1.25);
\draw[blue!65!black] (0,1.1)--(0,1.4) (11,1.1)--(11,1.4);
\node[above,align=center,text width=10cm] at (5.5,1.35)
 {Every replenishment arriving in this window was ordered by its start};
\node[align=center,text width=3cm,anchor=north] at (0,-.8)
 {Starting inventory\\mean $H$};
\node[align=center,text width=3cm,anchor=north] at (5.5,-.8)
 {Retained midpoint\\mean $H$};
\node[align=center,text width=3cm,anchor=north] at (11,-.8)
 {Terminal inventory\\mean $H$};
\draw[->,gray!70,thick] (.4,.5)--(5.1,.5);
\draw[->,gray!70,thick] (5.9,.5)--(10.6,.5);
\end{tikzpicture}
\end{adjustbox}
\caption{In the two-block lost-sales bound, replenishments arriving within
the lead-time window are fixed at its start. Averaging over window starts
gives the same mean inventory at $0$, $L/2$, and $L$.}
\label{fig:mean-window}
\end{figure}
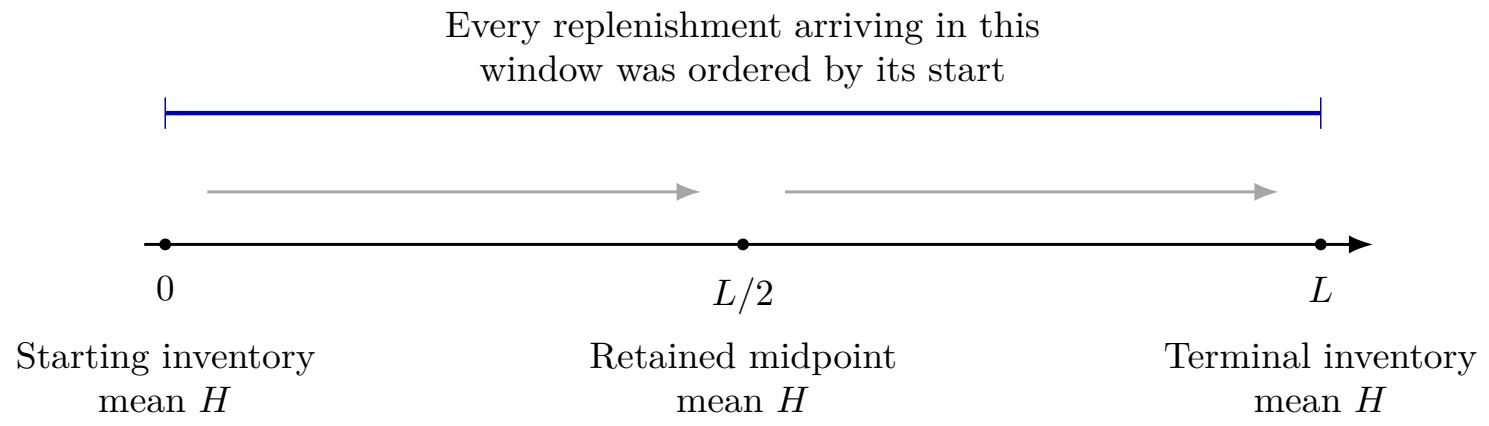

Average the start of a lead-time window over a long operating horizon.
Shifting the time integral by a fixed offset changes only its boundary
terms, so the limiting mean inventory is $H$ at each grid date.
Replenishments arriving during the window were ordered by its start.
Their schedule is independent of subsequent Poisson demand.
\Cref{fig:mean-window} shows the three dates used in the two-block bound.

Scale the window to $[0,1]$ and define
$u=H/\sqrt{\lambda L}$ and $v=d\sqrt{L/\lambda}$. Since $L=\alpha p$,
\[
 \frac{H+pd}{\sqrt{\lambda p}}=\sqrt\alpha\,u+\frac{v}{\sqrt\alpha}.
\]
Thus $u$ measures average holding inventory and $v$ measures lost demand on
the diffusion scale. Centered Poisson demand deficit converges to a standard
Brownian motion $W_s$. The averaged arrival rate is $\lambda-d$, producing
drift $-v$ after centering. Lost demand supplies an adapted, nondecreasing
upward correction $\mathcal A_s$. This correction belongs to the relaxed
comparison; it need not be an implementable replenishment rule. For integrable random variables, \emph{convex order}, written $X\le_{\rm cx}Y$,
means $\E\left[f(X)\right]\le\E\left[f(Y)\right]$ for every convex $f$ for which the expectations
exist; it entails equal means. The \emph{increasing convex order}, written $\le_{\rm icx}$, restricts
the comparison to increasing convex functions. For equal means, convex order is
equivalent to comparing the \emph{call functions}, the expected excesses
above each threshold: $\E\left[(X-q)^+\right]\le\E\left[(Y-q)^+\right]$ for every real $q$
\citep{ShakedShanthikumar2007}.

The first operation is \emph{minimal reflection}: add only the cumulative
correction needed to prevent a path from going below zero. For a block of
length $s>0$, initial level $z\ge0$, and input path $w$ starting at zero, let
\begin{equation*}
 \Gamma_s(z,w;v)=\max\left\{z+w_s-vs,
             \sup_{0\le r\le s}(w_s-w_r-v(s-r))\right\}.
\end{equation*}
The endpoint $\Gamma_s$ is nondecreasing in $z$ and jointly convex in
$(z,v)$ because it is a supremum of affine functions nondecreasing in $z$.
It is also $2$-\emph{Lipschitz} in the uniform norm of $w$. For an independent
Brownian block, the expected \emph{regulator}, the cumulative minimal upward correction,
from $z\ge0$ is
\begin{equation*}
 g_s(z,v)=\E\left[\left(\sup_{0\le r\le s}(W_r+vr)-z\right)^+\right].
\end{equation*}
Here $W$ starts at zero, and its symmetry allows the sign reversal in
the formula for $g_s$. An uncorrected path starting at $2$, reaching a
minimum of $-3$, and ending at $1$ requires a cumulative correction of
$3$. Its reflected endpoint is $4$.

\begin{lemma}
\label{lem:reduction}
For any fixed finite grid $0=t_0<t_1<\cdots<t_K=1$, every subsequential
normalized lost-sales mean pair $(u,v)$ from a sequence of causal policies with uniformly bounded normalized costs is feasible for
\begin{equation*}
 X_s=u+W_s-vs+\mathcal A_s\ge0,\qquad \mathcal A_0=0,\quad\mathcal A\text{ adapted and nondecreasing},
 \qquad \E\left[X_{t_i}\right]=u.
\end{equation*}
Feasibility here means existence of a relaxed process, not convergence of the
original controlled processes. In particular, $\E\left[\mathcal A_{t_i}\right]=vt_i$.
\end{lemma}

\begin{proof}
Condition on demand within the window and average over its initial
inventory and delivery schedule. Independent Poisson increments keep
the averaged lost-demand control adapted. At fixed model parameters,
time averaging gives a deterministic initial state, a constant mean
arrival rate up to a uniformly vanishing error, and grid means tending
to a common value. \Cref{app:averaging} removes the arrival-profile error
and keeps the varying targets in the fixed-grid recursion. Continuity of
that recursion in the targets and input law permits the diffusion passage
by the Poisson \emph{functional central limit theorem} (CLT).
\end{proof}

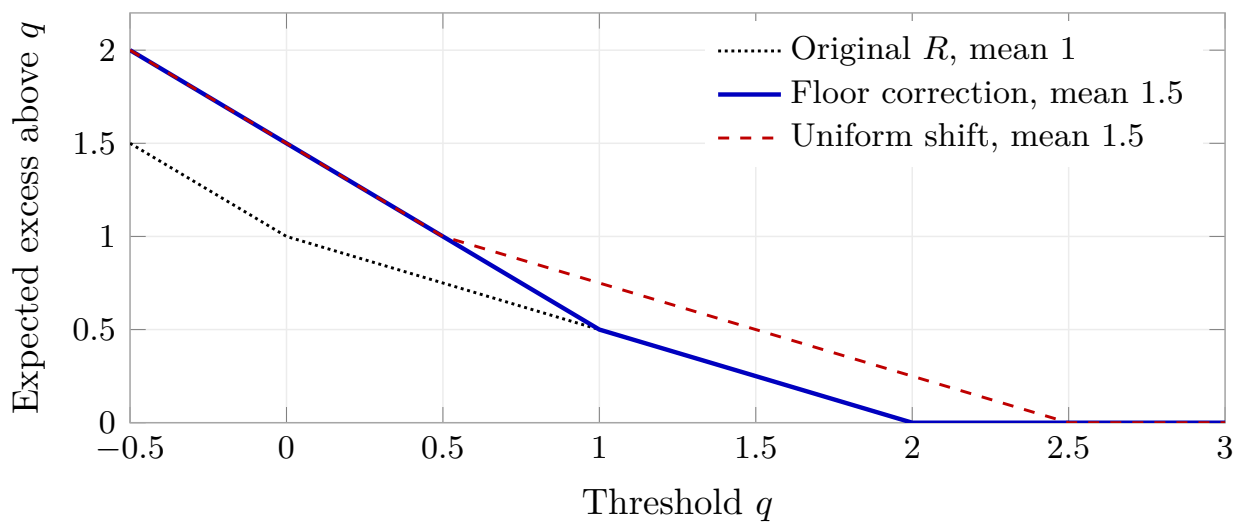
\begin{figure}[htbp]
\centering
\begin{adjustbox}{max width=.5\linewidth}
\begin{tikzpicture}
\begin{axis}[width=.73\linewidth,height=5.5cm,xlabel={Threshold $q$},
 ylabel={Expected excess above $q$},xmin=-.5,xmax=3,ymin=0,ymax=2.2,
 legend style={at={(.98,.98)},anchor=north east},legend cell align=left]
\addplot[black,densely dotted,thick] table[x=q,y=R,col sep=comma,row sep=newline]{
q,R,floor,shift
-0.5,1.5,2.0,2.0
-0.45,1.45,1.95,1.95
-0.4,1.4,1.9,1.9
-0.35,1.35,1.85,1.85
-0.3,1.3,1.8,1.8
-0.25,1.25,1.75,1.75
-0.2,1.2,1.7,1.7
-0.15,1.15,1.65,1.65
-0.1,1.1,1.6,1.6
-0.05,1.05,1.55,1.55
0.0,1.0,1.5,1.5
0.05,0.975,1.45,1.45
0.1,0.95,1.4,1.4
0.15,0.925,1.35,1.35
0.2,0.9,1.3,1.3
0.25,0.875,1.25,1.25
0.3,0.85,1.2,1.2
0.35,0.825,1.15,1.15
0.4,0.8,1.1,1.1
0.45,0.775,1.05,1.05
0.5,0.75,1.0,1.0
0.55,0.725,0.95,0.975
0.6,0.7,0.9,0.95
0.65,0.675,0.85,0.925
0.7,0.65,0.8,0.9
0.75,0.625,0.75,0.875
0.8,0.6,0.7,0.85
0.85,0.575,0.65,0.825
0.9,0.55,0.6,0.8
0.95,0.525,0.55,0.775
1.0,0.5,0.5,0.75
1.05,0.475,0.475,0.725
1.1,0.45,0.45,0.7
1.15,0.425,0.425,0.675
1.2,0.4,0.4,0.65
1.25,0.375,0.375,0.625
1.3,0.35,0.35,0.6
1.35,0.325,0.325,0.575
1.4,0.3,0.3,0.55
1.45,0.275,0.275,0.525
1.5,0.25,0.25,0.5
1.55,0.225,0.225,0.475
1.6,0.2,0.2,0.45
1.65,0.175,0.175,0.425
1.7,0.15,0.15,0.4
1.75,0.125,0.125,0.375
1.8,0.1,0.1,0.35
1.85,0.075,0.075,0.325
1.9,0.05,0.05,0.3
1.95,0.025,0.025,0.275
2.0,0.0,0.0,0.25
2.05,0.0,0.0,0.225
2.1,0.0,0.0,0.2
2.15,0.0,0.0,0.175
2.2,0.0,0.0,0.15
2.25,0.0,0.0,0.125
2.3,0.0,0.0,0.1
2.35,0.0,0.0,0.075
2.4,0.0,0.0,0.05
2.45,0.0,0.0,0.025
2.5,0.0,0.0,0.0
2.55,0.0,0.0,0.0
2.6,0.0,0.0,0.0
2.65,0.0,0.0,0.0
2.7,0.0,0.0,0.0
2.75,0.0,0.0,0.0
2.8,0.0,0.0,0.0
2.85,0.0,0.0,0.0
2.9,0.0,0.0,0.0
2.95,0.0,0.0,0.0
3.0,0.0,0.0,0.0
};\addlegendentry{Original $R$, mean $1$}
\addplot[blue!75!black,very thick] table[x=q,y=floor,col sep=comma,row sep=newline]{
q,R,floor,shift
-0.5,1.5,2.0,2.0
-0.45,1.45,1.95,1.95
-0.4,1.4,1.9,1.9
-0.35,1.35,1.85,1.85
-0.3,1.3,1.8,1.8
-0.25,1.25,1.75,1.75
-0.2,1.2,1.7,1.7
-0.15,1.15,1.65,1.65
-0.1,1.1,1.6,1.6
-0.05,1.05,1.55,1.55
0.0,1.0,1.5,1.5
0.05,0.975,1.45,1.45
0.1,0.95,1.4,1.4
0.15,0.925,1.35,1.35
0.2,0.9,1.3,1.3
0.25,0.875,1.25,1.25
0.3,0.85,1.2,1.2
0.35,0.825,1.15,1.15
0.4,0.8,1.1,1.1
0.45,0.775,1.05,1.05
0.5,0.75,1.0,1.0
0.55,0.725,0.95,0.975
0.6,0.7,0.9,0.95
0.65,0.675,0.85,0.925
0.7,0.65,0.8,0.9
0.75,0.625,0.75,0.875
0.8,0.6,0.7,0.85
0.85,0.575,0.65,0.825
0.9,0.55,0.6,0.8
0.95,0.525,0.55,0.775
1.0,0.5,0.5,0.75
1.05,0.475,0.475,0.725
1.1,0.45,0.45,0.7
1.15,0.425,0.425,0.675
1.2,0.4,0.4,0.65
1.25,0.375,0.375,0.625
1.3,0.35,0.35,0.6
1.35,0.325,0.325,0.575
1.4,0.3,0.3,0.55
1.45,0.275,0.275,0.525
1.5,0.25,0.25,0.5
1.55,0.225,0.225,0.475
1.6,0.2,0.2,0.45
1.65,0.175,0.175,0.425
1.7,0.15,0.15,0.4
1.75,0.125,0.125,0.375
1.8,0.1,0.1,0.35
1.85,0.075,0.075,0.325
1.9,0.05,0.05,0.3
1.95,0.025,0.025,0.275
2.0,0.0,0.0,0.25
2.05,0.0,0.0,0.225
2.1,0.0,0.0,0.2
2.15,0.0,0.0,0.175
2.2,0.0,0.0,0.15
2.25,0.0,0.0,0.125
2.3,0.0,0.0,0.1
2.35,0.0,0.0,0.075
2.4,0.0,0.0,0.05
2.45,0.0,0.0,0.025
2.5,0.0,0.0,0.0
2.55,0.0,0.0,0.0
2.6,0.0,0.0,0.0
2.65,0.0,0.0,0.0
2.7,0.0,0.0,0.0
2.75,0.0,0.0,0.0
2.8,0.0,0.0,0.0
2.85,0.0,0.0,0.0
2.9,0.0,0.0,0.0
2.95,0.0,0.0,0.0
3.0,0.0,0.0,0.0
};\addlegendentry{Floor correction, mean $1.5$}
\addplot[red!75!black,dashed,thick] table[x=q,y=shift,col sep=comma,row sep=newline]{
q,R,floor,shift
-0.5,1.5,2.0,2.0
-0.45,1.45,1.95,1.95
-0.4,1.4,1.9,1.9
-0.35,1.35,1.85,1.85
-0.3,1.3,1.8,1.8
-0.25,1.25,1.75,1.75
-0.2,1.2,1.7,1.7
-0.15,1.15,1.65,1.65
-0.1,1.1,1.6,1.6
-0.05,1.05,1.55,1.55
0.0,1.0,1.5,1.5
0.05,0.975,1.45,1.45
0.1,0.95,1.4,1.4
0.15,0.925,1.35,1.35
0.2,0.9,1.3,1.3
0.25,0.875,1.25,1.25
0.3,0.85,1.2,1.2
0.35,0.825,1.15,1.15
0.4,0.8,1.1,1.1
0.45,0.775,1.05,1.05
0.5,0.75,1.0,1.0
0.55,0.725,0.95,0.975
0.6,0.7,0.9,0.95
0.65,0.675,0.85,0.925
0.7,0.65,0.8,0.9
0.75,0.625,0.75,0.875
0.8,0.6,0.7,0.85
0.85,0.575,0.65,0.825
0.9,0.55,0.6,0.8
0.95,0.525,0.55,0.775
1.0,0.5,0.5,0.75
1.05,0.475,0.475,0.725
1.1,0.45,0.45,0.7
1.15,0.425,0.425,0.675
1.2,0.4,0.4,0.65
1.25,0.375,0.375,0.625
1.3,0.35,0.35,0.6
1.35,0.325,0.325,0.575
1.4,0.3,0.3,0.55
1.45,0.275,0.275,0.525
1.5,0.25,0.25,0.5
1.55,0.225,0.225,0.475
1.6,0.2,0.2,0.45
1.65,0.175,0.175,0.425
1.7,0.15,0.15,0.4
1.75,0.125,0.125,0.375
1.8,0.1,0.1,0.35
1.85,0.075,0.075,0.325
1.9,0.05,0.05,0.3
1.95,0.025,0.025,0.275
2.0,0.0,0.0,0.25
2.05,0.0,0.0,0.225
2.1,0.0,0.0,0.2
2.15,0.0,0.0,0.175
2.2,0.0,0.0,0.15
2.25,0.0,0.0,0.125
2.3,0.0,0.0,0.1
2.35,0.0,0.0,0.075
2.4,0.0,0.0,0.05
2.45,0.0,0.0,0.025
2.5,0.0,0.0,0.0
2.55,0.0,0.0,0.0
2.6,0.0,0.0,0.0
2.65,0.0,0.0,0.0
2.7,0.0,0.0,0.0
2.75,0.0,0.0,0.0
2.8,0.0,0.0,0.0
2.85,0.0,0.0,0.0
2.9,0.0,0.0,0.0
2.95,0.0,0.0,0.0
3.0,0.0,0.0,0.0
};\addlegendentry{Uniform shift, mean $1.5$}
\end{axis}
\end{tikzpicture}
\end{adjustbox}
\caption{A two-point illustration of \cref{lem:floor}. The original state is $0$ or $2$
with equal probabilities. Raising its lower value to $1$ restores mean $1.5$ and has
smaller call expectations than adding $0.5$ to both values. The blue curve is the
maximum of $1.5-q$ and the original call curve. }
\label{fig:floor}
\end{figure}

Minimal reflection can leave the endpoint mean below its target. If $R$
equals $0$ or $2$ with probability $1/2$ each, adding $1/2$ to each outcome
gives mean $3/2$ and variance $1$. Raising only the zero outcome to $1$
gives the same mean and variance $1/4$. \Cref{fig:floor} compares the
call functions of the two corrected inventories.

\begin{lemma}
\label{lem:floor}
Let $R$ be integrable and $\E\left[R\right]\le m$. If $\E\left[R\right]<m$, there is a unique
effective \emph{common floor} $k$ such that $Y=\max(k,R)$ has mean $m$. If $\E\left[R\right]=m$,
take $Y=R$. For every integrable $X\ge R$ with $\E\left[X\right]=m$,
$Y\le_{\rm cx}X$. More explicitly, for every $q\in\R$,
\begin{equation*}
 \E\left[(Y-q)^+\right]=\max\{m-q,\E\left[(R-q)^+\right]\}.
\end{equation*}
The correction is continuous in $L^1$ under $L^1$ convergence of $R$ and
convergence of the feasible target means.
\end{lemma}

\begin{proof}
The function $k\mapsto\E\left[\max(k,R)\right]$ is continuous, tends to $\E\left[R\right]$ at
$-\infty$, and tends to infinity at $+\infty$. 
Above the essential lower
endpoint it is strictly increasing wherever its value exceeds $\E\left[R\right]$.
If $\E\left[R\right]=m$, Jensen's inequality gives
$\E\left[(R-q)^+\right]\ge m-q$, so $Y=R$ satisfies the call identity.
If $\E\left[R\right]<m$, the call of $Y$ equals $m-q$ for $q<k$ and
$\E\left[(R-q)^+\right]$ for $q\ge k$. In each case this is the larger of
the two terms in the call identity.
Every $X$ in the statement has calls at least both terms, by Jensen and
$X\ge R$.
Equal means and the call characterization of convex order prove
the claim; the same characterization follows by integrating the second
derivative measure of a convex piecewise-linear approximation.

If the limiting mean exceeds $\E\left[R\right]$, the unique finite root is continuous
by monotonicity and convergence of the mean functions, which are
$1$-Lipschitz in the underlying random variable. If the limiting mean equals
$\E\left[R\right]$, the expected added mass is $m_n-\E\left[R_n\right]\to0$, so
$\E\left[|Y_n-R_n|\right]\to0$. This also covers floors tending to $-\infty$.
\end{proof}

At each grid date, test the reflected endpoint against the prescribed mean.
If the mean is feasible, use the common floor to obtain the starting law
for the next block.

\begin{theorem}
\label{thm:grid}
For $u,v\ge0$, set $Y_0=u$ and $\Delta_i=t_i-t_{i-1}$. Using independent Brownian blocks,
form $R_i=\Gamma_{\Delta_i}(Y_{i-1},W^{(i)};v)$. 
If $\E\left[R_i\right]>u$ at any
step, declare $(u,v)$ infeasible. If $\E\left[R_i\right]=u$, set $Y_i=R_i$.
If $\E\left[R_i\right]<u$, use \cref{lem:floor} to set
$Y_i=\max(k_i,R_i)$ with mean $u$. This test is necessary and sufficient
for feasibility of the mean-calibrated Brownian relaxation.
At each date, $Y_i$ is least in convex
order among feasible states. Grid refinement can only remove feasible pairs.
\end{theorem}

\begin{proof}
Given a feasible process, its state after a block dominates minimal reflection
from its previous state. If the latter dominates $Y_{i-1}$ in convex order,
then independent input and the increasing convex reflection map imply
$R_i\le_{\rm icx}X_{t_i}$ before correction. In particular $\E\left[R_i\right]\le u$.
By the call identity, restoring the smaller state to mean $u$ preserves
the call inequalities, and gives $Y_i\le_{\rm cx}X_{t_i}$. Induction proves
necessity and minimality. Conversely, within each block apply minimal
reflection, then add the floor correction at its end. This is adapted,
nondecreasing control, is nonnegative throughout each block, and has the
specified means. Refinement adds mean constraints, giving the last assertion.
\end{proof}

Let $\mathcal T=\{t_0,\ldots,t_K\}$ denote the grid, and let
$D_{\mathcal T}(\alpha)$ be the infimum of
$\sqrt\alpha\,u+v/\sqrt\alpha$ over grid-feasible pairs. By
\cref{lem:reduction}, it lower-bounds the limiting normalized optimal cost.
It is at most $\sqrt2$: a \emph{stationary} reflected Brownian motion, whose marginal law is constant over time, with drift
$-v$ has mean $u=1/(2v)$. Average its independent stationary initial state
conditional on the Brownian path. Conditional averaging preserves nonnegativity,
adaptation, and every mean, while the initial state becomes deterministic $u$.
Minimizing $\sqrt\alpha/(2v)+v/\sqrt\alpha$ gives $\sqrt2$.
The averaged control need not satisfy a contact condition at zero; the relaxation
requires no such condition.

At a positive pair on the boundary of the endpoint constraint,
restoring a suitable intermediate mean uses an expected correction of
$e>0$. It reduces the expected regulator in the next block by strictly
less than $e$.

\begin{lemma}
\label{lem:strict}
Suppose $u,v>0$ and $\E\left[\Gamma_1(u,W;v)\right]=u$. Some two-date grid
$\{0,t,1\}$ excludes $(u,v)$. For every fixed $\alpha>0$, a finite dyadic
refinement has a strictly larger minimum cost than the endpoint-only relaxation.
\end{lemma}

\begin{proof}
For small $t>0$, writing $R_t=\Gamma_t(u,W;v)$ gives
$\E\left[R_t\right]=u-vt+o(t)<u$, since a path starting strictly above zero hits the
boundary with probability tending rapidly to zero. Put $e=u-\E\left[R_t\right]>0$.
The reflection semigroup and endpoint equality imply
$\E\left[g_{1-t}(R_t,v)\right]=v(1-t)+e$. Raising $R_t$ to its floor-corrected law
adds mean $e$. For finite $z>0$, $-1<\partial_z g_{1-t}(z,v)<0$;
therefore it reduces that future regulator expectation by strictly less than
$e$. The repaired state needs more than the remaining budget $v(1-t)$,
contradicting feasibility.

The endpoint feasible set is closed. Its bounded cost sublevels are compact
and stay away from the axes: for $u=0$ or $v=0$ the reflected endpoint has
mean strictly greater than $u$. Every minimum lies on the endpoint boundary,
since a strict slack permits a small decrease in $u$. A neighborhood of each
minimum is excluded by the preceding strict violation; continuity permits a
dyadic choice of $t$. Compactness gives a finite cover, and a common dyadic
refinement excludes it. The refined feasible set is closed and nonempty by
the preceding stationary comparison, so its minimum is strictly larger.
\end{proof}

\subsection{Supporting inequalities and the uniform factor}
\label{sec:lost-sales-certificate}

Use the grid $\{0,1/2,1\}$. Given a feasible midpoint mean $u$, use the
common floor and test whether the expected regulator on the second block
is at most $v/2$.
The resulting supporting inequalities give a factor below $5/4$.

Put $t=\Delta=1/2$ and $R_t=\Gamma_t(u,W;v)$.
For $z\ge0$ the reflected midpoint CDF and
\emph{future-maximum tail} are
\begin{align*}
 F(z;u,v)&=\Phi\!\left(\frac{z-u+vt}{\sqrt t}\right)
       -e^{-2vz}\Phi\!\left(\frac{-z-u+vt}{\sqrt t}\right),\\
 M(z;v)&=\Phi\!\left(\frac{v\Delta-z}{\sqrt\Delta}\right)
       +e^{2vz}\Phi\!\left(\frac{-z-v\Delta}{\sqrt\Delta}\right).
\end{align*}
To verify the first formula, reverse Brownian increments within the block.
The endpoint has the law of
$\max\{u+W_t-vt,\sup_{s\le t}(W_s-vs)\}$.
For $x\le z$, reflection at level $z$ and Gaussian change of drift give the
killed endpoint density
\[
 \frac1{\sqrt t}\phi\!\left(\frac{x+vt}{\sqrt t}\right)
 -\frac{e^{-2vz}}{\sqrt t}\phi\!\left(\frac{x-2z+vt}{\sqrt t}\right).
\]
Integrating to $x=z-u$ gives $F$. Integrating the corresponding density
with drift $+v$ to its barrier $z$ and taking the complement gives $M$.
Both formulas include their zero-drift limits. Provided $\E\left[R_t\right]\le u$,
the first floor solves
$\E\left[R_t\right]+\int_0^kF(z;u,v)\,dz=u$, and the second feasibility condition is
$\int_k^\infty M(z;v)F(z;u,v)\,dz\le v\Delta$.

A feasible midpoint state $X_t$ satisfies $\E\left[X_t\right]=u$ and
$\E\left[g_\Delta(X_t,v)\right]\le v\Delta$. Weight the mean constraint
by a fixed \emph{multiplier} to obtain a convex necessary condition in
$(u,v)$.
Choose any positive trial $(u_0,v_0,k_0)$ and set
$\theta=M(k_0;v_0)\in(0,1)$. Keep $\theta$ fixed as $(u,v)$ varies. Define
\begin{equation}
 \mathcal W_\theta(u,v)=
 \E\left[\inf_{a\ge\Gamma_t(u,W;v)}\{g_\Delta(a,v)+\theta a\}\right]
       -\theta u-v\Delta.
 \label{eq:convex-dual}
\end{equation}

\begin{lemma}
\label{lem:support}
Every feasible midpoint pair satisfies $\mathcal W_\theta(u,v)\le0$.
This function is jointly convex. At a differentiability point let
$A=-\partial_u\mathcal W_\theta$, $C=-\partial_v\mathcal W_\theta$, and
$K=\mathcal W_\theta+Au+Cv$. Then
$Au+Cv\ge K$ is necessary for feasibility. The trial need not be feasible.
\end{lemma}

\begin{proof}
For a feasible midpoint state $X_t$, we have $X_t\ge\Gamma_t(u,W;v)$,
$\E\left[X_t\right]=u$, and $\E\left[g_\Delta(X_t,v)\right]\le v\Delta$. Substituting $a=X_t$
in the infimum proves the first assertion. The regulator cost is jointly convex
as an expected maximum of affine functions. Partial minimization over $a\ge r$
is jointly convex in $(r,v)$ and nondecreasing in $r$. Composition with the
jointly convex reflection endpoint preserves convexity. A tangent is below
$\mathcal W_\theta$, so it is nonpositive at every feasible pair, proving
the necessary halfspace.
\end{proof}

At the trial, the minimizing state is $\max(k_0,R_t)$ because
$\partial_a g_\Delta=-M$ and $M(k_0;v_0)=\theta$. Let $\bar F=1-F$ and suppress
the trial subscripts. With $a_z=(z-u+vt)/\sqrt t$,
$p_z=(-z-u+vt)/\sqrt t$, and $e_z=(-z-v\Delta)/\sqrt\Delta$, put
\[
 \bar F_u=\frac{\phi(a_z)-e^{-2vz}\phi(p_z)}{\sqrt t},\quad
 F_v=t\bar F_u+2ze^{-2vz}\Phi(p_z),\quad
 M_v=2ze^{2vz}\Phi(e_z).
\]
The coefficients are
\begin{align*}
 \mathcal W_\theta(u,v)&=\theta(k-u)+\int_k^\infty(\theta\bar F+MF)\,dz-v\Delta,\\
 A&=\theta-\int_k^\infty(\theta-M)\bar F_u\,dz,\\
 C&=\Delta-\int_k^\infty\left(M_vF-(\theta-M)F_v\right)\,dz,
 \qquad K=\mathcal W_\theta(u,v)+Au+Cv.
\end{align*}
For fixed $\theta$, the future maximum has a continuous positive density on
$(0,\infty)$, so the local cutoff is unique and varies continuously with $v$.
The midpoint has a continuous law. The integrands and their local parameter
derivatives are bounded by integrable Gaussian tails. Differentiation under the
integrals is therefore valid. With $\theta$ fixed, the boundary terms
from $\theta k$ and the integral starting at $k$ cancel because
$M(k)=\theta$. The derivatives give the tangent for the same fixed
multiplier in \cref{lem:support}.

Let
$
 \mathcal P=\{(u,v)\in\mathbb R_{\ge0}^2:
             A_i u+C_i v\ge K_i,\ i=0,\ldots,11\},
$
where \Cref{tab:planes} gives the rounded coefficients. Every limiting
mean pair lies in $\mathcal P$. Thus the \emph{linear program} with value
$L_{\rm LP}(\alpha):=\min_{(u,v)\in\mathcal P}(\sqrt\alpha\,u+v/\sqrt\alpha)$
lower-bounds the limiting normalized optimal cost. Its minimum is attained
at a vertex. To certify the cost quotient on a cell, \cref{lem:ls-certificate}
checks every vertex at both cell endpoints.

\begin{table}[htbp]
\centering\small
\caption{Printed necessary inequalities $A_i u+C_i v\ge K_i$ for the two-block relaxation. Coefficients are exact decimal rationals; $A_i,C_i$ are rounded upward and $K_i$ downward. Trial coordinates and integration errors are in \Cref{app:certificates}.}
\label{tab:planes}
\begin{adjustbox}{max width=.5\linewidth}
\begin{tabular}{@{}rrrr@{}}
\toprule
Index & $A_i$ & $C_i$ & $K_i$ \\\midrule
0 & 0.001114010228 & 0.556959770364 & 0.003904304426 \\
1 & 0.002848542103 & 0.569684220377 & 0.009232773469 \\
2 & 0.005807962486 & 0.580781376253 & 0.017582056348 \\
3 & 0.011851505119 & 0.592566085591 & 0.033164050179 \\
4 & 0.024137117742 & 0.603422374085 & 0.061600772897 \\
5 & 0.048773185288 & 0.609661627279 & 0.111528142283 \\
6 & 0.090823406828 & 0.605487652789 & 0.184262518259 \\
7 & 0.147420086572 & 0.589679512796 & 0.266646826737 \\
8 & 0.224450161710 & 0.561125127264 & 0.359373948280 \\
9 & 0.293055911851 & 0.532829102188 & 0.427869548807 \\
10 & 0.334733688784 & 0.514975386930 & 0.464141261235 \\
11 & 0.353650779651 & 0.506758108227 & 0.479403324607 \\
\bottomrule
\end{tabular}
\end{adjustbox}
\end{table}

\begin{lemma}
\label{lem:ls-certificate}
The twelve rational necessary halfspaces in \Cref{tab:planes} have thirteen
vertices in the nonnegative quadrant. Let
$L_{\rm LP}(\alpha)=\min_{(u,v)\text{ vertex}}(\sqrt\alpha\,u+v/\sqrt\alpha)$,
$\alpha_s=1/500$, and $\alpha_0=69787/100000$. Then
\[
 \sup_{\alpha_s\le\alpha\le\alpha_0}
       \frac{\U(\alpha)}{L_{\rm LP}(\alpha)}<1.248976725, ~~
 \sup_{0<\alpha\le\alpha_s}\frac{\U(\alpha)}{\V(\alpha)}
       <1.231626783, ~~
 \frac{\sqrt2}{L_{\rm LP}(\alpha_0)}<1.248825987.
\]
\end{lemma}

\begin{proof}
\Cref{app:certificates} specifies all inputs and error bounds. Coefficients
$A,C$ are rounded upward and $K$ downward; since $u,v\ge0$ this only weakens
the necessary halfspace. Positive objectives attain their minima at vertices
of the resulting nonempty polyhedron. For a rational BS trial $\beta$, its
cost divided by the cost of a vertex is
\begin{equation*}
 \frac{(\beta+\psi(-\beta))\alpha+\psi(-\beta)}{u\alpha+v}.
\end{equation*}
This is fractional linear with a positive denominator. Its maximum on a cell
is at an endpoint. The one hundred cells and trials in
\Cref{tab:bs-trials} therefore prove the first assertion by rational interval
evaluation at both endpoints for all thirteen vertices.
For the small region use monotonicity of $\U/\V$ on $(0,\alpha_0]$,
recorded in \Cref{app:normal}, and trial $\beta=3.629625443$ at $\alpha_s$.
With the coefficients rounded to twelve decimal places, the vertex
checks give upper bounds below $1.248976724300$ on the compact region
and $1.248825986315$ on the tail.
\end{proof}

\begin{proof}[Proof of the lost-sales guarantee in \Cref{thm:main}]
For each fixed $\alpha$, \cref{lem:reduction,lem:support} give the normalized
optimal-cost lower bound $L_{\rm LP}(\alpha)$, while
\cref{lem:policy-facts} gives upper bound $\min\{\U(\alpha),\sqrt2\}$.
Use the first two parts of \cref{lem:ls-certificate} on
$(0,\alpha_0]$. For $\alpha\ge\alpha_0$, compare the actual optimal cost
to that with lead time $\alpha_0 p$ and apply the CO bound $\sqrt2$.
This uses monotonicity of the physical optimum; it does not assume monotonicity
of $L_{\rm LP}$. All three bounds are strictly below $r_{\rm LS}$.
\end{proof}

\paragraph{The bound at the cutoff.}
At $\alpha_0=0.69787$, the BS upper bound is $\U(\alpha_0)\simeq1.414212$.
Dividing by the newsvendor bound $\V(\alpha_0)\simeq0.790576$ gives
approximately $1.788838$. The twelve inequalities give
$L_{\rm LP}(\alpha_0)\simeq1.132434$ and a ratio of approximately $1.248825$.
The uniform certificate also covers the rest of the final compact cell.

\section{The dual-sourcing approximation}
\label{sec:dual-sourcing-analysis}

We keep separate charges for on-hand inventory, backlog, and express
orders. The lower bound must control the backlog at the date selected by
the demand path. For large backorder penalties, signed mean calibration
relates the resulting costs to the lost-sales bound.

\subsection{Conditional backlog charges}
\label{sec:dual-sourcing}

A date chosen from the demand path may depend on the same demand that
created its backlog. Suppose the selection event $E$ is contained in an
event $F$ determined by subsequent demand. Independent increments make
the backlog $B_0\ge0$ independent of $F$. Hence
\[
\E\left[B_0\mathbf1_E\right]
\le\E\left[B_0\mathbf1_F\right]=\Prb(F)\E\left[B_0\right].
\]
For independent standard normals $Z_1,Z_2$, take
$B_0=2\mathbf1_{\{Z_2<0\}}$ and $E=\{Z_2<0,Z_1>0\}$.
Then $\E\left[B_0\right]=1$, $\Prb(E)=1/4$, and
$\E\left[B_0\mathbf1_E\right]=1/2$. Taking $F=\{Z_1>0\}$ gives the
coefficient $1/2$.

\begin{figure}[htbp]
\centering
\begin{adjustbox}{max width=.6\linewidth}
\begin{tikzpicture}[>=Latex,font=\small]
\coordinate (A) at (0,0);
\coordinate (Mid) at (4.6,0);
\coordinate (T) at (10.5,0);
\draw[->,thick] (-.2,0)--(10.9,0);
\draw[gray!65,very thick] (0,1.5)--(10.5,1.5);
\draw[gray!65] (0,1.35)--(0,1.65) (10.5,1.35)--(10.5,1.65);
\node[above] at (5.25,1.5) {Demand available to the selection rule $E_j^{(c)}$};
\draw[green!45!black,very thick] (4.6,.65)--(10.5,.65);
\draw[green!45!black] (4.6,.5)--(4.6,.8) (10.5,.5)--(10.5,.8);
\node[above,text=green!35!black] at (7.55,.65) {Demand defining $F_j^{(c)}$};
\foreach \P in {A,Mid,T}{\fill (\P) circle (1.5pt);}
\node[below=5pt] at (A) {$t-\delta c$};
\node[below=5pt] at (Mid) {$t-t_jc$};
\node[below=5pt] at (T) {$t$};
\node[anchor=north,align=center,text width=4.5cm] at (0,-.7)
 {Regular arrivals\\are fixed for the full window};
\node[anchor=north,align=center,text width=5.6cm] at (4.6,-.7)
 {Interior backlog\\$I(t-t_jc)^-$ is determined};
\node[anchor=north,align=center,text width=4.2cm] at (10.5,-.7)
 {Terminal inventory comparison};
\end{tikzpicture}
\end{adjustbox}
\caption{Information used in the conditional backlog charge for an interior
branch $j<J$. The finite selection event $E_j^{(c)}$ may use demand over the full window.
Its containing event $F_j^{(c)}$ uses only the subsequent suffix and is
independent of the interior backlog. Its probability gives the limiting
coefficient $\Prb(F_j)$ in \Cref{lem:conditional}.}
\label{fig:conditional-information}
\end{figure}
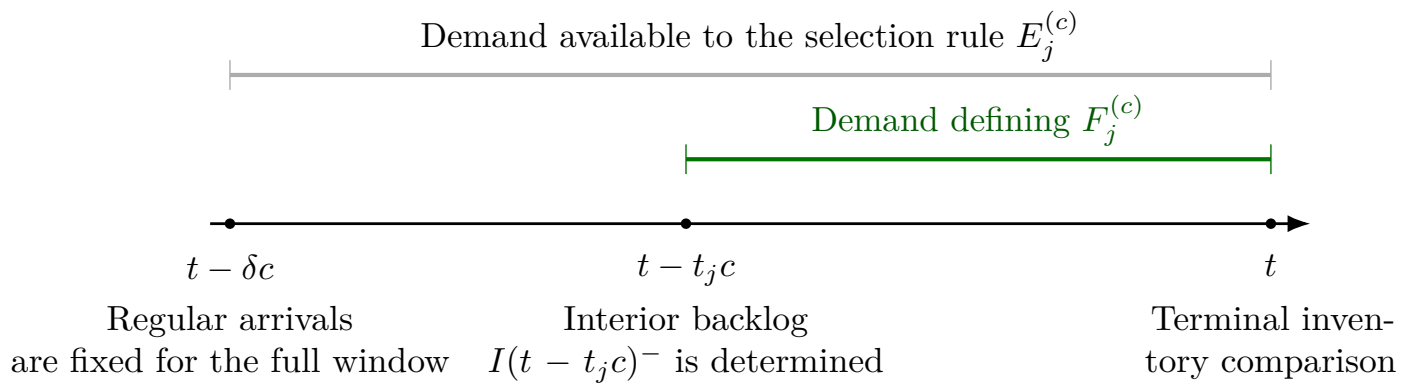

Normalize dual-sourcing holding, backlog, and express rate by
$x=H/\sqrt{\lambda c}$, $z=\bar B/\sqrt{\lambda c}$, and
$y=d\sqrt{c/\lambda}$. The quantity $x+pz+y$ is a lower comparison for normalized cost.
Fix comparison dates $0<t_1<\cdots<t_J=\delta\le\alpha$ and a finite set
$\mathcal S\subset(0,\delta]$ containing these dates. At terminal time $t$,
write
\[
 W_s^{(c)}(t)=\frac{\lambda sc-D(t-sc,t]}{\sqrt{\lambda c}},
 \qquad s\in\mathcal S,
\]
for the standardized demand deficits. Let $W$ be standard Brownian motion.
Choose disjoint Borel sets $\mathsf E_1,\ldots,\mathsf E_J$ in
$\R^{\mathcal S}$. A vector outside their union selects no branch. For
$j<J$, choose a Borel set $\mathsf F_j\supseteq\mathsf E_j$ whose membership
uses only coordinates $s\le t_j$. Let $E_j^{(c)}$ and $F_j^{(c)}$ be the
corresponding events for the vector $\bigl(W_s^{(c)}(t)\bigr)_{s\in\mathcal S}$.
Use $E_j$ and $F_j$ for the events obtained from the same sets with
$(W_s)_{s\in\mathcal S}$.
\Cref{fig:conditional-information} shows this information restriction.
Write $\mathcal E=(E_1,\ldots,E_J)$ for the event rule and define
\begin{equation*}
 p_j=\Prb(E_j),\quad a_{\mathcal E}=1-p_J,\quad d_{\mathcal E}=\sum_jt_jp_j,
 \quad m_{\mathcal E}=\E\left[\sum_jW_{t_j}\mathbf1_{E_j}\right],\quad
 \chi=p_J+\sum_{j<J}\Prb(F_j).
\end{equation*}
The balance inequality weights positive inventory, express flow, and
backlog by $a_{\mathcal E}$, $d_{\mathcal E}$, and $\chi$, respectively.
The quantity $m_{\mathcal E}$ is the expected demand deficit at the
selected date, with zero contribution when no date is selected.

\begin{lemma}
\label{lem:conditional}
Suppose the boundaries of all $\mathsf E_j$ and $\mathsf F_j$ have
probability zero under the law of $(W_s)_{s\in\mathcal S}$. Every limiting normalized mean triple of causal dual-sourcing policies with
uniformly bounded normalized costs satisfies
$a_{\mathcal E}x+d_{\mathcal E}y+\chi z\ge m_{\mathcal E}$.
Consequently, if $p\ge p_{\min}>0$ and $m_{\mathcal E}>0$, normalized optimal cost has
liminf at least
$\frac{m_{\mathcal E}}{\max\{a_{\mathcal E},d_{\mathcal E},\chi/p_{\min}\}}$.
This comparison requires $\delta\le\alpha$, but not $\delta p\ge1$.
\end{lemma}

The denominator is the largest amount the inequality credits to one unit
of total cost. Dividing the selected deficit moment by that coefficient
therefore gives a valid cost lower bound.

\begin{proof}
Let $A_R(s,t]$ count regular arrivals in $(s,t]$.
All regular arrivals over a window of length $\delta c\le L$ are measurable
at its start. 
Dropping nonnegative express arrivals in the inventory balance
gives, for the end branch and an interior suffix respectively,
\begin{align*}
 I(t)^+&\ge I(t-\delta c)+A_R(t-\delta c,t]-D(t-\delta c,t],\\
 I(t)^+&\ge A_R(t-t_jc,t]-D(t-t_jc,t]-I(t-t_jc)^-.
\end{align*}
Multiply these inequalities by the disjoint finite selection indicators
$\mathbf1_{E_j^{(c)}}$ and sum. Initial net inventory and regular arrivals
factor against these indicators because they are measurable at the start of
the full window. For $j<J$, containment and independent increments give
\[
 \E\left[I(t-t_jc)^-\mathbf1_{E_j^{(c)}}\right]
 \le\E\left[I(t-t_jc)^-\mathbf1_{F_j^{(c)}}\right]
 =\Prb(F_j^{(c)})\E\left[I(t-t_jc)^-\right].
\]
Write $p_j^{(c)}=\Prb(E_j^{(c)})$. Time averaging as in \cref{app:averaging}
uses regular mean arrival rate $\lambda-d$ and removes all shifts and
endpoints. Centering the demand terms gives
\begin{align*}
 (1-p_J^{(c)})H + dc\sum_j t_j p_j^{(c)}
  +\left(p_J^{(c)}+\sum_{j<J}\Prb(F_j^{(c)})\right)\bar B
 &\ge \sqrt{\lambda c}\,\E\left[\sum_j W_{t_j}^{(c)}(t)
                    \mathbf1_{E_j^{(c)}}\right].
\end{align*}
Divide by $\sqrt{\lambda c}$ and apply the finite-vector Poisson CLT.
The boundary assumption gives convergence of the event probabilities.
Since $\E[(W_{t_j}^{(c)}(t))^2]=t_j$, uniform integrability gives
convergence of the selected moments as well. Finally,
$a_{\mathcal E}x+d_{\mathcal E}y+\chi z\le\max\{a_{\mathcal E},d_{\mathcal E},\chi/p\}(x+y+pz)$.
\end{proof}

\subsubsection{Two dates}
\label{subsec:quarter}

Take $t_1=\delta/4$ and $t_2=\delta$. For rational selector parameters
$\xi,\zeta>0$, put $s_1=\sqrt\delta/2$, $h_0=s_1\xi$, and
$y_0=\zeta/s_1$. 
Select the largest of
$0$, $W_{\delta/4}-y_0\delta/4$, and $h_0+W_\delta-y_0\delta$.
Break ties toward the earliest branch, including zero. Put
$Z_1=W_{\delta/4}/s_1$ and
$Z_2=(W_\delta-W_{\delta/4})/(\sqrt3\,s_1)$. These are independent standard
normals, and the same transformations of the finite deficits define the
Poisson rule. The interior event is $\{Z_1>\zeta,Z_2\le e\}$, where
$e=(3\zeta-\xi)/\sqrt3$. Only $Z_1$ uses demand after the interior backlog
is determined, so $F_1=\{Z_1\ge\zeta\}$ is valid.

For this rule, write $q=p_1$ for the interior-branch probability and
$\varpi=p_2$ for the end-branch probability. The statistics needed by
\cref{lem:conditional} are
\begin{align*}
 q&=\Q(\zeta)\Phi(e),\qquad
 \varpi=\int_{3\zeta-\xi}^{\infty}
        \frac{\phi(s/\sqrt3)}{\sqrt3}\Phi(s-4\zeta+\xi)\,ds,\\
 \frac{m_{\mathcal E}}{\sqrt\delta}
 &=\phi((4\zeta-\xi)/2)\Phi(\xi/(2\sqrt3))
       +\frac{\sqrt3}{2}\phi(e)\Q(\zeta)
       +\frac12\phi(\zeta)\Phi(e),\\
 a_{\mathcal E}&=1-\varpi,\qquad d_{\mathcal E}=\delta(\varpi+q/4),\qquad
 \chi=\varpi+\Q(\zeta).
\end{align*}
The end event is
$\{Z_2>e,\ s_1Z_1+\sqrt{3\delta/4}Z_2>y_0\delta-h_0\}$.
Integrating out $Z_1$ and substituting $s=\sqrt3 Z_2$ gives $\varpi$.
Put $s_2=\sqrt3s_1$, $k_G=4\zeta-\xi$, and $A(t)=k_G-\sqrt3t$.
Since $A(e)=\zeta$, the combined interior and end moments equal
\[
 m_{\mathcal E}=s_1\phi(\zeta)\Phi(e)+
 \int_e^\infty\left(s_1\phi(A(t))+s_2t\Q(A(t))\right)\phi(t)\,dt.
\]
Integration by parts using $t\phi(t)=-\phi'(t)$ gives
\[
 m_{\mathcal E}=s_1\phi(\zeta)\Phi(e)+4s_1 I_G+s_2\phi(e)\Q(\zeta),\qquad
 I_G=\int_e^\infty\phi(t)\phi(k_G-\sqrt3t)\,dt.
\]
Completing the square yields
$I_G=\tfrac12\phi(k_G/2)\Phi(\xi/(2\sqrt3))$.
Dividing by $\sqrt\delta$ gives the stated moment. At the interior
date, only $Z_1$ describes subsequent demand. Thus $F_1=\{Z_1\ge\zeta\}$
gives the coefficient $\Q(\zeta)$. The selection probability
$q=\Q(\zeta)\Phi(e)$ also uses $Z_2$.

\paragraph{A numerical example of the backlog charge.}
For $\delta=0.69787$, $\xi=3.199890$, $\zeta=0.667859$, and $p=2$,
the rule gives $a_{\mathcal E}\simeq d_{\mathcal E}\simeq0.41737$,
$\chi\simeq0.83474$, and $m_{\mathcal E}\simeq0.35458$. The terms
$a_{\mathcal E}$, $d_{\mathcal E}$, and $\chi/p$ are nearly equal,
giving a normalized lower bound of about $0.84956$ for $\alpha\ge\delta$.
Using coefficient one for the interior backlog would give $\chi=1+\varpi$
and a lower bound of about $0.44809$.

\subsubsection{Four dates and finite rectangle rules}
\label{subsec:rectangles}

Use $t_j=j\delta/4$ and independent standard normal increments $Z_1,\ldots,Z_4$.
\emph{Quantize} each increment, that is, map it to a bin center, using centers $-3,-3+\eta,\ldots,3$, with midpoint
boundaries and unbounded extreme bins. Each finite boundary belongs to the bin on its right.
We use $\eta=1/2$ or $1/4$, giving $n=13$ or $25$ bins. Write $Q_i$ for the quantized values.
Select the largest
of
\[
 0,\quad \sum_{i=1}^jQ_i-j\zeta\quad(1\le j<4),\quad
 \xi+\sum_{i=1}^4Q_i-4\zeta,
\]
breaking ties in favor of the earliest branch, including zero.
This gives disjoint demand events to which \cref{lem:conditional} applies.
At a selected interior branch, the quantized score is a \emph{strict
record}: it exceeds every earlier prefix score. This record event
supplies $F_j$.

For a bin with endpoints $l_i,r_i$ set
$P_i=\Phi(r_i)-\Phi(l_i)$ and $C_i=\phi(l_i)-\phi(r_i)$, using zero densities
at infinity. A rectangle indexed by $\boldsymbol i$ has probability
$\prod_kP_{i_k}$. If it selects branch $j$, its unscaled Gaussian moment is
$\sum_{k\le j}C_{i_k}\prod_{l\ne k}P_{i_l}$.
Let $\varpi$ be the total end-branch probability, $R=\sum_jjp_j$, $M_0$ the total
unscaled selected Gaussian moment, and $\chi=\varpi+\sum_{j<4}\Prb(F_j)$.
Then \cref{lem:conditional} gives
\begin{equation*}
 \mathcal B_{\xi,\zeta,\eta}(\delta,p_{\min})=
 \frac{\sqrt{\delta/4}\,M_0}
      {\max\{1-\varpi,\delta R/4,\chi/p_{\min}\}}.
\end{equation*}
The outer bins extend to infinity in both directions. The two rules
enumerate $13^4=28{,}561$ and $25^4=390{,}625$ rectangles, respectively.

The construction also extends to finer grids. The record estimate in
\Cref{lem:record} bounds the total backlog
coefficient by $2\sqrt J$ for $J$ equally spaced dates and nonnegative selector
drift, including symmetric finite quantizers with earliest-branch ties.

\subsubsection{Upper-policy refinements and coverage of all parameters}

In the diffusion limit, lowering the DI express target reduces holding
cost and incurs backlog when truncated demand exceeds the new target.

\begin{lemma}
\label{lem:di-shift}
Put $\kappa=14/75$. For $p\le5$ and $\alpha\le\alpha_0$, DI has
normalized upper cost at most $\U(\alpha)-\kappa\sqrt\alpha$.
\end{lemma}

\begin{proof}
Let $Z_\beta$ have law $(Z-\beta\mid Z\le\beta)$. Formula
\eqref{eq:di-exact}, with $U=\lfloor\lambda L+\beta\sqrt{\lambda L}\rfloor$
and $S=U+s\sqrt{\lambda L}+O(1)$, has normalized limit
\begin{equation*}
 \sqrt\alpha\left[\frac{\psi(-\beta)}\alpha+
        \E\left[(s-Z_\beta)^++p(Z_\beta-s)^+\right]\right].
\end{equation*}
The truncated Poisson CLT and bounded conditional second moments give the
necessary uniform integrability. For $\beta=\beta(\alpha)$ and $s=-1/3$,
the difference from $s=0$ is
$\sqrt\alpha[-1/3+(1+p)\E\left[(Z_\beta+1/3)^+\right]]$.
\Cref{app:normal} gives $\beta(\alpha)\ge\beta(\alpha^*)>0.8619684164042$,
where $\U(\alpha^*)=\sqrt2$ and $\alpha_0<\alpha^*$.
On $[-1/3,0]$, the density of $Z_\beta$ is bounded by
$\phi(\beta(\alpha^*)-1/3)/\Phi(\beta(\alpha^*))<11/25$.
Thus $\E\left[(Z_\beta+1/3)^+\right]\le11/450$. Since $1+p\le6$, the difference is
at most $-14\sqrt\alpha/75$.
\end{proof}

For small backorder penalties, TBS can carry shortages cheaply. The next
comparison covers this region using the inherited newsvendor bound, so no
additional event rule is needed there.

\begin{lemma}
\label{lem:small-backlog}
For $0<p\le1$, $F_{\rm TBS}(p)/\V(1/p)\le3/2$.
\end{lemma}

\begin{proof}
Put $q=p/(1+p)=\Q(t)$. The ratio equals
$\sqrt{2\log(1/q)}/\psi(t)$, with $\psi=\phi/\Q$.
For $t\ge0$, the normal bounds in \Cref{app:normal} give
$2\log(1/\Q(t))\le t^2+8t/5+7/5$ and
$\psi(t)\ge3t/4+1/\sqrt2$.
If $p\le1/6$, then $q\le1/7<\Q(1)$, so $t>1$. The difference
\[
 \frac94(3t/4+7/10)^2-(t^2+8t/5+7/5)
 =17t^2/64+61t/80-119/400
\]
is positive for $t\ge1$. This proves the claim in that range.
For $p\in[j/60,(j+1)/60]$, $j=10,\ldots,59$, monotonicity of $F_{\rm TBS}$ and
of $\V(1/p)$ gives the upper bound
$F_{\rm TBS}((j+1)/60)/\V(60/j)$. The fifty exact interval checks described in
\Cref{app:certificates} give a maximum below $1.475695<3/2$.
\end{proof}

For a fixed two- or four-date rule, put
$\bar d_{\mathcal E}=d_{\mathcal E}/\delta$ and
$\bar m_{\mathcal E}=m_{\mathcal E}/\sqrt\delta$.
These constants, $a_{\mathcal E}$, and $\chi$ do not depend on the window
length $\delta$. Taking $\delta=\alpha$ gives the lower comparison
\[
 \mathcal B(\alpha,p_{\min})=
 \frac{\sqrt\alpha\,\bar m_{\mathcal E}}
 {\max\{a_{\mathcal E},\alpha\bar d_{\mathcal E},\chi/p_{\min}\}}.
\]
The function $\U(\alpha)/\sqrt\alpha$ decreases: it is the infimum of
$\beta+\psi(-\beta)+\psi(-\beta)/\alpha$. The function
$\sqrt\alpha\U(\alpha)$ increases, by positivity and monotonicity of $\U$.
Consequently
\begin{equation*}
 \sup_{l\le\alpha\le r}\frac{\U(\alpha)}{\mathcal B(\alpha,p_{\min})}
 \le\max\left\{\frac{\U(l)}{\mathcal B(l,p_{\min})},
                \frac{\U(r)}{\mathcal B(r,p_{\min})}\right\}.
\end{equation*}
Rational BS trials upper-bound $\U$ at each endpoint.

The newsvendor bound covers small $\alpha$, and TBS covers small $p$.
The shifted DI target improves the upper bound for moderate $p$. The
conditional backlog inequalities strengthen the lower bound at larger
penalties.

\begin{proof}[Proof of the dual-sourcing guarantee in \Cref{thm:main}]
We cover the feasible domain $\alpha p\ge1$. The exact numerical comparisons
in this proof are specified in \Cref{app:certificates}; all are strict
comparisons with $r_{\rm DS}$.

If $\alpha\le3/10$, use $\U/\V\le\U(3/10)/\V(3/10)<1.595$.
If $p<1$, shorten the regular lead-time difference to $c/p$ and use the
one-date comparison $\V(1/p)$ and \cref{lem:small-backlog}. This gives at
most $3/2$.

For $1\le p\le5$ and $3/10<\alpha\le1/2$, use
$\U(1/2)/\V(1/2)-\kappa\sqrt{3/10}/\V(1)$.
For $\alpha\in[l,r]$ with consecutive endpoints
$1/2,11/20,3/5,13/20,27/40,\alpha_0$, use
$(\U(r)-\kappa\sqrt l)/\V(l)$. At $r=\alpha_0$, replace $\U(r)$
by $\sqrt2$; elsewhere a specified rational BS trial suffices.

For $\alpha\ge\alpha_0$ and $1\le p\le8/5$, take the shorter window
$\delta=\max\{\alpha_0,1/p\}\in[\alpha_0,1]$. It satisfies
$\delta\le\alpha$ and $\delta p\ge1$. Its lower comparison is at least
$\V(\alpha_0)$, whereas TBS costs at most $F_{\rm TBS}(8/5)$.
For $8/5\le p\le2$, apply a fixed quarter-window rule at $\alpha_0$
with $p_{\min}=8/5$ and upper bound $F_{\rm TBS}(2)$. For $2\le p\le5$, use the
second fixed rule with $p_{\min}=2$ and upper bound $F_{\rm TBS}(5)$.

For $p\ge5$, the 22 cells in \Cref{tab:dual-witnesses} cover
$[3/10,\alpha_0]$. Apply the endpoint argument to each cell.
For $\alpha\ge\alpha_0$, use the fine four-date rule at
$\delta=\alpha_0$ and the TBS upper bound $\sqrt2$. These event
rules apply whether or not $\delta p\ge1$.
The largest of the 106 case bounds is at the right endpoint of
cell 16 in \Cref{tab:dual-witnesses}, using zero-based indices.
It is less than $1.599527279<r_{\rm DS}$.
\end{proof}

\subsection{Signed calibration and large backorder penalties}
\label{sec:signed}

A uniform bound on the normalized cost $x+pz+y$ gives $z=O(1/p)$
as $p\to\infty$. The inventory can still be negative at each finite
$p$. We keep its signed mean at each grid date and let the mean backlog
vanish as the penalty grows.

For dual sourcing, put $u=H/\sqrt{\lambda L}$, $z_L=\bar B/\sqrt{\lambda L}$,
$\nu=u-z_L$, and $v=d\sqrt{L/\lambda}$. 
A signed relaxation removes
nonnegativity but retains the same upward control and mean at grid dates.
For a uniform grid with $K$ steps and $\Delta=1/K$, define
\begin{equation*}
 Z_0=0,\qquad R_i=Z_{i-1}+G_i-v\Delta,\qquad
 Z_i=\begin{cases}
    \max(k_i,R_i),& v>0,\\
    R_i,& v=0,
 \end{cases}
 \qquad G_i\stackrel{\rm iid}{\sim}N(0,\Delta).
\end{equation*}
Here each normal increment $G_i$ is independent of the preceding state.
For $v>0$, $k_i$ is the real floor from \cref{lem:floor} giving
$\E\left[Z_i\right]=0$. For $v=0$, the same mean follows from the centered
increments.

Without nonnegativity, a block needs no reflection. The common floor still
minimizes dispersion at its prescribed mean, and convex holding-backlog
costs can only increase under a more dispersed state law.

\begin{lemma}
\label{lem:signed}
The chain satisfies $Z_i\le_{\rm cx}Z_{i+1}$. Every limiting normalized dual-sourcing mean triple arising from causal
policies with uniformly bounded normalized costs satisfies, for every fixed $K$,
\begin{equation*}
 \E\left[(\nu+Z_K(v))^+\right]\le u,\qquad
 \E\left[(\nu+Z_K(v))^-\right]\le z_L.
\end{equation*}
In particular, with $f_p(w)=w^++pw^-$,
\begin{equation*}
 \liminf_{c\to\infty}\frac{\OPT_{\rm DS}}{\sqrt{\lambda c}}
 \ge\inf_{\nu\in\R,v\ge0}
       \{\sqrt\alpha\,\E\left[f_p(\nu+Z_K(v))\right]+v/\sqrt\alpha\}.
\end{equation*}
\end{lemma}

\begin{proof}
Conditional averaging of prewindow history fixes regular arrivals and gives
an adapted nondecreasing express-arrival control. Ignoring the express lead-time
constraint only enlarges the relaxation. Without reflection, each state is
at least its preceding state plus the new signed input increment. The signed
form of \cref{lem:floor} and independent increments make the recursion least
in convex order at every date. Conditional Jensen bounds its positive and
negative parts by the original mean holding and backlog. The time-average
and diffusion passages, including unequal intermediate means, are proved in
\cref{app:averaging}.

For a uniform grid the same transition-and-correction map preserves convex
order. Since zero is least in convex order among mean-zero laws,
$Z_0\le_{\rm cx}Z_1$; induction gives the indexwise ordering.
The two component inequalities imply the signed lower bound after adding
the rate cost. For $v>0$, each floor is negative: a nonnegative floor and
the positive upper tail of $R_i$ would give a positive mean. Flooring therefore
cannot increase the square when it acts. Independent centered increments and
$\E\left[Z_{i-1}\right]=0$ give
$\E\left[R_i^2\right]=\E\left[Z_{i-1}^2\right]+\Delta+v^2\Delta^2$. For $v=0$, no floor is
applied and the same bound holds. Thus
$\E\left[Z_i^2\right]\le i\Delta+i v^2\Delta^2\le1+v^2/K$.
\end{proof}

With one date, the floor and inventory target reduce the signed problem
to a normal newsvendor calculation. Extra dates retain additional mean
requirements and can therefore only strengthen that lower comparison.

\begin{lemma}
\label{lem:one-date}
If $\alpha p\ge1$, the right side of the signed lower bound with $K=1$
equals $\V(\alpha)$. Every finer signed grid is at least as strong.
\end{lemma}

\begin{proof}
For $K=1$ and $v>0$, write $G=G_1\sim N(0,1)$, $k=k_1$, and $q=k+v$.
The mean-zero floor condition gives
$k=-n(-q)$ and $v=n(q)$. Let $q_p=\Phi^{-1}(1/(1+p))$.
The minimizing shift places zero at a $1/(1+p)$ quantile of the
corrected normal law. Its atom at the floor has probability $\Phi(q)$.
If $q\ge q_p$, the quantile is this atom, so $\nu=-k$: the shifted
state is $(G-q)^+$ and its inventory cost is $n(-q)$.
If $q<q_p$, the optimal shift is $v-q_p$. The uncorrected newsvendor cost
is $(1+p)\phi(q_p)$, and flooring increases only negative states, reducing
that cost by $pv$. Thus the full objective on this branch is
$\sqrt\alpha(1+p)\phi(q_p)+(\alpha^{-1/2}-p\sqrt\alpha)n(q)$,
which is nonincreasing in $q$ under $\alpha p\ge1$.
On $q\ge q_p$ it is $\sqrt\alpha n(-q)+n(q)/\sqrt\alpha$; its derivative
vanishes at $\Phi(q)=\alpha/(1+\alpha)$. This point lies in the branch
exactly when $\alpha p\ge1$. Substitution yields $\V(\alpha)$.
The endpoint $v=0$ is included by $q\to-\infty$.
Every finer-grid terminal state dominates $G-v$ pathwise before upward
corrections and has mean zero. \Cref{lem:floor} makes its law at least as
dispersed as the one-date corrected law. Apply each shifted convex cost
and minimize over shift and drift.
\end{proof}

Take $c\to\infty$ at fixed $p$ on each fixed grid. Then send $p$ to
infinity, which makes the mean backlog vanish. As the grid is refined,
Brownian path continuity gives the continuous-reflection inequalities
for the two-block certificate.

\begin{proof}[Proof of the large-$p$ statement in \cref{thm:main}]
First consider $\alpha\in[\alpha_s,\alpha_0]$. Define
\[
 \mathcal D_p(\alpha)=\liminf_{c\to\infty}
 \frac{\OPT_{\rm DS}(\alpha c,c,p,L_E)}{\sqrt{\lambda c}},
\]
with $p$ and $\alpha$ fixed in this definition. If the ratio assertion fails
for arbitrarily large fixed backorder penalties, \cref{lem:policy-facts} gives
$p^{(j)}\to\infty$ and $\alpha_j$ in this compact interval such that
$\mathcal D_{p^{(j)}}(\alpha_j)<\U(\alpha_j)/r_{\rm LS}$.
For each fixed $(p^{(j)},\alpha_j)$, choose near-optimal policies along a
diffusion subsequence realizing $\mathcal D_{p^{(j)}}(\alpha_j)$. Their normalized costs
are bounded. Thus $u_j,v_j$ remain bounded and $z_{L,j}=O(1/p^{(j)})$.
Pass to $\alpha_j\to\bar\alpha$, $\nu_j=u_j-z_{L,j}\to u\ge0$, and $v_j\to v$.
The cost-component inequality gives
$\liminf_j \mathcal D_{p^{(j)}}(\alpha_j)\ge\sqrt{\bar\alpha}u+v/\sqrt{\bar\alpha}$.
The component inequalities in \cref{lem:signed} hold for all fixed dyadic
$K$; no $K$ is allowed to grow during the diffusion limit.

Continuity of the finite recursion and $z_{L,j}\to0$ give
$u+Z_K(v)\ge0$ almost surely. Indexwise convex ordering gives the same
property at every earlier date. The signed recursion realizes an adapted
process nonnegative at all grid dates, with mean $u$ there and mean control
increment $v$ times the elapsed time.

For each even dyadic $K$, reflect minimally only at its grid dates within
each half-window. The preceding process dominates those discrete reflections.
Apply the fixed-multiplier argument of \cref{lem:support} to obtain the
discrete counterpart of $\mathcal W_\theta(u,v)\le0$. Its difference from
the continuous counterpart is bounded by
\[
 (1+\theta)\left(\E\left[\omega_W(1/K)\right]+v/K\right),
 \qquad
 \omega_W(\varepsilon)=\sup_{|s-t|\le\varepsilon}|W_s-W_t|.
\]
Indeed, grid and continuous maxima differ by at most the path modulus plus
$v/K$. The partial infimum in \eqref{eq:convex-dual} is nondecreasing and
$\theta$-Lipschitz in its lower constraint; changing its regulator cost by
a uniform error changes the infimum by no more than that error.
Brownian continuity and the integrable supremum imply
$\E\left[\omega_W(1/K)\right]\to0$. Therefore all twelve continuous supporting
inequalities hold for $(u,v)$.

The supporting inequalities and continuity of $\U$ now imply
\[
 L_{\rm LP}(\bar\alpha)
 \le\sqrt{\bar\alpha}u+v/\sqrt{\bar\alpha}
 \le\liminf_j \mathcal D_{p^{(j)}}(\alpha_j)
 \le\limsup_j \mathcal D_{p^{(j)}}(\alpha_j)
 \le\U(\bar\alpha)/r_{\rm LS}.
\]
This contradicts the strict compact-cell bound
$\U(\bar\alpha)/L_{\rm LP}(\bar\alpha)<r_{\rm LS}$ in
\cref{lem:ls-certificate}. Sequential compactness therefore gives a
uniform finite threshold on $[\alpha_s,\alpha_0]$.

For $\alpha\le\alpha_s$ with $\alpha p\ge1$, use
\cref{lem:one-date} and the small-region bound of
\cref{lem:ls-certificate}, which is independent of $p$. For
$\alpha\ge\alpha_0$, compare the actual dual optimum to its value at
$\alpha_0$. The same limiting-plane argument gives
$\liminf_{p\to\infty}\mathcal D_p(\alpha_0)\ge L_{\rm LP}(\alpha_0)$,
and TBS costs at most $\sqrt2$.
The strict tail margin in \cref{lem:ls-certificate} permits a further finite
increase of the threshold. Make it at least $1/\alpha_0$ so the comparison
is feasible. These three regions prove the assertion. The proof keeps $p$ fixed during each $c\to\infty$ limit.
\end{proof}

\section{Policy parameters and numerical implications}
\label{sec:implementation}

For BS, compute Erlang blocking probabilities by
$B(j,q)=qB(j-1,q)/(j+qB(j-1,q))$ starting from $B(0,q)=1$.
Never ordering costs $\lambda p$, and BS holding costs at least
$S-\lambda L$. Hence a search for the best BS target can stop at
$\lceil\lambda(L+p)\rceil$. The enumeration is
\emph{pseudo-polynomial} in the parameter magnitudes. For the asymptotic
guarantee, it suffices to compare the normal BS trial with the CO trial
in \cref{lem:policy-facts} and the option of never ordering.

The following trials make the policy choice explicit. With
$\beta=\beta(\alpha)$, put
$S_L=\max\{0,\lfloor\lambda L+\beta\sqrt{\lambda L}\rfloor\}$.
The lost-sales candidates are BS at $S_L$, CO at
$r=\lambda-\sqrt{\lambda/(2 p)}$ when this is positive, and never ordering.
For dual sourcing, use DI with $U=S_L$ and $S=U$, adding
$S=U-\lfloor\sqrt{\lambda L}/3\rfloor$ when $p\le5$.
For TBS put
$\theta_p=\sqrt{p\log(1+1/p)/2}$,
$r=\lambda-\theta_p\sqrt{\lambda/c}$, and
$S=-\lfloor k_p\sqrt{\lambda c}\rfloor$, where
$k_p=\log(1+1/p)/(2\theta_p)$.
All rate trials are used only when $0<r<\lambda$, which holds eventually at
fixed parameters. Integer rounding changes none of the normalized limits.
Compare the trial policies by their exact costs or by certified
evaluations with additive error $o(\sqrt p)$ for lost sales and
$o(\sqrt c)$ for dual sourcing. These errors vanish after normalization.
In \eqref{eq:di-exact}, the law of $N_1$
has finite support, and the tail bound in \Cref{app:finite-costs}
controls truncation of $N_2$. The capped classes of Xin~\citep{Xin2022}
contain the respective policy menus in their closures and inherit the
bounds for their cost infima.

The direct enumeration of an $n$-bin rule on $J$ dates takes
$O(Jn^J)$ integer operations. Each supporting plane uses one-dimensional
quadrature with explicit derivative and tail bounds. Floating-point
optimization proposes the witnesses, and rational enclosures certify
the theorem comparisons.

\subsection{What strengthens the guarantees}
\label{subsec:numerical}

\begin{figure}[htbp]
\centering
\begin{adjustbox}{max width=.7\linewidth}
\begin{tikzpicture}
\begin{groupplot}[group style={group size=2 by 1,horizontal sep=1.25cm},
 width=.47\linewidth,height=6.2cm,legend cell align=left]
\nextgroupplot[xmode=log,xmin=.002,xmax=.69787,ymin=1.05,ymax=1.85,
 xlabel={Lead-time ratio $\alpha$},ylabel={Comparison quotient},
 title={Lost-sales compact cells},
 legend style={at={(.02,.98)},anchor=north west}]
\addplot[black,dashed,thick] table[x=alpha,y=baseline,col sep=comma,row sep=newline]{
alpha,new,baseline
0.002000000000,1.117621787,1.233227181
0.002120594000,1.117621787,1.233227181
0.002120594000,1.120549315,1.234811819
0.002248459000,1.120549315,1.234811819
0.002248459000,1.123056222,1.236420918
0.002384033000,1.123056222,1.236420918
0.002384033000,1.125156178,1.238055091
0.002527783000,1.125156178,1.238055091
0.002527783000,1.126862803,1.239714918
0.002680200000,1.126862803,1.239714918
0.002680200000,1.128189719,1.241401040
0.002841808000,1.128189719,1.241401040
0.002841808000,1.129150409,1.243114076
0.003013159000,1.129150409,1.243114076
0.003013159000,1.129758233,1.244854719
0.003194843000,1.129758233,1.244854719
0.003194843000,1.130026320,1.246623643
0.003387482000,1.130026320,1.246623643
0.003387482000,1.130027261,1.248421553
0.003591736000,1.130027261,1.248421553
0.003591736000,1.129968518,1.250249196
0.003808307000,1.129968518,1.250249196
0.003808307000,1.129595547,1.252107305
0.004037935000,1.129595547,1.252107305
0.004037935000,1.128920663,1.253996685
0.004281410000,1.128920663,1.253996685
0.004281410000,1.127955837,1.255918129
0.004539565000,1.127955837,1.255918129
0.004539565000,1.126712701,1.257872480
0.004813286000,1.126712701,1.257872480
0.004813286000,1.125210038,1.259860603
0.005103512000,1.125210038,1.259860603
0.005103512000,1.128121235,1.261883385
0.005411237000,1.128121235,1.261883385
0.005411237000,1.130564572,1.263941756
0.005737517000,1.130564572,1.263941756
0.005737517000,1.132554078,1.266036674
0.006083471000,1.132554078,1.266036674
0.006083471000,1.134103902,1.268169127
0.006450285000,1.134103902,1.268169127
0.006450285000,1.135228242,1.270340136
0.006839216000,1.135228242,1.270340136
0.006839216000,1.135941276,1.272550769
0.007251599000,1.135941276,1.272550769
0.007251599000,1.136257082,1.274802119
0.007688847000,1.136257082,1.274802119
0.007688847000,1.136258211,1.277095328
0.008152460000,1.136258211,1.277095328
0.008152460000,1.136190739,1.279431570
0.008644027000,1.136190739,1.279431570
0.008644027000,1.135753708,1.281812071
0.009165234000,1.135753708,1.281812071
0.009165234000,1.134960607,1.284238098
0.009717869000,1.134960607,1.284238098
0.009717869000,1.135057097,1.286710955
0.010303825000,1.135057097,1.286710955
0.010303825000,1.138384300,1.289232007
0.010925112000,1.138384300,1.289232007
0.010925112000,1.141176069,1.291802666
0.011583861000,1.141176069,1.291802666
0.011583861000,1.143446465,1.294424395
0.012282331000,1.143446465,1.294424395
0.012282331000,1.145209892,1.297098707
0.013022916000,1.145209892,1.297098707
0.013022916000,1.146481011,1.299827180
0.013808156000,1.146481011,1.299827180
0.013808156000,1.147274635,1.302611447
0.014640744000,1.147274635,1.302611447
0.014640744000,1.147605641,1.305453199
0.015523533000,1.147605641,1.305453199
0.015523533000,1.147606992,1.308354203
0.016459553000,1.147606992,1.308354203
0.016459553000,1.147490273,1.311316279
0.017452011000,1.147490273,1.311316279
0.017452011000,1.146940598,1.314341326
0.018504311000,1.146940598,1.314341326
0.018504311000,1.145972615,1.317431313
0.019620061000,1.145972615,1.317431313
0.019620061000,1.147017804,1.320588289
0.020803088000,1.147017804,1.320588289
0.020803088000,1.150882453,1.323814378
0.022057448000,1.150882453,1.323814378
0.022057448000,1.154124756,1.327111784
0.023387441000,1.154124756,1.327111784
0.023387441000,1.156758178,1.330482802
0.024797628000,1.156758178,1.330482802
0.024797628000,1.158796889,1.333929819
0.026292846000,1.158796889,1.333929819
0.026292846000,1.160255625,1.337455309
0.027878220000,1.160255625,1.337455309
0.027878220000,1.161149576,1.341061846
0.029559187000,1.161149576,1.341061846
0.029559187000,1.161494267,1.344752108
0.031341511000,1.161494267,1.344752108
0.031341511000,1.161495904,1.348528875
0.033231303000,1.161495904,1.348528875
0.033231303000,1.161307119,1.352395040
0.035235044000,1.161307119,1.352395040
0.035235044000,1.160600724,1.356353608
0.037359604000,1.160600724,1.356353608
0.037359604000,1.159392661,1.360407704
0.039612268000,1.159392661,1.360407704
0.039612268000,1.161734507,1.364560578
0.042000760000,1.161734507,1.364560578
0.042000760000,1.166290234,1.368815609
0.044533271000,1.166290234,1.368815609
0.044533271000,1.170111809,1.373176307
0.047218484000,1.170111809,1.373176307
0.047218484000,1.173210981,1.377646327
0.050065607000,1.173210981,1.377646327
0.050065607000,1.175600756,1.382229463
0.053084403000,1.175600756,1.382229463
0.053084403000,1.177295237,1.386929661
0.056285222000,1.177295237,1.386929661
0.056285222000,1.178309470,1.391751024
0.059679040000,1.178309470,1.391751024
0.059679040000,1.178659299,1.396697815
0.063277494000,1.178659299,1.396697815
0.063277494000,1.178661287,1.401774466
0.067092923000,1.178661287,1.401774466
0.067092923000,1.178363240,1.406985583
0.071138411000,1.178363240,1.406985583
0.071138411000,1.177434315,1.412335949
0.075427829000,1.177434315,1.412335949
0.075427829000,1.175891950,1.417830534
0.079975884000,1.175891950,1.417830534
0.079975884000,1.179470659,1.423474501
0.084798173000,1.179470659,1.423474501
0.084798173000,1.184365553,1.429273211
0.089911231000,1.184365553,1.429273211
0.089911231000,1.188398715,1.435232226
0.095332589000,1.188398715,1.435232226
0.095332589000,1.191579757,1.441357322
0.101080838000,1.191579757,1.441357322
0.101080838000,1.193920262,1.447654493
0.107175688000,1.193920262,1.447654493
0.107175688000,1.195433600,1.454129950
0.113638037000,1.195433600,1.454129950
0.113638037000,1.196134757,1.460790142
0.120490046000,1.196134757,1.460790142
0.120490046000,1.196137080,1.467641743
0.127755208000,1.196137080,1.467641743
0.127755208000,1.196042494,1.474691676
0.135458437000,1.196042494,1.474691676
0.135458437000,1.195169822,1.481947105
0.143626145000,1.195169822,1.481947105
0.143626145000,1.193537845,1.489415448
0.152286341000,1.193537845,1.489415448
0.152286341000,1.198904737,1.497104379
0.161468718000,1.198904737,1.497104379
0.161468718000,1.203391628,1.505021831
0.171204762000,1.203391628,1.505021831
0.171204762000,1.206890844,1.513176008
0.181527859000,1.206890844,1.513176008
0.181527859000,1.209411490,1.521575379
0.192473406000,1.209411490,1.521575379
0.192473406000,1.210965413,1.530228688
0.204078934000,1.210965413,1.530228688
0.204078934000,1.211567008,1.539144953
0.216384238000,1.211567008,1.539144953
0.216384238000,1.211569541,1.548333474
0.229431512000,1.211569541,1.548333474
0.229431512000,1.211235553,1.557803829
0.243265495000,1.211235553,1.557803829
0.243265495000,1.212471794,1.567565881
0.257933623000,1.212471794,1.567565881
0.257933623000,1.217951415,1.577629772
0.273486192000,1.217951415,1.577629772
0.273486192000,1.222315337,1.588005930
0.289976530000,1.222315337,1.588005930
0.289976530000,1.225568095,1.598705066
0.307461184000,1.225568095,1.598705066
0.307461184000,1.227718124,1.609738169
0.326000105000,1.227718124,1.609738169
0.326000105000,1.228777564,1.621116511
0.345656864000,1.228777564,1.621116511
0.345656864000,1.228780108,1.632851642
0.366498862000,1.228780108,1.632851642
0.366498862000,1.228764574,1.644955385
0.388597566000,1.228764574,1.644955385
0.388597566000,1.229449963,1.657439834
0.412028751000,1.229449963,1.657439834
0.412028751000,1.233820295,1.670317351
0.436872760000,1.233820295,1.670317351
0.436872760000,1.236985765,1.683600563
0.463214784000,1.236985765,1.683600563
0.463214784000,1.238954198,1.697302353
0.491145147000,1.238954198,1.697302353
0.491145147000,1.239738000,1.711435858
0.520759621000,1.239738000,1.711435858
0.520759621000,1.239740360,1.726014463
0.552159753000,1.239740360,1.726014463
0.552159753000,1.243113526,1.741051797
0.585453212000,1.243113526,1.741051797
0.585453212000,1.245286508,1.756561728
0.620754160000,1.245286508,1.756561728
0.620754160000,1.246762066,1.772558354
0.658183641000,1.246762066,1.772558354
0.658183641000,1.248976725,1.789056004
0.697870000000,1.248976725,1.789056004
};\addlegendentry{Newsvendor comparison}
\addplot[blue!75!black,thick] table[x=alpha,y=new,col sep=comma,row sep=newline]{
alpha,new,baseline
0.002000000000,1.117621787,1.233227181
0.002120594000,1.117621787,1.233227181
0.002120594000,1.120549315,1.234811819
0.002248459000,1.120549315,1.234811819
0.002248459000,1.123056222,1.236420918
0.002384033000,1.123056222,1.236420918
0.002384033000,1.125156178,1.238055091
0.002527783000,1.125156178,1.238055091
0.002527783000,1.126862803,1.239714918
0.002680200000,1.126862803,1.239714918
0.002680200000,1.128189719,1.241401040
0.002841808000,1.128189719,1.241401040
0.002841808000,1.129150409,1.243114076
0.003013159000,1.129150409,1.243114076
0.003013159000,1.129758233,1.244854719
0.003194843000,1.129758233,1.244854719
0.003194843000,1.130026320,1.246623643
0.003387482000,1.130026320,1.246623643
0.003387482000,1.130027261,1.248421553
0.003591736000,1.130027261,1.248421553
0.003591736000,1.129968518,1.250249196
0.003808307000,1.129968518,1.250249196
0.003808307000,1.129595547,1.252107305
0.004037935000,1.129595547,1.252107305
0.004037935000,1.128920663,1.253996685
0.004281410000,1.128920663,1.253996685
0.004281410000,1.127955837,1.255918129
0.004539565000,1.127955837,1.255918129
0.004539565000,1.126712701,1.257872480
0.004813286000,1.126712701,1.257872480
0.004813286000,1.125210038,1.259860603
0.005103512000,1.125210038,1.259860603
0.005103512000,1.128121235,1.261883385
0.005411237000,1.128121235,1.261883385
0.005411237000,1.130564572,1.263941756
0.005737517000,1.130564572,1.263941756
0.005737517000,1.132554078,1.266036674
0.006083471000,1.132554078,1.266036674
0.006083471000,1.134103902,1.268169127
0.006450285000,1.134103902,1.268169127
0.006450285000,1.135228242,1.270340136
0.006839216000,1.135228242,1.270340136
0.006839216000,1.135941276,1.272550769
0.007251599000,1.135941276,1.272550769
0.007251599000,1.136257082,1.274802119
0.007688847000,1.136257082,1.274802119
0.007688847000,1.136258211,1.277095328
0.008152460000,1.136258211,1.277095328
0.008152460000,1.136190739,1.279431570
0.008644027000,1.136190739,1.279431570
0.008644027000,1.135753708,1.281812071
0.009165234000,1.135753708,1.281812071
0.009165234000,1.134960607,1.284238098
0.009717869000,1.134960607,1.284238098
0.009717869000,1.135057097,1.286710955
0.010303825000,1.135057097,1.286710955
0.010303825000,1.138384300,1.289232007
0.010925112000,1.138384300,1.289232007
0.010925112000,1.141176069,1.291802666
0.011583861000,1.141176069,1.291802666
0.011583861000,1.143446465,1.294424395
0.012282331000,1.143446465,1.294424395
0.012282331000,1.145209892,1.297098707
0.013022916000,1.145209892,1.297098707
0.013022916000,1.146481011,1.299827180
0.013808156000,1.146481011,1.299827180
0.013808156000,1.147274635,1.302611447
0.014640744000,1.147274635,1.302611447
0.014640744000,1.147605641,1.305453199
0.015523533000,1.147605641,1.305453199
0.015523533000,1.147606992,1.308354203
0.016459553000,1.147606992,1.308354203
0.016459553000,1.147490273,1.311316279
0.017452011000,1.147490273,1.311316279
0.017452011000,1.146940598,1.314341326
0.018504311000,1.146940598,1.314341326
0.018504311000,1.145972615,1.317431313
0.019620061000,1.145972615,1.317431313
0.019620061000,1.147017804,1.320588289
0.020803088000,1.147017804,1.320588289
0.020803088000,1.150882453,1.323814378
0.022057448000,1.150882453,1.323814378
0.022057448000,1.154124756,1.327111784
0.023387441000,1.154124756,1.327111784
0.023387441000,1.156758178,1.330482802
0.024797628000,1.156758178,1.330482802
0.024797628000,1.158796889,1.333929819
0.026292846000,1.158796889,1.333929819
0.026292846000,1.160255625,1.337455309
0.027878220000,1.160255625,1.337455309
0.027878220000,1.161149576,1.341061846
0.029559187000,1.161149576,1.341061846
0.029559187000,1.161494267,1.344752108
0.031341511000,1.161494267,1.344752108
0.031341511000,1.161495904,1.348528875
0.033231303000,1.161495904,1.348528875
0.033231303000,1.161307119,1.352395040
0.035235044000,1.161307119,1.352395040
0.035235044000,1.160600724,1.356353608
0.037359604000,1.160600724,1.356353608
0.037359604000,1.159392661,1.360407704
0.039612268000,1.159392661,1.360407704
0.039612268000,1.161734507,1.364560578
0.042000760000,1.161734507,1.364560578
0.042000760000,1.166290234,1.368815609
0.044533271000,1.166290234,1.368815609
0.044533271000,1.170111809,1.373176307
0.047218484000,1.170111809,1.373176307
0.047218484000,1.173210981,1.377646327
0.050065607000,1.173210981,1.377646327
0.050065607000,1.175600756,1.382229463
0.053084403000,1.175600756,1.382229463
0.053084403000,1.177295237,1.386929661
0.056285222000,1.177295237,1.386929661
0.056285222000,1.178309470,1.391751024
0.059679040000,1.178309470,1.391751024
0.059679040000,1.178659299,1.396697815
0.063277494000,1.178659299,1.396697815
0.063277494000,1.178661287,1.401774466
0.067092923000,1.178661287,1.401774466
0.067092923000,1.178363240,1.406985583
0.071138411000,1.178363240,1.406985583
0.071138411000,1.177434315,1.412335949
0.075427829000,1.177434315,1.412335949
0.075427829000,1.175891950,1.417830534
0.079975884000,1.175891950,1.417830534
0.079975884000,1.179470659,1.423474501
0.084798173000,1.179470659,1.423474501
0.084798173000,1.184365553,1.429273211
0.089911231000,1.184365553,1.429273211
0.089911231000,1.188398715,1.435232226
0.095332589000,1.188398715,1.435232226
0.095332589000,1.191579757,1.441357322
0.101080838000,1.191579757,1.441357322
0.101080838000,1.193920262,1.447654493
0.107175688000,1.193920262,1.447654493
0.107175688000,1.195433600,1.454129950
0.113638037000,1.195433600,1.454129950
0.113638037000,1.196134757,1.460790142
0.120490046000,1.196134757,1.460790142
0.120490046000,1.196137080,1.467641743
0.127755208000,1.196137080,1.467641743
0.127755208000,1.196042494,1.474691676
0.135458437000,1.196042494,1.474691676
0.135458437000,1.195169822,1.481947105
0.143626145000,1.195169822,1.481947105
0.143626145000,1.193537845,1.489415448
0.152286341000,1.193537845,1.489415448
0.152286341000,1.198904737,1.497104379
0.161468718000,1.198904737,1.497104379
0.161468718000,1.203391628,1.505021831
0.171204762000,1.203391628,1.505021831
0.171204762000,1.206890844,1.513176008
0.181527859000,1.206890844,1.513176008
0.181527859000,1.209411490,1.521575379
0.192473406000,1.209411490,1.521575379
0.192473406000,1.210965413,1.530228688
0.204078934000,1.210965413,1.530228688
0.204078934000,1.211567008,1.539144953
0.216384238000,1.211567008,1.539144953
0.216384238000,1.211569541,1.548333474
0.229431512000,1.211569541,1.548333474
0.229431512000,1.211235553,1.557803829
0.243265495000,1.211235553,1.557803829
0.243265495000,1.212471794,1.567565881
0.257933623000,1.212471794,1.567565881
0.257933623000,1.217951415,1.577629772
0.273486192000,1.217951415,1.577629772
0.273486192000,1.222315337,1.588005930
0.289976530000,1.222315337,1.588005930
0.289976530000,1.225568095,1.598705066
0.307461184000,1.225568095,1.598705066
0.307461184000,1.227718124,1.609738169
0.326000105000,1.227718124,1.609738169
0.326000105000,1.228777564,1.621116511
0.345656864000,1.228777564,1.621116511
0.345656864000,1.228780108,1.632851642
0.366498862000,1.228780108,1.632851642
0.366498862000,1.228764574,1.644955385
0.388597566000,1.228764574,1.644955385
0.388597566000,1.229449963,1.657439834
0.412028751000,1.229449963,1.657439834
0.412028751000,1.233820295,1.670317351
0.436872760000,1.233820295,1.670317351
0.436872760000,1.236985765,1.683600563
0.463214784000,1.236985765,1.683600563
0.463214784000,1.238954198,1.697302353
0.491145147000,1.238954198,1.697302353
0.491145147000,1.239738000,1.711435858
0.520759621000,1.239738000,1.711435858
0.520759621000,1.239740360,1.726014463
0.552159753000,1.239740360,1.726014463
0.552159753000,1.243113526,1.741051797
0.585453212000,1.243113526,1.741051797
0.585453212000,1.245286508,1.756561728
0.620754160000,1.245286508,1.756561728
0.620754160000,1.246762066,1.772558354
0.658183641000,1.246762066,1.772558354
0.658183641000,1.248976725,1.789056004
0.697870000000,1.248976725,1.789056004
};\addlegendentry{Two-block certificate}
\addplot[red!70!black,dotted,domain=.002:.69787]{1.248977};
\addlegendentry{Factor $1.248977$}
\nextgroupplot[xmode=log,xmin=2,xmax=20,ymin=1.3,ymax=6.5,
 xtick={2,5,10,20},xticklabels={2,5,10,20},
 xlabel={Backlog cost $p$},title={One fixed dual-sourcing rule},
 legend style={at={(.98,.98)},anchor=north east}]
\addplot[black,dashed,thick] table[x=p,y=unconditional,col sep=comma,row sep=newline]{
p,conditional,unconditional
2.0,2.723174048,6.255984722
2.1,2.593499093,5.958080687
2.2,2.475612771,5.687258838
2.3,2.367977433,5.439986715
2.4,2.269311706,5.213320602
2.5,2.178539238,5.004787777
2.6,2.094749268,4.812295940
2.7,2.017165961,4.634062757
2.8,1.945124320,4.468560516
2.9,1.878051068,4.314472222
3.0,1.815449365,4.170656481
3.1,1.756886483,4.036119175
3.2,1.701983780,3.909990451
3.3,1.650408514,3.791505892
3.4,1.601867087,3.679991013
3.5,1.564172169,3.574848413
3.6,1.564172169,3.475547068
3.7,1.564172169,3.381613363
3.8,1.564172169,3.292623538
3.9,1.564172169,3.208197293
4.0,1.564172169,3.127992361
4.1,1.564172169,3.051699865
4.2,1.564172169,2.979040344
4.3,1.564172169,2.909760336
4.4,1.564172169,2.843629419
4.5,1.564172169,2.780437654
4.6,1.564172169,2.719993358
4.7,1.564172169,2.662121158
4.8,1.564172169,2.606660301
4.9,1.564172169,2.553463152
5.0,1.564172169,2.502393889
5.1,1.564172169,2.453327342
5.2,1.564172169,2.406147970
5.3,1.564172169,2.360748952
5.4,1.564172169,2.317031379
5.5,1.564172169,2.274903535
5.6,1.564172169,2.234280258
5.7,1.564172169,2.195082359
5.8,1.564172169,2.157236111
5.9,1.564172169,2.120672787
6.0,1.564172169,2.085328241
6.1,1.564172169,2.051142532
6.2,1.564172169,2.018059588
6.3,1.564172169,1.986026896
6.4,1.564172169,1.954995226
6.5,1.564172169,1.924918376
6.6,1.564172169,1.895752946
6.7,1.564172169,1.867458126
6.8,1.564172169,1.839995507
6.9,1.564172169,1.813328905
7.0,1.564172169,1.787424207
7.1,1.564172169,1.762249218
7.2,1.564172169,1.737773534
7.3,1.564172169,1.713968417
7.4,1.564172169,1.690806682
7.5,1.564172169,1.668262593
7.6,1.564172169,1.646311769
7.7,1.564172169,1.624931097
7.8,1.564172169,1.604098647
7.9,1.564172169,1.583793601
8.0,1.564172169,1.564172169
8.1,1.564172169,1.564172169
8.2,1.564172169,1.564172169
8.3,1.564172169,1.564172169
8.4,1.564172169,1.564172169
8.5,1.564172169,1.564172169
8.6,1.564172169,1.564172169
8.7,1.564172169,1.564172169
8.8,1.564172169,1.564172169
8.9,1.564172169,1.564172169
9.0,1.564172169,1.564172169
9.1,1.564172169,1.564172169
9.2,1.564172169,1.564172169
9.3,1.564172169,1.564172169
9.4,1.564172169,1.564172169
9.5,1.564172169,1.564172169
9.6,1.564172169,1.564172169
9.7,1.564172169,1.564172169
9.8,1.564172169,1.564172169
9.9,1.564172169,1.564172169
10.0,1.564172169,1.564172169
10.1,1.564172169,1.564172169
10.2,1.564172169,1.564172169
10.3,1.564172169,1.564172169
10.4,1.564172169,1.564172169
10.5,1.564172169,1.564172169
10.6,1.564172169,1.564172169
10.7,1.564172169,1.564172169
10.8,1.564172169,1.564172169
10.9,1.564172169,1.564172169
11.0,1.564172169,1.564172169
11.1,1.564172169,1.564172169
11.2,1.564172169,1.564172169
11.3,1.564172169,1.564172169
11.4,1.564172169,1.564172169
11.5,1.564172169,1.564172169
11.6,1.564172169,1.564172169
11.7,1.564172169,1.564172169
11.8,1.564172169,1.564172169
11.9,1.564172169,1.564172169
12.0,1.564172169,1.564172169
12.1,1.564172169,1.564172169
12.2,1.564172169,1.564172169
12.3,1.564172169,1.564172169
12.4,1.564172169,1.564172169
12.5,1.564172169,1.564172169
12.6,1.564172169,1.564172169
12.7,1.564172169,1.564172169
12.8,1.564172169,1.564172169
12.9,1.564172169,1.564172169
13.0,1.564172169,1.564172169
13.1,1.564172169,1.564172169
13.2,1.564172169,1.564172169
13.3,1.564172169,1.564172169
13.4,1.564172169,1.564172169
13.5,1.564172169,1.564172169
13.6,1.564172169,1.564172169
13.7,1.564172169,1.564172169
13.8,1.564172169,1.564172169
13.9,1.564172169,1.564172169
14.0,1.564172169,1.564172169
14.1,1.564172169,1.564172169
14.2,1.564172169,1.564172169
14.3,1.564172169,1.564172169
14.4,1.564172169,1.564172169
14.5,1.564172169,1.564172169
14.6,1.564172169,1.564172169
14.7,1.564172169,1.564172169
14.8,1.564172169,1.564172169
14.9,1.564172169,1.564172169
15.0,1.564172169,1.564172169
15.1,1.564172169,1.564172169
15.2,1.564172169,1.564172169
15.3,1.564172169,1.564172169
15.4,1.564172169,1.564172169
15.5,1.564172169,1.564172169
15.6,1.564172169,1.564172169
15.7,1.564172169,1.564172169
15.8,1.564172169,1.564172169
15.9,1.564172169,1.564172169
16.0,1.564172169,1.564172169
16.1,1.564172169,1.564172169
16.2,1.564172169,1.564172169
16.3,1.564172169,1.564172169
16.4,1.564172169,1.564172169
16.5,1.564172169,1.564172169
16.6,1.564172169,1.564172169
16.7,1.564172169,1.564172169
16.8,1.564172169,1.564172169
16.9,1.564172169,1.564172169
17.0,1.564172169,1.564172169
17.1,1.564172169,1.564172169
17.2,1.564172169,1.564172169
17.3,1.564172169,1.564172169
17.4,1.564172169,1.564172169
17.5,1.564172169,1.564172169
17.6,1.564172169,1.564172169
17.7,1.564172169,1.564172169
17.8,1.564172169,1.564172169
17.9,1.564172169,1.564172169
18.0,1.564172169,1.564172169
18.1,1.564172169,1.564172169
18.2,1.564172169,1.564172169
18.3,1.564172169,1.564172169
18.4,1.564172169,1.564172169
18.5,1.564172169,1.564172169
18.6,1.564172169,1.564172169
18.7,1.564172169,1.564172169
18.8,1.564172169,1.564172169
18.9,1.564172169,1.564172169
19.0,1.564172169,1.564172169
19.1,1.564172169,1.564172169
19.2,1.564172169,1.564172169
19.3,1.564172169,1.564172169
19.4,1.564172169,1.564172169
19.5,1.564172169,1.564172169
19.6,1.564172169,1.564172169
19.7,1.564172169,1.564172169
19.8,1.564172169,1.564172169
19.9,1.564172169,1.564172169
20.0,1.564172169,1.564172169
};\addlegendentry{Unconditional charge}
\addplot[blue!75!black,thick] table[x=p,y=conditional,col sep=comma,row sep=newline]{
p,conditional,unconditional
2.0,2.723174048,6.255984722
2.1,2.593499093,5.958080687
2.2,2.475612771,5.687258838
2.3,2.367977433,5.439986715
2.4,2.269311706,5.213320602
2.5,2.178539238,5.004787777
2.6,2.094749268,4.812295940
2.7,2.017165961,4.634062757
2.8,1.945124320,4.468560516
2.9,1.878051068,4.314472222
3.0,1.815449365,4.170656481
3.1,1.756886483,4.036119175
3.2,1.701983780,3.909990451
3.3,1.650408514,3.791505892
3.4,1.601867087,3.679991013
3.5,1.564172169,3.574848413
3.6,1.564172169,3.475547068
3.7,1.564172169,3.381613363
3.8,1.564172169,3.292623538
3.9,1.564172169,3.208197293
4.0,1.564172169,3.127992361
4.1,1.564172169,3.051699865
4.2,1.564172169,2.979040344
4.3,1.564172169,2.909760336
4.4,1.564172169,2.843629419
4.5,1.564172169,2.780437654
4.6,1.564172169,2.719993358
4.7,1.564172169,2.662121158
4.8,1.564172169,2.606660301
4.9,1.564172169,2.553463152
5.0,1.564172169,2.502393889
5.1,1.564172169,2.453327342
5.2,1.564172169,2.406147970
5.3,1.564172169,2.360748952
5.4,1.564172169,2.317031379
5.5,1.564172169,2.274903535
5.6,1.564172169,2.234280258
5.7,1.564172169,2.195082359
5.8,1.564172169,2.157236111
5.9,1.564172169,2.120672787
6.0,1.564172169,2.085328241
6.1,1.564172169,2.051142532
6.2,1.564172169,2.018059588
6.3,1.564172169,1.986026896
6.4,1.564172169,1.954995226
6.5,1.564172169,1.924918376
6.6,1.564172169,1.895752946
6.7,1.564172169,1.867458126
6.8,1.564172169,1.839995507
6.9,1.564172169,1.813328905
7.0,1.564172169,1.787424207
7.1,1.564172169,1.762249218
7.2,1.564172169,1.737773534
7.3,1.564172169,1.713968417
7.4,1.564172169,1.690806682
7.5,1.564172169,1.668262593
7.6,1.564172169,1.646311769
7.7,1.564172169,1.624931097
7.8,1.564172169,1.604098647
7.9,1.564172169,1.583793601
8.0,1.564172169,1.564172169
8.1,1.564172169,1.564172169
8.2,1.564172169,1.564172169
8.3,1.564172169,1.564172169
8.4,1.564172169,1.564172169
8.5,1.564172169,1.564172169
8.6,1.564172169,1.564172169
8.7,1.564172169,1.564172169
8.8,1.564172169,1.564172169
8.9,1.564172169,1.564172169
9.0,1.564172169,1.564172169
9.1,1.564172169,1.564172169
9.2,1.564172169,1.564172169
9.3,1.564172169,1.564172169
9.4,1.564172169,1.564172169
9.5,1.564172169,1.564172169
9.6,1.564172169,1.564172169
9.7,1.564172169,1.564172169
9.8,1.564172169,1.564172169
9.9,1.564172169,1.564172169
10.0,1.564172169,1.564172169
10.1,1.564172169,1.564172169
10.2,1.564172169,1.564172169
10.3,1.564172169,1.564172169
10.4,1.564172169,1.564172169
10.5,1.564172169,1.564172169
10.6,1.564172169,1.564172169
10.7,1.564172169,1.564172169
10.8,1.564172169,1.564172169
10.9,1.564172169,1.564172169
11.0,1.564172169,1.564172169
11.1,1.564172169,1.564172169
11.2,1.564172169,1.564172169
11.3,1.564172169,1.564172169
11.4,1.564172169,1.564172169
11.5,1.564172169,1.564172169
11.6,1.564172169,1.564172169
11.7,1.564172169,1.564172169
11.8,1.564172169,1.564172169
11.9,1.564172169,1.564172169
12.0,1.564172169,1.564172169
12.1,1.564172169,1.564172169
12.2,1.564172169,1.564172169
12.3,1.564172169,1.564172169
12.4,1.564172169,1.564172169
12.5,1.564172169,1.564172169
12.6,1.564172169,1.564172169
12.7,1.564172169,1.564172169
12.8,1.564172169,1.564172169
12.9,1.564172169,1.564172169
13.0,1.564172169,1.564172169
13.1,1.564172169,1.564172169
13.2,1.564172169,1.564172169
13.3,1.564172169,1.564172169
13.4,1.564172169,1.564172169
13.5,1.564172169,1.564172169
13.6,1.564172169,1.564172169
13.7,1.564172169,1.564172169
13.8,1.564172169,1.564172169
13.9,1.564172169,1.564172169
14.0,1.564172169,1.564172169
14.1,1.564172169,1.564172169
14.2,1.564172169,1.564172169
14.3,1.564172169,1.564172169
14.4,1.564172169,1.564172169
14.5,1.564172169,1.564172169
14.6,1.564172169,1.564172169
14.7,1.564172169,1.564172169
14.8,1.564172169,1.564172169
14.9,1.564172169,1.564172169
15.0,1.564172169,1.564172169
15.1,1.564172169,1.564172169
15.2,1.564172169,1.564172169
15.3,1.564172169,1.564172169
15.4,1.564172169,1.564172169
15.5,1.564172169,1.564172169
15.6,1.564172169,1.564172169
15.7,1.564172169,1.564172169
15.8,1.564172169,1.564172169
15.9,1.564172169,1.564172169
16.0,1.564172169,1.564172169
16.1,1.564172169,1.564172169
16.2,1.564172169,1.564172169
16.3,1.564172169,1.564172169
16.4,1.564172169,1.564172169
16.5,1.564172169,1.564172169
16.6,1.564172169,1.564172169
16.7,1.564172169,1.564172169
16.8,1.564172169,1.564172169
16.9,1.564172169,1.564172169
17.0,1.564172169,1.564172169
17.1,1.564172169,1.564172169
17.2,1.564172169,1.564172169
17.3,1.564172169,1.564172169
17.4,1.564172169,1.564172169
17.5,1.564172169,1.564172169
17.6,1.564172169,1.564172169
17.7,1.564172169,1.564172169
17.8,1.564172169,1.564172169
17.9,1.564172169,1.564172169
18.0,1.564172169,1.564172169
18.1,1.564172169,1.564172169
18.2,1.564172169,1.564172169
18.3,1.564172169,1.564172169
18.4,1.564172169,1.564172169
18.5,1.564172169,1.564172169
18.6,1.564172169,1.564172169
18.7,1.564172169,1.564172169
18.8,1.564172169,1.564172169
18.9,1.564172169,1.564172169
19.0,1.564172169,1.564172169
19.1,1.564172169,1.564172169
19.2,1.564172169,1.564172169
19.3,1.564172169,1.564172169
19.4,1.564172169,1.564172169
19.5,1.564172169,1.564172169
19.6,1.564172169,1.564172169
19.7,1.564172169,1.564172169
19.8,1.564172169,1.564172169
19.9,1.564172169,1.564172169
20.0,1.564172169,1.564172169
};\addlegendentry{Conditional charge}
\end{groupplot}
\end{tikzpicture}
\end{adjustbox}
\caption{Cost-comparison upper bounds from the finite certificates. The left panel
covers the compact lost-sales region. The right panel uses one four-date rule at
$\delta=0.69787$ and applies when $\alpha\ge\delta$ and $p\ge2$.
The plotted values are ratios of policy-cost upper bounds to optimal-cost
lower bounds, rounded upward to nine decimal places.}
\label{fig:certificate-comparisons}
\end{figure}
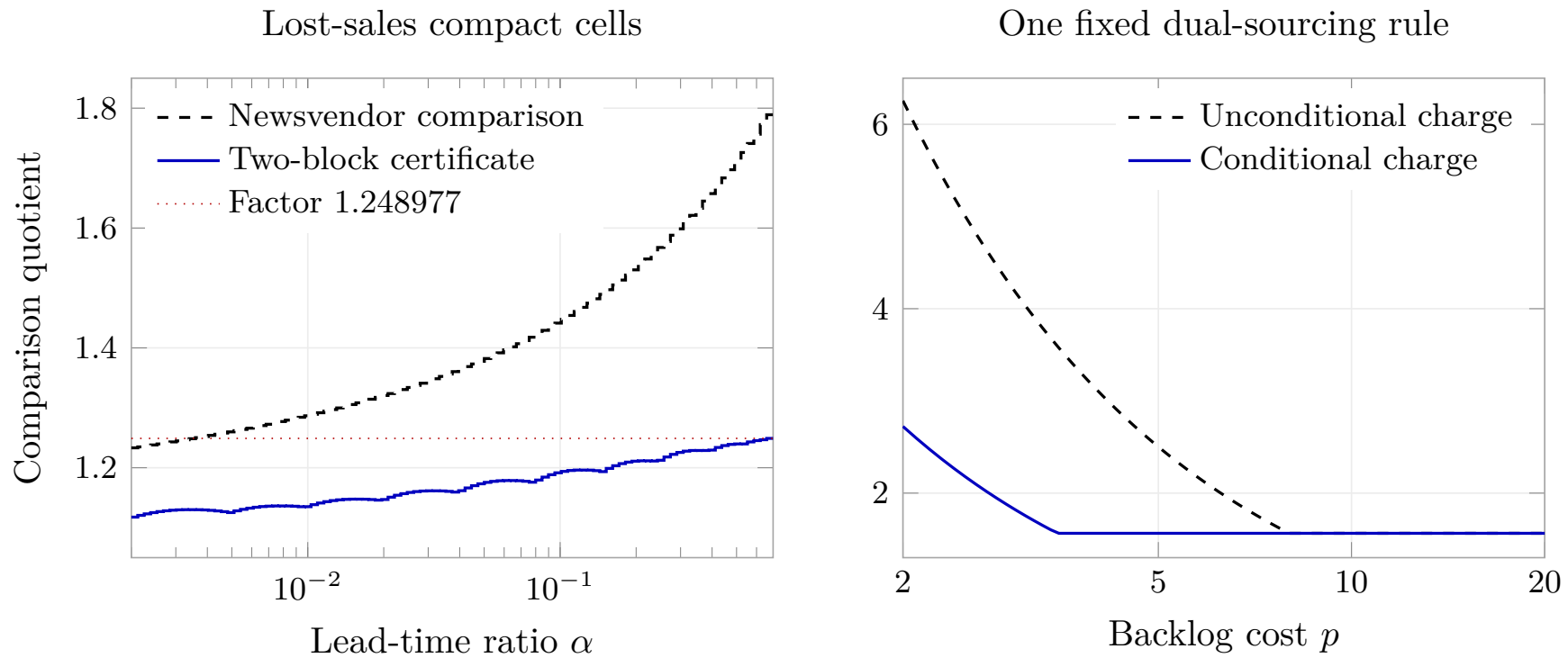

\Cref{fig:certificate-comparisons} shows how the lower-bound constructions affect
the cost comparisons. The lost-sales panel plots upper envelopes for all one hundred
compact cells. The baseline uses the newsvendor lower comparison $\V$; the other
curve uses the twelve supporting inequalities. For each cell, the plotted value
is rounded upward from its rational certificate. For a cell $[l,r]$,
monotonicity of $\U/\V$ recorded in \Cref{app:normal} gives
$\U(\alpha)/\V(\alpha)\le F_{\mathrm{upper},\beta_j}(r)/\V(r)$ throughout the cell.
Thus the right-endpoint baseline is a whole-cell upper bound, even though
the BS trial itself is fixed only for that endpoint comparison.
The maximum of the two-block curve
is below $1.248977$. The small-parameter region and lead-time tail are handled
separately in \cref{lem:ls-certificate}.

For the dual-sourcing panel, fix the fine four-date rule at
$\delta=\alpha_0$, $\xi=1.311002$, and $\zeta=0.12719$. The curves divide
the policy bound $\sqrt2$ by the lower bound with backlog coefficient
$\chi$ or $3+\varpi$. The conditional coefficient yields a useful
comparison at smaller backorder penalties.

With the same BS trials, nested subsets of three, six, and twelve planes
give compact upper bounds of $1.391503589$, $1.252099509$, and
$1.248976725$, respectively. Rounding the twelve planes to eight decimal
places still certifies $1.248977$. \Cref{app:precision} gives both checks.
The pair $(u,v)=(0.921702,0.279154)$ satisfies the endpoint constraint.
The interval certificate gives $Au+Cv-K<-0.01$ for supporting
inequality 11, so the two-block relaxation excludes the pair.

\paragraph{Finite-parameter illustrations.}
\Cref{app:finite-study} evaluates 27 lost-sales and 100 dual-sourcing
instances. For lost sales we compute the BS/CO menu costs and use
newsvendor lower bounds to certify instance-specific gaps. For dual
sourcing we optimize the integer targets at fixed DI target differences
and TBS rates and compare those costs with the limits of the same rules.
The accompanying code recomputes all instances.

\section{Conclusion}
\label{sec:conclusion}

We improve the asymptotic approximation guarantees established by
Xin~\citep{Xin2022} under the same proportional-growth assumptions.
The factors decrease from 1.79 to 1.248977 for the better of base-stock
and constant-order policies in the lost-sales model, and to 1.599528
for the better of dual-index and tailored base-surge policies in the
dual-sourcing model. For sufficiently large fixed backorder penalties,
the dual-sourcing factor also becomes 1.248977. These improvements follow
from stronger lower bounds on the unrestricted optimal cost. For lost
sales, constraints on mean inventory at intermediate dates and the
common-floor correction give an exact feasibility test for a finite-grid
relaxation. For dual sourcing, events determined only by subsequent demand
give valid backlog bounds, while signed calibration transfers the stronger
lost-sales comparison to the large-penalty regime.

Future research can pursue three concrete objectives. First, strengthen
the lost-sales lower bound by dividing the lead-time window into four,
eight, and progressively more blocks, retaining the mean-inventory
constraint at every division point. Characterizing the limiting lower
bound would determine the best guarantee this relaxation can establish
for the base-stock and constant-order comparison. Second, derive cost
bounds for capped base-stock and capped dual-index policies that account
jointly for their inventory targets and order caps. Such bounds could
exploit policy choices beyond the two simpler families used in each
comparison and yield smaller approximation factors. Third, obtain
explicit finite-parameter guarantees approaching the present asymptotic
factors. For fixed scaling parameters and any prescribed tolerance,
the goal is to compute thresholds ensuring a cost ratio at most the
corresponding factor plus that tolerance. This quantitative analysis
should also identify an explicit backorder-penalty threshold for the
1.248977 dual-sourcing guarantee.

\section*{Acknowledgments}

\noindent\textbf{Competing interests.} The authors have no relevant financial or
non-financial interests to disclose.


\vspace{3pt}
\noindent\textbf{AI-assisted technologies.} OpenAI ChatGPT was used to
help write and refine the code, prepare computational checks, and draft
and revise the manuscript. The authors independently verified the
mathematical claims, proofs, computational results, and references,
and take full responsibility for the accuracy, originality, and final
content of the paper.

\vspace{3pt}
\noindent\textbf{Data statement}: No real-world data were used in this study.
The certificate inputs and numerical instances are specified in
\Cref{app:certificates,app:finite-study}. All code can be made available upon request.


\appendix
\crefalias{section}{appendix}
\crefalias{subsection}{subappendix}

\section{Time averaging, varying means, and the diffusion passage}
\label{app:averaging}

We prove the reductions for unrestricted policies in
\cref{lem:reduction,lem:conditional,lem:signed}. At fixed model
parameters, choose a time-average subsequence with vanishing
boundary terms and apply the finite-input comparison on the fixed grid.
Then take the diffusion limit.

\subsection{Lost sales at fixed parameters}

Let $I(t)$ be on-hand inventory, $A$ cumulative arrivals, and $Y$ cumulative
lost demand under a causal policy of finite limsup average cost $C$.
Write $g(t)=\E\left[I(t)\right]$. Then $\int_0^Tg(t)\,dt=O(T)$. On $[n,2n]$ both
$g(t)$ and $\int_{t-L}^tg(s)\,ds$ have integrals $O(n)$, the latter by
Fubini. Each exceeds $\sqrt n$ on a set of length $O(\sqrt n)$.
Hence we can choose $T_n\in[n,2n]$ for which
\begin{equation*}
 \E\left[I(T_n)\right]+\int_{T_n-L}^{T_n}\E\left[I(t)\right]\,dt=O(\sqrt n)=o(T_n).
\end{equation*}
Finite initial intervals cause no difficulty by local integrability.

Let $S_n$ be independent uniform on $[0,T_n-L]$. Conditional on any
deterministic anchor $S_n=s$, the relative future Poisson path has the same
law and is independent of the history at $s$. It is therefore independent
of the pair consisting of the random anchor and that history. Every arrival
in $(S_n,S_n+L]$ is measurable in this prewindow information, since its order
was placed no later than $S_n$.

Average the inventory balance over the prewindow information and independent
policy randomization conditional on the relative future demand path. With
$D_s$ a fresh rate-$\lambda$ Poisson process, this gives
\begin{equation*}
 X_s^{(n)}=H_n+a_n(s)-D_s+\mathcal A_s^{(n)}\ge0,
 \qquad 0\le s\le L,
\end{equation*}
where $H_n=(T_n-L)^{-1}\int_0^{T_n-L}g(t)\,dt$ and $a_n(s)$ is deterministic.
The averaged cumulative lost demand $\mathcal A^{(n)}$ is adapted and nondecreasing.
Indeed, lost demand through a date is a measurable function of prewindow
information and the demand prefix through that date. Integrating the former
cannot introduce dependence on later increments. A \emph{càdlàg} version, right-continuous with left limits, is obtained
by summing the conditional loss probabilities at the finitely many demand
epochs of a bounded window. The averaged control is not required to act only
when $X^{(n)}=0$.

Put $r_n=\E\left[A(0,T_n]\right]/T_n$ and $d_n=\E\left[Y(T_n)\right]/T_n$.
For every $s\in[0,L]$, Fubini bounds the difference between
$a_n(s)$ and $r_ns$ by a constant multiple of
\[
 \frac{L}{T_n-L}\left(\E\left[A(0,L]\right]+\E\left[A(T_n-L,T_n]\right]\right)+O(T_n^{-1}).
\]
The inventory balance over the last window gives
$A(T_n-L,T_n]\le D(T_n-L,T_n]+I(T_n)$ pathwise, because sales do not
exceed demand and initial inventory is nonnegative. Thus the displayed error
vanishes uniformly in $s$. Similarly,
\begin{equation*}
 \sup_{s\le L}|\E\left[X_s^{(n)}\right]-H_n|\to0,
 \qquad r_n+d_n\to\lambda.
\end{equation*}
The first bound follows by shifting the inventory integral: its difference
uses only fixed initial and selected terminal windows. The second follows
from total inventory balance and the endpoint bound. Lost-demand
boundary terms are at most the expected demand in a fixed window, so
$\E\left[\mathcal A_s^{(n)}\right]=d_ns+o(1)$ uniformly as well. The selected means are bounded,
and after a subsequence $H_n\to H$, $d_n\to d\in[0,\lambda]$.
The policy's limsup cost is at least $H+ p d$; replacing the inventory
integral up to $T_n$ by the one up to $T_n-L$ changes it by $o(T_n)$.

To apply the constant-mean recursion, first remove the deterministic
arrival-profile error. Define
$e_n(s)=a_n(s)-r_ns$ and $\varepsilon_n=\sup_{s\le L}|e_n(s)|\to0$.
Then
\[
 \widetilde X_s^{(n)}=X_s^{(n)}-e_n(s)+\varepsilon_n
 =H_n+\varepsilon_n+r_ns-D_s+\mathcal A_s^{(n)}\ge0.
\]
The input is homogeneous. The grid means
$m_{i,n}=\E\left[\widetilde X_{Lt_i}^{(n)}\right]$ may differ at finite
$n$, and each tends to $H$.

\subsection{Grid recursion with different target means}

Fix independent input blocks with finite expected suprema, an initial
deterministic mean $m_0$, and feasible target means $m_1,\ldots,m_K$.
In the nonnegative problem, first apply reflection from the previous
corrected state. If the resulting mean is below $m_i$, floor-correct to
$m_i$. If the means are equal, leave the reflected state as is. The induction in
\cref{thm:grid} applies without change:
the previous compared states have
the common mean $m_{i-1}$, their reflected states are ordered increasingly
convexly, and the call identity compares corrections to the common
mean $m_i$. Physical feasibility implies that the minimal pre-correction
mean is at most $m_i$ at every step.

If input paths converge in expected uniform distance, starting values converge
in $L^1$, and target means converge, reflection is continuous in $L^1$ by
its Lipschitz property. \Cref{lem:floor} gives continuity of each correction.
Induction proves continuity of the whole fixed finite recursion, including
the boundary where the pre-correction mean equals the target mean. Closedness
of each inequality $\E\left[R_i\right]\le m_i$ then proves feasibility of the limiting
recursion. Apply this argument at fixed Poisson parameters to the preceding
$m_{i,n}$ and $r_n$. We obtain the exact constant-mean finite-Poisson
necessary grid test with initial mean $H$ and arrival rate $\lambda-d$.

For the signed problem replace reflection by addition of the independent
input block. The same call comparison and continuity argument applies. The
comparison is also valid against an actual state with extra dependence on
independent past information, because its next demand block remains independent
of that information and of its current state.

\subsection{Dual sourcing at fixed parameters}

At fixed $p>0$, finite average cost implies
$\int_0^T\E\left[|I(t)|\right]\,dt=O(T)$. For a fixed finite collection of window
lengths and shifts, Markov's inequality and Fubini select $T_n\in[n,2n]$
such that the endpoint means $\E\left[|I(T_n-s)|\right]$ at all needed shifts and the
terminal window integrals are $o(T_n)$. There are only finitely many bad
sets of length $O(\sqrt n)$ to exclude. Let $A_R$ and $A_E$ be cumulative regular and express arrivals.
Total-arrival balance gives
\[
 A_R(0,T_n]+A_E(0,T_n]=D(0,T_n]+I(T_n)-I(0),
\]
so subsequential regular rate is $\lambda-d$, where $d$ is express rate.
If premiums are charged on placement, the express-arrival count is at most
the express-order count plus the initial pipeline. Hence the normalized
arrival-rate cost used below is bounded above by the actual premium cost
along the selected subsequence. This inequality suffices for a lower bound.
For any boundary window $[a,a+q]$, total arrivals are at most
$D(a,a+q]+|I(a+q)|+|I(a)|$. The endpoint selection therefore controls
regular and express arrival boundary errors separately. Shifted time averages
of $I^+$ and $I^-$ have the same limiting means $H,\bar B$.

On a length-$L$ window, regular arrivals are prewindow measurable because
$L\le L_E+L$. Conditional averaging gives a signed state starting at
mean $H-\bar B$ and an adapted nondecreasing express-arrival control. A version
of this conditional cumulative measure is obtained from rational intervals
and right limits; local finiteness and finite expected mass give a càdlàg
version. Its deterministic regular-arrival profile differs uniformly from
$(\lambda-d)s$ by a vanishing error. Subtract that error, allow the
corresponding varying grid means, and use the signed continuity argument.
The endpoint selection can be made using shifts $0,L$ and the terminal
length-$L$ integral. The arrival-profile and shifted-mean errors are then uniform
over $0\le s\le L$, by the same boundary-window estimates as in the lost-sales
argument. Thus one subsequential triple $(H,\bar B,d)$ supplies the comparison for
every fixed finite grid. No new triple is chosen when the number of grid dates
changes. Conditional Jensen gives the positive-part bound by $H$ and the negative-part
bound by $\bar B$ before the limit. The drift correction is vanishing and both
part functions are $1$-Lipschitz, so both bounds survive. This proves the
finite-Poisson precursor of \cref{lem:signed}.

For \cref{lem:conditional}, independence is used at each deterministic
terminal time before averaging. 
Regular arrivals and initial inventory are
independent of the complete future window. The containing event $F_j^{(c)}$ uses only demand after the interior
backlog time, so it is independent of that backlog. Average those valid
inequalities over terminal time and use the same boundary bounds.
This
proves the finite-parameter linear comparison without a stationarity assumption.

\subsection{Poisson to Brownian input}

Now let $L\to\infty$ with $\lambda$ fixed, after completing the long-run
argument at every finite parameter. The centered process
$M_L(t)=(\lambda Lt-D_{Lt})/\sqrt{\lambda L}$ is a square-integrable
martingale with predictable quadratic variation $t$ and jumps of absolute
size $1/\sqrt{\lambda L}$. Thus the predictable-quadratic-variation and
vanishing-jump conditions of Whitt~\citep{Whitt2007} hold.
It converges in the \emph{Skorokhod topology}, which allows small time changes of càdlàg paths, to standard Brownian motion.
Since that limit is continuous, it admits a representation with convergence
in the uniform norm in probability. Doob's inequality gives
\[
 \E\left[\sup_{t\le1}|M_L(t)|^2\right]\le4.
\]
The same bound gives uniform integrability and hence convergence of expected
uniform distances under such a coupling. For bounded subsequential $u,v$,
the finite-grid continuity argument transfers the Poisson comparison to its
Brownian counterpart. In the nonnegative case floors are at most the target
mean, giving local moment bounds; in the signed case the explicit bound in
\cref{lem:signed} supplies them. The positive and negative part bounds pass
in $L^1$. Finite-vector event comparisons in \cref{lem:conditional} follow
from the same input convergence, null event boundaries, and the displayed
second-moment bound.

Finally, apply these arguments to a sequence of policies with costs approaching
the finite-parameter optimum. Whenever normalized costs have a bounded liminf,
their nonnegative cost components admit bounded subsequences (with fixed
$p>0$ in dual sourcing). Extract the needed mean subsequences. If the costs
are unbounded, every stated lower comparison is automatic. This proves the
reductions for the full infimum benchmark. At no point is a joint diffusion
limit for a controlled policy asserted.

\section{Extensions to finer grids}
\label{app:grid-extensions}

\subsection{A convex dual on any fixed grid}
\label{app:multistage}

The two-block dual is part of a general necessary hierarchy. On a grid
$0=t_0<\cdots<t_K=1$, choose \emph{dual weights} $\omega_i\ge0$, $1\le i<K$, with
$\sum_i\omega_i<1$, and put $f_K(z,v)=z$ for $z\ge0$. Recursively define
\[
 f_i(r,v)=\inf_{a\ge r}\left\{-\omega_i a+
       \E\left[f_{i+1}(\Gamma_{t_{i+1}-t_i}(a,W;v),v)\right]\right\}.
\]
Then $f_i$ is jointly convex and nondecreasing in $r$, and every grid-feasible
pair satisfies
\[
 \mathcal F_{\boldsymbol\omega}(u,v):=
 \E\left[f_1(\Gamma_{t_1}(u,W;v),v)\right]
       -(1-\textstyle\sum_i\omega_i)u\le0.
\]
For finiteness, put $c_i=1-\sum_{j=i}^{K-1}\omega_j>0$ and verify by
backward induction that $|f_i(r,v)-c_i r|\le B_i(v)$ for some finite $B_i(v)$.
The claim holds at $i=K$ with $B_K(v)=0$. For $s=t_{i+1}-t_i$,
\[
 a-vs\le\E\left[\Gamma_s(a,W;v)\right]
 \le a+2\E\left[\sup_{0\le b\le s}|W_b|\right]+|v|s.
\]
The lower estimate bounds the minimized expression below by
$c_i a-c_{i+1}vs-B_{i+1}(v)$ for every $a\ge r$. The upper estimate at
$a=r$ bounds it above by $c_i r$ plus a finite term independent of $r$.
These estimates prove the induction. Convexity follows by composition
with the reflected endpoint, using monotonicity of $f_{i+1}$ in its first
argument, then taking expectation and minimizing over $a\ge r$. Increasing the lower constraint restricts the
minimization domain, so $f_i$ is nondecreasing in $r$.

For a feasible process, let $R_i$ be the minimal reflected endpoint of block
$i$ starting from its actual preceding state $X_{t_{i-1}}$. The actual next
state dominates $R_i$. Monotonicity of $f_{i+1}$ and substitution of the actual
state in the infimum give
$\E\left[f_i(R_i,v)\right]\le-\omega_i\E\left[X_{t_i}\right]+\E\left[f_{i+1}(R_{i+1},v)\right]$.
Sum these inequalities backward and use the common mean $u$ at every grid
date. This proves the necessary inequality. Its left side is jointly convex,
so every subgradient yields a global supporting halfspace. For $K=2$ and
$\omega_1=1-\theta$, the identity
$\E\left[\Gamma_\Delta(a,W;v)\right]=a-v\Delta+g_\Delta(a,v)$ reduces it exactly
to \eqref{eq:convex-dual}. No strong-duality theorem or infinite-grid
convergence rate is needed for the finite certificate in this paper.

\subsection{Record charges on finer grids}
\label{app:record-charges}
On a finer grid, an interior date is charged only when the corresponding
future-demand score sets a strict record. Record probabilities decrease with
the number of dates, so their sum grows only on the square-root scale.
\begin{lemma}
\label{lem:record}
For $J$ equally spaced dates and nonnegative selector drift, define the
future-only strict record events from the corresponding Gaussian scores.
Then, with the \emph{strict-survival probability} $q_j=4^{-j}\binom{2j}{j}$ and $q_0=1$,
\[
 \Prb(F_j)\le q_j,\qquad
 \chi\le(2J-1)q_{J-1}\le(2J-1)/\sqrt J<2\sqrt J.
\]
The same bound holds for symmetric finite quantizers with earliest-branch ties.
\end{lemma}

\begin{proof}
Reverse the first $j$ iid increments. A positive record requires every reversed
partial sum after subtraction of nonnegative drift to be positive. Dropping
the drift enlarges the event. For continuous symmetric iid increments let
$q_j$ be the probability that all first $j$ partial sums are positive.
The unique maximum of $S_0,\ldots,S_n$ occurs at $k$ with probability
$q_kq_{n-k}$: the reversed prefix is positive, the independent suffix is
negative, and symmetry identifies its probability. Hence
$\sum_{k=0}^nq_kq_{n-k}=1$. Squaring the generating series gives
$(\sum q_jz^j)^2=(1-z)^{-1}$, proving the displayed binomial expression.
This is the classical symmetric-walk identity associated with
Sparre Andersen~\citep{Andersen1953}.
The recurrence $q_{j+1}/q_j=(2j+1)/(2j+2)$ gives
$\sum_{j=0}^{J-1}q_j=(2J-1)q_{J-1}$ and, by induction,
$q_j\le1/\sqrt{j+1}$. Use $p_J\le1$ to bound $\chi$.
If a symmetric finite quantizer has no strictly positive sum of at most
$J$ increments, all strict record events are empty. Otherwise, let
$\varepsilon_Q>0$ be the minimum of these finitely many positive sums.
Independent continuous symmetric jitter of absolute size less than
$\varepsilon_Q/J$ preserves every strict-survival path. The perturbed
increments are continuous, iid, and symmetric; their survival probability
is $q_j$. This proves the stated bound for any fixed symmetric finite quantizer.
\end{proof}

\section{Exact certificates and their reproducible inputs}
\label{app:certificates}

This appendix records the normal comparisons and the certificate
computations used in the two approximation guarantees.
The supporting inequalities and the parameter partition were proved above;
the computations establish their numerical margins. Decimal inputs in the
tables are exact rationals. All displayed decimal upper bounds below are
rounded upward from rational enclosures.

\subsection{Normal comparisons used in the parameter partition}
\label{app:normal}

With the definitions of $\psi$, $\beta$, $\U$, and $\V$ in \Cref{sec:models},
Xin~\citep{Xin2022} establishes that the root
$\beta(\alpha)$ is unique and decreasing and that $\U$ is increasing.
Xin~\citep{Xin2022} proves that $\U/\V$ is nondecreasing
on $(0,\alpha^*]$. The theorem applies directly: our functions and crossing
$\alpha^*$ are identical, and the rational cutoff satisfies $\alpha_0<\alpha^*$.
Normal symmetry gives $\V(\alpha)=\V(1/\alpha)$, and
Xin~\citep{Xin2022} records that $\V$ increases up to one
and decreases thereafter.

For $t>0$, the normal-tail inequalities in
Lin and Bai~\citep{LinBai2010}, also
stated in Xin~\citep{Xin2022}, give
\[
 \frac{t\phi(t)}{1+t^2}<\Q(t)<\frac{\phi(t)}t,
 \qquad \psi(t)\ge H_M(t):=\frac{3t+\sqrt{t^2+8}}4.
\]
Only nonnegative arguments are used in the small-backlog bounds below;
the value at zero follows by continuity. The normal hazard also satisfies
$0<\psi'(t)<1$: its derivative is $\psi(t)(\psi(t)-t)>0$ and the
variance of a normal variable conditioned to exceed $t$ is $1-\psi'(t)>0$.
Alternating \emph{Mills expansions} used by the interval arithmetic follow from
finite integration by parts, retaining the sign of the remainder.

Rational interval bisection, with root signs checked, gives
\begin{equation*}
 0.69787463947<\alpha^*<0.69787463949,~~
 0.8619684164042<\beta(\alpha^*)<0.8619684164279.
\end{equation*}
In particular, $\alpha_0<\alpha^*$ and $\beta(\alpha^*)>0.86$.
The same arithmetic checks the density cap in \Cref{lem:di-shift}.

\subsubsection{The small-backlog analytic inequalities}

For $t\ge0$, integrate $\psi(s)\le s+\psi(0)$, using
$\psi'<1$, $\psi(0)<4/5$, and $\log2<7/10$. Since
$d\log(1/\Q(t))/dt=\psi(t)$, this gives
$2\log(1/\Q(t))\le t^2+8t/5+7/5$.
The sharper Mills bound also gives
$\psi(t)\ge3t/4+1/\sqrt2$. The exact CDF evaluation
$\Phi(1)<6/7$ supplies the cutoff $1/7<\Q(1)$ used in
\Cref{lem:small-backlog}. These facts prove its analytic tail; its remaining
fifty cells involve arguments between zero and four, so the CDF enclosures
in the next subsection apply directly.

\subsection{Arithmetic primitives}

\emph{Interval arithmetic} encloses each exact value between rational endpoints, rounded outward after each operation
to the lattice $10^{-40}\mathbb Z$. Square roots use integer square-root
bounds at that scale. Machin's identity
$\pi=16\arctan(1/5)-4\arctan(1/239)$, with eighty alternating terms in each
arctangent and its next-term error, encloses $\pi$. For $0\le w\le10$,
the degree-100 Taylor polynomial for $e^{-w}$ has error at most
$w^{101}/101!$. Larger nonnegative arguments are halved until they lie in
that domain, then evaluated and squared back. Positive exponentials use
reciprocals of positive intervals. Monotone endpoint evaluation avoids
dependency inflation on wide intervals.

For $0\le x\le4$, integrate the density expansion to obtain
\[
 \Phi(x)=\frac12+\frac1{\sqrt{2\pi}}
   \sum_{k=0}^{100}\frac{(-1)^kx^{2k+1}}{2^kk!(2k+1)}+R,
 \quad
 |R|\le\frac{x^{203}}{\sqrt{2\pi}\,2^{101}101!\,203}.
\]
Symmetry covers negative arguments. For $x>4$, four and five integration-by-parts
terms give, respectively, the lower and upper tail bounds
\[
 \phi(x)(x^{-1}-x^{-3}+3x^{-5}-15x^{-7})
 \le\Q(x)\le
 \phi(x)(x^{-1}-x^{-3}+3x^{-5}-15x^{-7}+105x^{-9}).
\]
The sign of each remainder follows from the remaining positive Gaussian
integral. For a CDF interval, midpoint evaluation plus the global Lipschitz
constant $2/5$ encloses its whole range. A density interval uses the nearest
and farthest distances from zero. Quantile bisection preserves a validated
initial bracket and updates an endpoint only when the CDF interval proves
the required sign. In particular, the small-$\alpha$ quantile has bracket
$[2,4]$, and all other normal quantiles used here have bracket $[0,4]$.
The actual brackets satisfy their CDF inequalities. For $x>0$, logarithms use
\[
 \log(1+x)=2\sum_{j=0}^{N-1}\frac{z^{2j+1}}{2j+1}+R_N,
 \quad z=\frac{x}{2+x},\quad
 0\le R_N\le\frac{2z^{2N+1}}{(2N+1)(1-z^2)}.
\]
These formulas suffice for every primitive in the certificates.

\subsection{Supporting-plane integration}

For each trial in \Cref{tab:plane-trials}, integrate the three
coefficient functions from $k$ to $Z=u+6$ by \emph{Simpson quadrature}
with panel widths at most $1/10$. For a panel of width $h$, the error
is at most $h^5\sup|f^{(4)}|/2880$. If $M_4$ bounds
$|f^{(4)}|/4!$ on the panel, the same bound is $h^5 M_4/120$.
Compute the interval Taylor coefficients through degree four, using
convolution for products and derivatives of $\phi'=-x\phi$ and
$\Phi'=\phi$ for the normal functions.

\begin{table}[htbp]
\centering\small
\caption{Trial coordinates and safe necessary inequalities $Au+Cv\ge K$. All decimals are exact. The stage fraction is $1/2$ and Simpson mesh is at most $1/10$.}
\label{tab:plane-trials}
\begin{adjustbox}{max width=.8\linewidth}
\begin{tabular}{@{}rrrrrrr@{}}
\toprule
Index & $u_0$ & $v_0$ & $k_0$ & $A$ & $C$ & $K$ \\\midrule
0 & 3.227987051 & 0.000554641 & 1.111583460 & 0.001114010228 & 0.556959770364 & 0.003904304426 \\
1 & 2.942268682 & 0.001496171 & 1.031892433 & 0.002848542103 & 0.569684220377 & 0.009232773469 \\
2 & 2.708297177 & 0.003190906 & 0.966124384 & 0.005807962486 & 0.580781376253 & 0.017582056348 \\
3 & 2.456147504 & 0.006844776 & 0.894677524 & 0.011851505119 & 0.592566085591 & 0.033164050179 \\
4 & 2.183449959 & 0.014748673 & 0.816711216 & 0.024137117742 & 0.603422374085 & 0.061600772897 \\
5 & 1.889247315 & 0.031795919 & 0.731698878 & 0.048773185288 & 0.609661627279 & 0.111528142283 \\
6 & 1.606354958 & 0.063369139 & 0.648623871 & 0.090823406828 & 0.605487652789 & 0.184262518259 \\
7 & 1.370826392 & 0.109484866 & 0.577340494 & 0.147420086572 & 0.589679512796 & 0.266646826737 \\
8 & 1.157260453 & 0.177551237 & 0.509056925 & 0.224450161710 & 0.561125127264 & 0.359373948280 \\
9 & 1.018724803 & 0.242720284 & 0.461376359 & 0.293055911851 & 0.532829102188 & 0.427869548807 \\
10 & 0.949104557 & 0.284375782 & 0.435913981 & 0.334733688784 & 0.514975386930 & 0.464141261235 \\
11 & 0.920236850 & 0.303820697 & 0.424991246 & 0.353650779651 & 0.506758108227 & 0.479403324607 \\
\bottomrule
\end{tabular}
\end{adjustbox}
\end{table}

Let $s=\sqrt t$, $r=\sqrt\Delta$,
$a_Z=(Z-u+vt)/s$, and $q_Z=(Z+u-vt)/s$. The omitted tails obey
\begin{align*}
 \int_Z^\infty\bar F\,dz&\le s\phi(a_Z)+2s\phi(Z/s),&
 \int_Z^\infty M\,dz&\le2r\phi((Z-v\Delta)/r),\\
 \int_Z^\infty \bar F_u\,dz&\le\Q(a_Z),&
 \int_Z^\infty F_v\,dz&\le t\Q(a_Z)
       +2t\max\{1,Z/(Z+u-vt)\}\Q(q_Z),\\
 \int_Z^\infty M_v\,dz&\le\Delta M(Z).
\end{align*}
Here $\bar F=1-F$. The first inequality separates
the unreflected Gaussian tail and a negative-drift Brownian maximum bounded
by the zero-drift maximum. The maximum with positive drift is bounded by
$v\Delta+\sup W$, giving the second inequality. For the third,
$0\le \bar F_u\le\phi(a_z)/s$. For the fourth, use its explicit formula,
Mills' inequality, $e^{-2vz}\le1$, and
$z/(z+u-vt)\le\max\{1,Z/(Z+u-vt)\}$ for $z\ge Z$.
For the fifth, increasing the drift by $h$ raises a block maximum by at
most $h\Delta$. Differentiate its excess expectation to obtain the bound.
Every trial satisfies $Z>v\Delta$ and $Z+u-vt>0$.
The tail of the first integrand in the coefficient formulas lies
between zero and $\theta\int\bar F+\int M$; the second between zero and
$\theta\int \bar F_u$; the third between $-\theta\int F_v$ and $\int M_v$.
These signs are used before outward rounding.

All entries in the last three columns are necessary halfspace coefficients,
with $A,C$ rounded up and $K$ rounded down to twelve decimal places. Exact
intersection enumeration includes every pair of boundaries and both axes;
retain only nonnegative intersections satisfying all twelve inequalities.
There are thirteen vertices. The printed coefficients suffice for the factor.

For the one hundred BS cells, define $N_j$ as the nearest integer to
$2{,}000{,}000(69787/200)^{j/100}$ and $\alpha_j=N_j/10^9$, $0\le j\le100$.
The cell with index $j$ is $[\alpha_j,\alpha_{j+1}]$. The endpoints increase,
start at $\alpha_s$, and end at $\alpha_0$. Their integer specification is
certified without floating powers by comparing
$(N_j\pm1/2)^{100}$ with
$2{,}000{,}000^{100}(69787/200)^j$. No tie occurs. The trial on cell $j$
is the corresponding exact value in \cref{tab:bs-trials}.

\begin{table}[htbp]
\centering\small
\caption{BS trials for the one hundred lost-sales cells. Cell endpoints are given by the integer formula in the text; $j$ is zero-based.}
\label{tab:bs-trials}
\begin{adjustbox}{max width=.5\linewidth}
\begin{tabular}{@{}rrrrrrrr@{}}
\toprule
$j$ & $\beta_j$ & $j$ & $\beta_j$ & $j$ & $\beta_j$ & $j$ & $\beta_j$ \\\midrule
0 & 3.620905996 & 25 & 3.149785692 & 50 & 2.583055632 & 75 & 1.853455831 \\
1 & 3.603395241 & 26 & 3.129263772 & 51 & 2.557606810 & 76 & 1.819368038 \\
2 & 3.585787423 & 27 & 3.108590069 & 52 & 2.531899362 & 77 & 1.784821320 \\
3 & 3.568080848 & 28 & 3.087761581 & 53 & 2.505927338 & 78 & 1.749805352 \\
4 & 3.550273877 & 29 & 3.066775174 & 54 & 2.479684633 & 79 & 1.714309638 \\
5 & 3.532364891 & 30 & 3.045627643 & 55 & 2.453164978 & 80 & 1.678323512 \\
6 & 3.514352220 & 31 & 3.024315721 & 56 & 2.426361949 & 81 & 1.641836141 \\
7 & 3.496234103 & 32 & 3.002836042 & 57 & 2.399268957 & 82 & 1.604836529 \\
8 & 3.478008673 & 33 & 2.981185129 & 58 & 2.371879242 & 83 & 1.567313515 \\
9 & 3.459674168 & 34 & 2.959359440 & 59 & 2.344185873 & 84 & 1.529255781 \\
10 & 3.441228659 & 35 & 2.937355302 & 60 & 2.316181747 & 85 & 1.490651843 \\
11 & 3.422670252 & 36 & 2.915168948 & 61 & 2.287859586 & 86 & 1.451490062 \\
12 & 3.403996968 & 37 & 2.892796535 & 62 & 2.259211946 & 87 & 1.411758638 \\
13 & 3.385206737 & 38 & 2.870234081 & 63 & 2.230231196 & 88 & 1.371445612 \\
14 & 3.366297514 & 39 & 2.847477494 & 64 & 2.200909524 & 89 & 1.330538861 \\
15 & 3.347267114 & 40 & 2.824522575 & 65 & 2.171238946 & 90 & 1.289026100 \\
16 & 3.328113363 & 41 & 2.801365027 & 66 & 2.141211295 & 91 & 1.246894876 \\
17 & 3.308834016 & 42 & 2.778000424 & 67 & 2.110818214 & 92 & 1.204132563 \\
18 & 3.289426725 & 43 & 2.754424195 & 68 & 2.080051174 & 93 & 1.160726354 \\
19 & 3.269889115 & 44 & 2.730631665 & 69 & 2.048901456 & 94 & 1.116663257 \\
20 & 3.250218773 & 45 & 2.706618035 & 70 & 2.017360160 & 95 & 1.071930082 \\
21 & 3.230413179 & 46 & 2.682378360 & 71 & 1.985418207 & 96 & 1.026513429 \\
22 & 3.210469747 & 47 & 2.657907567 & 72 & 1.953066330 & 97 & 0.980399676 \\
23 & 3.190385844 & 48 & 2.633200441 & 73 & 1.920295087 & 98 & 0.933574966 \\
24 & 3.170158753 & 49 & 2.608251628 & 74 & 1.887094853 & 99 & 0.886025181 \\
\bottomrule
\end{tabular}
\end{adjustbox}
\end{table}

The largest compact upper bound with the printed planes is less than
$1.248976724300$. It occurs on cell 99 at its right endpoint and at the
first vertex in the lexicographic vertex list. The small-region trial
is $3.629625443$. Its quotient is less than $1.231626782744$; the CO tail
quotient at $\alpha_0$ is less than $1.248825986315$.

\subsection{Quarter-window probabilities and rectangle errors}

For the probability integral in the two-date statistics, use rational
cells of width at most $1/50$ from $3\zeta-\xi$ to $8$. Approximate its
second CDF by its affine Taylor polynomial at the midpoint. Gaussian cell
mass and first moment integrate that polynomial exactly. Since
$|\Phi''|\le\phi(1)<1/4$, the absolute remainder is at most the cell
mass times the squared cell width divided by $32$. Add the omitted tail
between zero and $\Q(8/\sqrt3)$. All listed lower endpoints are below $8$.

For rectangle rules set integer scale $A_{\rm int}=10^{12}$. Enclose bin
probabilities by $P_i\in[\widehat P_i,\widehat P_i+2/A_{\rm int}]$ and
signed moments by $|C_i-\widehat C_i|\le2/A_{\rm int}$, with integer
multiples $A_{\rm int}\widehat P_i,A_{\rm int}\widehat C_i$ and
$\widehat P_i\ge0$, $\sum_i\widehat P_i\le1$. The elementary enclosures
above check every bin, including the extreme bins. Ties are decided by the
rational center scores; they involve no numerical tolerance.

On $J$ dates, product telescoping bounds the total mass error by
$\varepsilon_P=2Jn/A_{\rm int}$. The errors of $R$ and $\chi$ are at
most $J\varepsilon_P$. For a moment term, write the difference using the
true signed moment measure and the approximate probability product. The
moment measure has total absolute mass less than one, its approximation
has $L^1$ error at most $2n/A_{\rm int}$, and probability product error is
at most $2(J-1)n/A_{\rm int}$. Summing over the at most $J$ contributing
increments gives at most $2J^2n/A_{\rm int}$; we use the conservative
bound $\varepsilon_M=4J^2n/A_{\rm int}$. The entire enumeration is then
integer arithmetic with denominator $A_{\rm int}^J$, followed by these
outward errors. The numerator of the four-date lower comparison is rounded
down and each possible denominator rounded up.

\Cref{tab:dual-witnesses} gives all large-$p$ compact inputs. A Q rule is
the two-date statistics; G13 and G25 are the four-date quantizers.
At $\alpha\ge\alpha_0$ use G25 with $(\xi,\zeta)=(1.311002,0.127190)$
and window $\delta=\alpha_0$, giving a ratio less than $1.564173$.
The two remaining quarter-window witnesses at $\alpha_0$ are
$(\xi,\zeta)=(6.328054,1.466928)$ for $p\in[8/5,2]$ and
$(3.199890,0.667859)$ for $p\in[2,5]$; their ratio bounds are less than
$1.579572$ and $1.589410$, respectively.

\begin{table}[htbp]
\centering\small
\caption{Complete compact certificate for $p\ge5$. Each row supplies its selection rule and the two BS endpoint trials. Decimal values are exact rationals.}
\label{tab:dual-witnesses}
\begin{adjustbox}{max width=.5\linewidth}
\begin{tabular}{@{}rrrcrrrr@{}}
\toprule
Cell & $l$ & $r$ & Rule & $\xi$ & $\zeta$ & $\beta_l$ & $\beta_r$ \\\midrule
0 & 0.3000000 & 0.3250870 & Q & 2.593017 & 0.242243 & 1.487524 & 1.433599 \\
1 & 0.3250870 & 0.3504126 & Q & 1.909239 & 0.080974 & 1.433599 & 1.382251 \\
2 & 0.3504126 & 0.3746898 & Q & 1.839920 & 0.085338 & 1.382251 & 1.335582 \\
3 & 0.3746898 & 0.3976703 & Q & 1.773150 & 0.088088 & 1.335582 & 1.293448 \\
4 & 0.3976703 & 0.4191664 & Q & 1.723810 & 0.093424 & 1.293448 & 1.255653 \\
5 & 0.4191664 & 0.4467602 & G13 & 2.656372 & 0.328369 & 1.255653 & 1.209201 \\
6 & 0.4467602 & 0.5027787 & G13 & 1.968872 & 0.156494 & 1.209201 & 1.121106 \\
7 & 0.5027787 & 0.5411561 & G13 & 1.812622 & 0.156494 & 1.121106 & 1.064882 \\
8 & 0.5411561 & 0.5661199 & G13 & 1.656372 & 0.156494 & 1.064882 & 1.029877 \\
9 & 0.5661199 & 0.5698815 & Q & 1.413969 & 0.123174 & 1.029877 & 1.024702 \\
10 & 0.5698815 & 0.5731199 & Q & 1.407633 & 0.123631 & 1.024702 & 1.020267 \\
11 & 0.5731199 & 0.5759046 & Q & 1.402482 & 0.124098 & 1.020267 & 1.016468 \\
12 & 0.5759046 & 0.5782950 & Q & 1.399486 & 0.124884 & 1.016468 & 1.013218 \\
13 & 0.5782950 & 0.5803460 & Q & 1.396002 & 0.125304 & 1.013218 & 1.010437 \\
14 & 0.5803460 & 0.5821038 & Q & 1.392118 & 0.125415 & 1.010437 & 1.008060 \\
15 & 0.5821038 & 0.5836082 & Q & 1.389641 & 0.125740 & 1.008060 & 1.006030 \\
16 & 0.5836082 & 0.5848960 & Q & 1.388504 & 0.126286 & 1.006030 & 1.004294 \\
17 & 0.5848960 & 0.6238037 & G13 & 2.312622 & 0.328369 & 1.004294 & 0.953155 \\
18 & 0.6238037 & 0.6580943 & G13 & 1.937622 & 0.234619 & 0.953155 & 0.910002 \\
19 & 0.6580943 & 0.6812146 & G13 & 1.625122 & 0.156494 & 0.910002 & 0.881833 \\
20 & 0.6812146 & 0.6835950 & G13 & 1.718872 & 0.234619 & 0.881833 & 0.878973 \\
21 & 0.6835950 & 0.6978700 & G25 & 1.154720 & 0.051787 & 0.878973 & 0.861974 \\
\bottomrule
\end{tabular}
\end{adjustbox}
\end{table}

For the moderate-$p$ compact cells in the proof, use $\beta=1.1$ at
$\alpha=1/2$, and the successive right-endpoint trials
$1.052347,0.984152,0.920036,0.889334$ at
$11/20,3/5,13/20,27/40$. The final endpoint uses $\sqrt2$.
The small-$\alpha$ comparison uses trial $1.487524$ at $3/10$.
The small-$p$ cells have endpoints $j/60$, $10\le j\le60$, as specified
in \Cref{lem:small-backlog}. Thus every finite comparison has explicit inputs.

There are 106 bounds in the final envelope: four region or tail bounds,
six moderate-$p$ compact bounds, two finite-$p$ end bounds, forty-four
large-$p$ endpoint bounds, and fifty small-$p$ cell bounds. Their largest
upper endpoint is exactly
\[
 \frac{399881819707912762723889056027176246537}
      {250000000000000000000000000000000000000}
 <1.599527279<r_{\rm DS}.
\]
The fraction is an upper enclosure, not the exact approximation ratio.
\Cref{tab:certificate-summary} records the checks that enter the main proof.

\begin{table}[htbp]
\centering\small
\caption{Conservative upper enclosures used in the approximation proof. These are certified comparison bounds, not observed policy ratios.}
\label{tab:certificate-summary}
\begin{adjustbox}{max width=.6\linewidth}
\begin{tabular}{@{}ll@{}}
\toprule
Check & Certified outcome \\\midrule
Normal ratio monotonicity & Xin~\citep[Theorem~4]{Xin2022} \\
Lost sales, compact region & $<1.248976725$ (12 planes, 13 vertices, 100 cells) \\
Lost sales, small region & $<1.231626783$ \\
Lost sales, CO tail & $<1.248825987$ \\
Dual sourcing, moderate-$p$ compact cases & $<1.597689$ \\
Dual sourcing, finite-$p$ end cases & $<1.589410$ \\
Dual sourcing, large-$p$ tail & $<1.564173$ \\
Dual sourcing, small-$p$ cells & $<1.475695$ \\
Dual sourcing, all 106 case bounds & $<1.599527279$ \\
\bottomrule
\end{tabular}
\end{adjustbox}
\end{table}

\subsection{Sensitivity to the number and precision of inequalities}
\label{app:precision}

\Cref{tab:plane-ablation} first compares nested subsets of three,
six, and twelve supporting inequalities, always with the same BS trials.
Removing an inequality enlarges the lower-bound feasible region; this weakens
the guarantee without changing the two-block relaxation or the policy menu.

\begin{table}[htbp]
\centering\small
\caption{Sensitivity to the number of retained supporting inequalities.
All rows use the same 100 BS cells and printed, safely rounded coefficients.
The entries are upper enclosures for the stated certificate comparisons;
they are not measured policy-to-optimum ratios.}
\label{tab:plane-ablation}
\begin{adjustbox}{max width=.6\linewidth}
\begin{tabular}{@{}lrrr@{}}
\toprule
Retained indices & Vertices & Compact upper bound & CO tail upper bound\\\midrule
0,5,11 & 4 & $1.391503589$ & $1.248825987$ \\
0,3,5,7,9,11 & 7 & $1.252099509$ & $1.248825987$ \\
0,1,2,3,4,5,6,7,8,9,10,11 & 13 & $1.248976725$ & $1.248825987$ \\
\bottomrule
\end{tabular}
\end{adjustbox}
\end{table}

Round $A,C$ upward and $K$ downward to six, eight, ten, and twelve
decimal places. For each precision, enumerate the vertices and
check the endpoints of the original one hundred cells with the same
BS trials. \Cref{tab:precision} gives the outward-rounded bounds.
Eight decimal places suffice for $1.248977$ with these trials.
At six decimal places the computed bounds exceed $1.248977$.

\begin{table}[htbp]
\centering\small
\caption{Effect of outward rounding of the twelve printed inequalities.
Every polyhedron has thirteen vertices. The small-$\alpha$ comparison is unchanged.}
\label{tab:precision}
\begin{adjustbox}{max width=.6\linewidth}
\begin{tabular}{@{}rrrr@{}}
\toprule
Decimal places & Compact upper bound & CO tail upper bound & Below $1.248977$\\\midrule
6 & $1.248979768$ & $1.248829030$ & No\\
8 & $1.248976741$ & $1.248826003$ & Yes\\
10 & $1.248976725$ & $1.248825987$ & Yes\\
12 & $1.248976725$ & $1.248825987$ & Yes\\
\bottomrule
\end{tabular}
\end{adjustbox}
\end{table}

\section{Finite-parameter computational study}
\label{app:finite-study}

\subsection{Finite lost-sales costs and gap bounds}
For $\lambda=1$, we evaluate 27 instances with
$\alpha\in\{0.1,0.69787,2\}$ and nine penalties from $20$ to $10{,}000$.
For each instance, we minimize the exact BS formula \eqref{eq:bs-exact} over the finite target
range and the CO formula over its rate. The CO root satisfies
$1-(1+w)e^{-w}=1/ p$, and its minimum cost is $ p(1-e^{-w})$.
In addition, we evaluate the finite newsvendor lower bound
\[
 B_{\rm NV}(L, p)=\inf_s\E\left[(s-D_L)^++( p/L)(D_L-s)^+\right]
 \le\OPT_{\rm LS}(L, p),
\]
imported from Xin~\citep{Xin2022}.
Integer interval recurrences enclose the BS minimum and Poisson lower bound;
rational root brackets enclose the CO minimum. \Cref{app:finite-costs}
gives the tail and rounding controls. Separate high-precision calculations check the newsvendor lower bounds.

The lead-time monotonicity in \Cref{lem:policy-facts} also permits
evaluating this lower bound at a shorter lead time. We use
\[
 \widetilde B_{\rm NV}(L, p)=
 \max\{B_{\rm NV}(L, p),B_{\rm NV}(\min\{L, p\}, p)\}
 \le\OPT_{\rm LS}(L, p).
\]
Indeed, the shorter-system newsvendor bound is at most its optimum, which
is at most the original-system optimum. This applies two existing comparisons;
it does not change the policy menu. Nine additional lower-bound evaluations
suffice for the $\alpha=2$ cases; the other eighteen cases are unchanged.

\begin{figure}[htbp]
\centering
\begin{adjustbox}{max width=.7\linewidth}
\begin{tikzpicture}
\begin{groupplot}[group style={group size=2 by 1,horizontal sep=1.55cm},
 width=.47\linewidth,height=5.7cm,xmode=log,xmin=20,xmax=10000,
 xlabel={Lost-sales penalty $ p$},
 legend style={at={(1.17,-.31)},anchor=north},legend columns=2,legend cell align=left]
\nextgroupplot[title={Finite policy costs},ylabel={Menu cost / $\sqrt p$}]
\addplot[blue!75!black,mark=*,thick] table[x=ell,y=normalized_menu,col sep=comma,row sep=newline]{
alpha,ell,BS_target,newsvendor_target,menu_cost,lower_bound,normalized_menu,normalized_lower,gap_upper,selected
0.1,20,5,4,3.807339449541,2.826551105909,0.851346982259,0.632036041469,1.346991195,BS
0.1,50,10,8,6.011151368516,4.343202218328,0.850105179083,0.614221548129,1.384036724,BS
0.1,100,17,14,8.424376274720,6.056308679938,0.842437627472,0.605630867994,1.391008405,BS
0.1,200,29,26,11.814780453276,8.405074604433,0.835431137674,0.594328524917,1.405672289,BS
0.1,500,65,60,18.547496786023,13.120059900664,0.829469272520,0.586746916135,1.413674704,BS
0.1,1000,121,113,26.148565659346,18.395640148046,0.826890250300,0.581721218847,1.421454511,BS
0.1,2000,230,219,36.888970460127,25.836692204334,0.824862455688,0.577726000826,1.427774507,BS
0.1,5000,548,530,58.204805752750,40.632857876422,0.823140256908,0.574635386868,1.432456608,BS
0.1,10000,1067,1042,82.205361321275,57.307828601925,0.822053613213,0.573078286019,1.434452558,BS
};
\addlegendentry{$\alpha=0.1$}
\addplot[red!75!black,mark=square*,thick] table[x=ell,y=normalized_menu,col sep=comma,row sep=newline]{
alpha,ell,BS_target,newsvendor_target,menu_cost,lower_bound,normalized_menu,normalized_lower,gap_upper,selected
0.69787,20,16,15,5.888376358713,3.563315601677,1.316680981518,0.796781591064,1.652499250,BS
0.69787,50,39,36,9.564936249489,5.610478285505,1.352686256726,0.793441448276,1.704834377,BS
0.69787,100,76,72,13.710311191647,7.943915860511,1.371031119165,0.794391586051,1.725888269,BS
0.69787,200,149,142,19.571363625543,11.208310390342,1.383904393669,0.792547228265,1.746147541,BS
0.69787,500,364,353,31.195576151940,17.708356653227,1.395108577460,0.791941784929,1.761630216,BS
0.69787,1000,720,704,44.296233295992,25.034330470521,1.400769889815,0.791655039842,1.769419532,BS
0.69787,2000,1427,1404,62.820040466606,35.388623014992,1.404698808326,0.791313666916,1.775147919,BS
0.69787,5000,3539,3502,99.575804658146,55.937872535330,1.408214534318,0.791080979898,1.780114262,BS
0.69787,10000,7050,6997,140.997036167028,79.092554203935,1.409970361670,0.790925542039,1.782684067,BS
};
\addlegendentry{$\alpha=0.69787$}
\addplot[black,mark=triangle*,thick] table[x=ell,y=normalized_menu,col sep=comma,row sep=newline]{
alpha,ell,BS_target,newsvendor_target,menu_cost,lower_bound,normalized_menu,normalized_lower,gap_upper,selected
2,20,38,37,5.981599853602,3.402353403618,1.337526388686,0.760789349397,1.758077174,CO
2,50,97,96,9.660791547984,5.414713665868,1.366242243042,0.765756150264,1.784174039,CO
2,100,197,194,13.804717763674,7.671062327225,1.380471776367,0.767106232722,1.799583575,CO
2,200,396,391,19.663812035649,10.866963236331,1.390441483439,0.768410339531,1.809503871,CO
2,500,994,986,31.287656125940,17.206686155231,1.399226519085,0.769506398212,1.818342931,CO
2,1000,1992,1981,44.386768899198,24.351609316168,1.403632876970,0.770065501297,1.822744786,CO
2,2000,3990,3973,62.911333966851,34.454701865146,1.406740193051,0.770430555150,1.825914333,CO
2,5000,9984,9957,99.666108126671,54.500386379092,1.409491618217,0.770751855719,1.828722965,CO
2,10000,19978,19939,141.087628577820,77.091626263727,1.410876285778,0.770916262637,1.830129100,CO
};
\addlegendentry{$\alpha=2$}
\addlegendimage{green!45!black,densely dashed,mark=diamond*,thick}
\addlegendentry{Shorter window, $\alpha=2$}
\nextgroupplot[title={Finite-instance gap bounds},ylabel={Menu cost / lower bound}]
\addplot[blue!75!black,mark=*,thick] table[x=ell,y=gap_upper,col sep=comma,row sep=newline]{
alpha,ell,BS_target,newsvendor_target,menu_cost,lower_bound,normalized_menu,normalized_lower,gap_upper,selected
0.1,20,5,4,3.807339449541,2.826551105909,0.851346982259,0.632036041469,1.346991195,BS
0.1,50,10,8,6.011151368516,4.343202218328,0.850105179083,0.614221548129,1.384036724,BS
0.1,100,17,14,8.424376274720,6.056308679938,0.842437627472,0.605630867994,1.391008405,BS
0.1,200,29,26,11.814780453276,8.405074604433,0.835431137674,0.594328524917,1.405672289,BS
0.1,500,65,60,18.547496786023,13.120059900664,0.829469272520,0.586746916135,1.413674704,BS
0.1,1000,121,113,26.148565659346,18.395640148046,0.826890250300,0.581721218847,1.421454511,BS
0.1,2000,230,219,36.888970460127,25.836692204334,0.824862455688,0.577726000826,1.427774507,BS
0.1,5000,548,530,58.204805752750,40.632857876422,0.823140256908,0.574635386868,1.432456608,BS
0.1,10000,1067,1042,82.205361321275,57.307828601925,0.822053613213,0.573078286019,1.434452558,BS
};
\addplot[red!75!black,mark=square*,thick] table[x=ell,y=gap_upper,col sep=comma,row sep=newline]{
alpha,ell,BS_target,newsvendor_target,menu_cost,lower_bound,normalized_menu,normalized_lower,gap_upper,selected
0.69787,20,16,15,5.888376358713,3.563315601677,1.316680981518,0.796781591064,1.652499250,BS
0.69787,50,39,36,9.564936249489,5.610478285505,1.352686256726,0.793441448276,1.704834377,BS
0.69787,100,76,72,13.710311191647,7.943915860511,1.371031119165,0.794391586051,1.725888269,BS
0.69787,200,149,142,19.571363625543,11.208310390342,1.383904393669,0.792547228265,1.746147541,BS
0.69787,500,364,353,31.195576151940,17.708356653227,1.395108577460,0.791941784929,1.761630216,BS
0.69787,1000,720,704,44.296233295992,25.034330470521,1.400769889815,0.791655039842,1.769419532,BS
0.69787,2000,1427,1404,62.820040466606,35.388623014992,1.404698808326,0.791313666916,1.775147919,BS
0.69787,5000,3539,3502,99.575804658146,55.937872535330,1.408214534318,0.791080979898,1.780114262,BS
0.69787,10000,7050,6997,140.997036167028,79.092554203935,1.409970361670,0.790925542039,1.782684067,BS
};
\addplot[black,mark=triangle*,thick] table[x=ell,y=gap_upper,col sep=comma,row sep=newline]{
alpha,ell,BS_target,newsvendor_target,menu_cost,lower_bound,normalized_menu,normalized_lower,gap_upper,selected
2,20,38,37,5.981599853602,3.402353403618,1.337526388686,0.760789349397,1.758077174,CO
2,50,97,96,9.660791547984,5.414713665868,1.366242243042,0.765756150264,1.784174039,CO
2,100,197,194,13.804717763674,7.671062327225,1.380471776367,0.767106232722,1.799583575,CO
2,200,396,391,19.663812035649,10.866963236331,1.390441483439,0.768410339531,1.809503871,CO
2,500,994,986,31.287656125940,17.206686155231,1.399226519085,0.769506398212,1.818342931,CO
2,1000,1992,1981,44.386768899198,24.351609316168,1.403632876970,0.770065501297,1.822744786,CO
2,2000,3990,3973,62.911333966851,34.454701865146,1.406740193051,0.770430555150,1.825914333,CO
2,5000,9984,9957,99.666108126671,54.500386379092,1.409491618217,0.770751855719,1.828722965,CO
2,10000,19978,19939,141.087628577820,77.091626263727,1.410876285778,0.770916262637,1.830129100,CO
};
\addplot[green!45!black,densely dashed,mark=diamond*,thick] table[x=ell,y=monotone_gap_upper,col sep=comma,row sep=newline]{
alpha,ell,shorter_alpha,raw_gap_upper,monotone_gap_upper,normalized_monotone_lower
2,20,1,1.758077174,1.683339473,0.794567233966
2,50,1,1.784174039,1.715186945,0.796555878458
2,100,1,1.799583575,1.731607194,0.797219936183
2,200,1,1.809503871,1.743386228,0.797552178431
2,500,1,1.818342931,1.753962681,0.797751591142
2,1000,1,1.822744786,1.759339534,0.797818073195
2,2000,1,1.825914333,1.763160835,0.797851316306
2,5000,1,1.828722965,1.766565215,0.797871262838
2,10000,1,1.830129100,1.768285931,0.797877911793
};
\end{groupplot}
\end{tikzpicture}
\end{adjustbox}
\caption{Finite lost-sales policy costs and certified instance-specific gaps
for $L=\alpha p$ and unit demand rate. The left panel uses the exact BS/CO
menu. Each right-panel point is rounded upward from a rational upper enclosure
of the menu cost divided by a finite newsvendor lower bound. Solid curves
use the original lead time. The dashed curve uses $\widetilde B_{\rm NV}$
for $\alpha=2$; for the other two values of $\alpha$ the bounds coincide.}
\label{fig:finite-costs}
\end{figure}
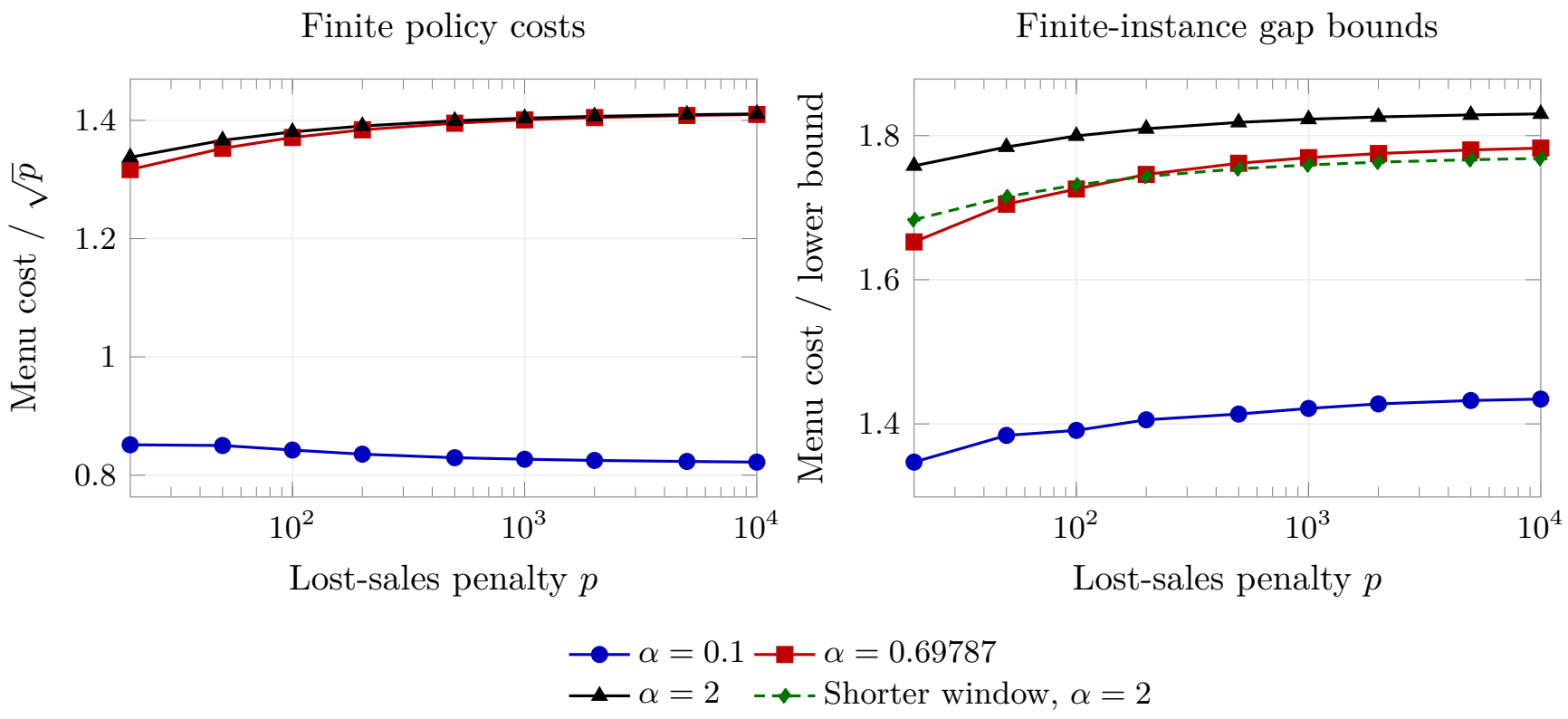

\Cref{fig:finite-costs} reports normalized policy costs and both sets
of instance-specific gap upper bounds. At $ p=10000$, the original bounds
are below $1.434453$, $1.782685$, and $1.830130$, respectively. For
$\alpha=2$, the shorter-window comparison reduces the bound to
$1.768285931$. 
Across all 27 instances, the strengthened bounds are below
$1.782684067$. These finite-instance certificates use newsvendor relaxations;
they do not provide a uniform finite-parameter counterpart of \Cref{thm:main}.


%

\subsection{Finite dual-sourcing costs}
\begin{figure}[htbp]
\centering
\begin{adjustbox}{max width=.7\linewidth}
\begin{tikzpicture}
\begin{groupplot}[group style={group size=2 by 1,horizontal sep=1.25cm},
 width=.47\linewidth,height=5.8cm,xmode=log,xmin=20,xmax=10000,
 xlabel={Express premium $c$},ylabel={Menu cost divided by $\sqrt c$},
 legend style={at={(.98,.98)},anchor=north east},legend cell align=left]
\nextgroupplot[title={DI selected: $\alpha=0.69787$, $p=20$},ymin=1.25,ymax=1.8]
\addplot[blue!75!black,mark=*,thick] table[x=c,y=normalized_menu,col sep=comma,row sep=newline]{
c,normalized_menu
20,1.3263141613812468
100,1.3637611485689392
500,1.3573694237642953
2000,1.3598706121638877
10000,1.358783331570794
};\addlegendentry{$L_E=0$}
\addplot[red!75!black,mark=square*,dashed,thick] table[x=c,y=normalized_menu,col sep=comma,row sep=newline]{
c,normalized_menu
20,1.690702245669056
100,1.4549208864786605
500,1.3758901953100346
2000,1.364649498264684
10000,1.3597425972247448
};\addlegendentry{$L_E=2$}
\addplot[black,dotted,thick,domain=20:10000]{1.358094223869};\addlegendentry{Trial limit}
\nextgroupplot[title={TBS selected: $\alpha=2$, $p=2$},ymin=1.15,ymax=1.45]
\addplot[blue!75!black,mark=*,thick] table[x=c,y=normalized_menu,col sep=comma,row sep=newline]{
c,normalized_menu
20,1.2174393293856616
100,1.2462244527124673
500,1.2613766618169915
2000,1.2674708918762747
10000,1.2708281486630395
};\addlegendentry{$L_E=0$}
\addplot[red!75!black,mark=square*,dashed,thick] table[x=c,y=normalized_menu,col sep=comma,row sep=newline]{
c,normalized_menu
20,1.3472796474792428
100,1.2720736691370351
500,1.2664742703241996
2000,1.2687487123950918
10000,1.2710819074961486
};\addlegendentry{$L_E=2$}
\addplot[black,dotted,thick,domain=20:10000]{1.273522843310};\addlegendentry{Trial limit}
\end{groupplot}
\end{tikzpicture}
\end{adjustbox}
\caption{Finite dual-sourcing costs for two representative parameter pairs.
Solid and dashed curves use zero and positive express lead times; dotted
lines give the limits of the specified policy rules. Each panel compares
fixed-$U$ DI with fixed-rate TBS after integer-target optimization.}
\label{fig:finite-dual}
\end{figure}
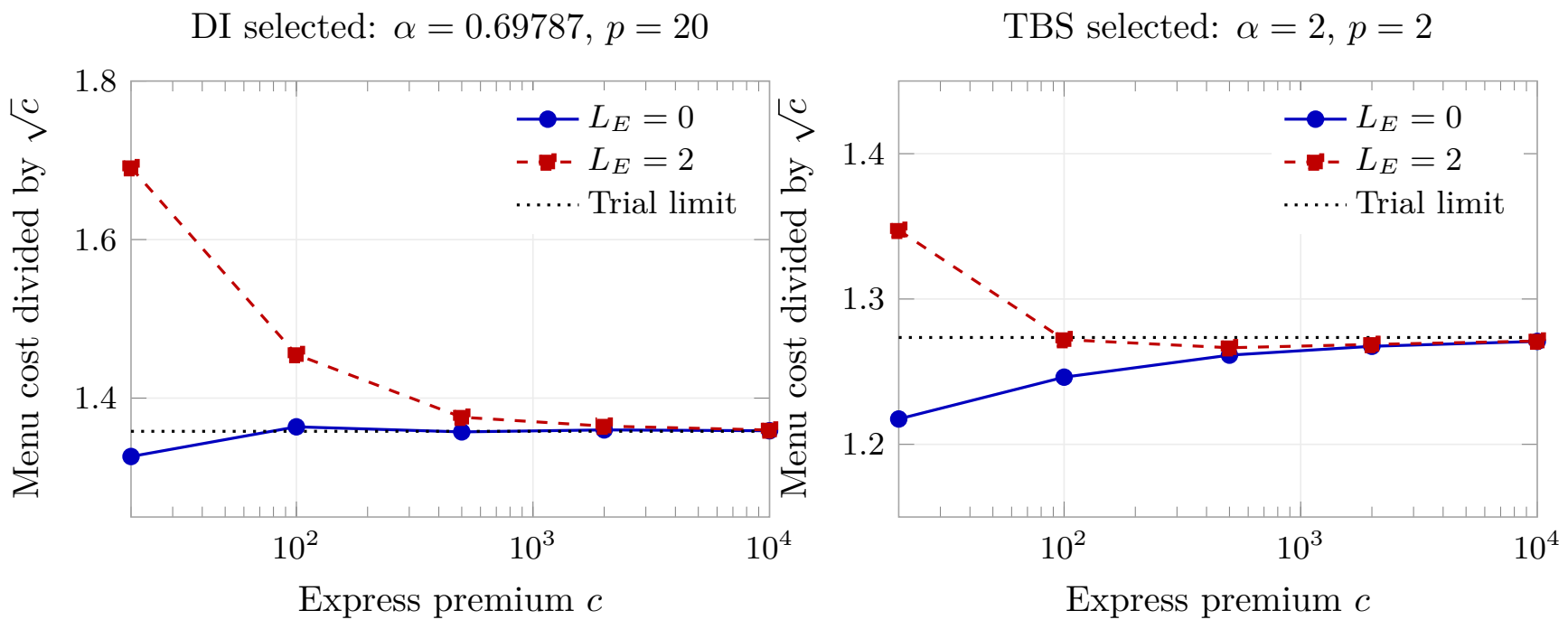

We next consider $\lambda=1$, $\alpha\in\{0.1,0.3,0.69787,2\}$,
$p\in\{0.5,2,5,20\}$ with $\alpha p\ge1$,
$c\in\{20,100,500,2000,10000\}$, and $L_E\in\{0,2\}$.
These choices give 100 instances. For DI we fix $U$, and for TBS
we fix the regular rate, at the trials in \Cref{sec:implementation}.
For each rule the optimal integer $S$ is its newsvendor quantile.
\Cref{app:finite-costs} gives the distribution formulas and tail bounds.
\Cref{tab:finite-dual} reports costs at $c=20$ and $c=10000$ and the
limits of the two trial rules after target optimization. DI is cheaper
in 60 instances and TBS in 40. \Cref{fig:finite-dual} shows one example
for each policy. In the table, the cost difference between $L_E=2$
and $L_E=0$ is smaller at $c=10000$ than at $c=20$. The accompanying
code recomputes the 100 instances at 95-digit working precision with
a Poisson truncation error below $4.641\times10^{-92}$.

\begin{table}[htbp]
\centering\small
\caption{Finite dual-sourcing menu costs divided by $\sqrt c$ and
rounded to four decimal places. The first two cost columns use $c=20$
and the next two use $c=10000$. The final column gives the limiting
menu cost with $U$ fixed for DI and the regular rate fixed for TBS,
after optimizing the integer targets.}
\label{tab:finite-dual}
\begin{adjustbox}{max width=.5\linewidth}
\begin{tabular}{@{}rrrrrrr@{}}
\toprule
 & & \multicolumn{2}{c}{$c=20$} & \multicolumn{2}{c}{$c=10000$} & \\
\cmidrule(lr){3-4}\cmidrule(lr){5-6}
$\alpha$ & $p$ & $L_E=0$ & $L_E=2$ & $L_E=0$ & $L_E=2$ & Trial limit\\\midrule
0.1 & 20 & 0.8001 & 1.1409 & 0.7145 & 0.7154 & 0.7070 \\
0.3 & 5 & 0.9849 & 1.1487 & 0.9278 & 0.9282 & 0.9252 \\
0.3 & 20 & 1.1077 & 1.4701 & 1.0739 & 1.0747 & 1.0708 \\
0.69787 & 2 & 1.0454 & 1.1292 & 1.0743 & 1.0745 & 1.0738 \\
0.69787 & 5 & 1.2168 & 1.3602 & 1.2322 & 1.2325 & 1.2316 \\
0.69787 & 20 & 1.3263 & 1.6907 & 1.3588 & 1.3597 & 1.3581 \\
2 & 0.5 & 1.0078 & 1.0354 & 1.0463 & 1.0464 & 1.0481 \\
2 & 2 & 1.2174 & 1.3473 & 1.2708 & 1.2711 & 1.2735 \\
2 & 5 & 1.3181 & 1.5418 & 1.3473 & 1.3480 & 1.3503 \\
2 & 20 & 1.3377 & 1.8317 & 1.3941 & 1.3969 & 1.3970 \\
\bottomrule
\end{tabular}
\end{adjustbox}
\end{table}

\subsection{Finite policy-cost evaluations}
\label{app:finite-costs}

The finite-parameter study uses the exact policy formulas in \Cref{lem:policy-facts}.

\paragraph{Lost-sales interval certificates.}
The lost-sales instances have $\lambda=1$. For $q=\lambda L$, put
$m=\lfloor q\rfloor$ and use relative Poisson weights
$w_m=1$, $w_{j-1}=jw_j/q$, and $w_{j+1}=qw_j/(j+1)$.
For a cutoff $M>q$, the omitted relative mass and first moment satisfy
\[
 T_M:=\sum_{j>M}w_j\le\frac{qw_M/(M+1)}{1-q/(M+2)},\qquad
 \sum_{j>M}jw_j=q(w_M+T_M).
\]
The first inequality bounds successive tail ratios by $q/(M+2)$; the second
is the Poisson moment recurrence. We use
$M=m+16(\lfloor\sqrt{\lceil q\rceil}\rfloor+1)+100$.
Each recurrence is rounded outward to $10^{-60}\mathbb Z$ using integer
division. Sum the relative weights and divide by an enclosing normalizer.
The smallest integer target whose CDF is at least $ p/(L+ p)$ minimizes
the newsvendor cost. We certify its CDF inequalities at that target and the
preceding integer. Lower cost sums omit the nonnegative tail; upper sums
include its moment bound. The resulting lower-bound intervals have width
less than $1.1\times10^{-54}$ in all 27 instances.

The BS minimum is enclosed by the monotone Erlang recursion over the finite
target range in \Cref{sec:implementation}. For CO, the function
$1-(1+w)e^{-w}$ increases strictly from zero to one, so rational bisection
brackets its unique optimizer when $ p>1$. Substitution of its root
equation gives the minimum $ p(1-e^{-w})$. Taking the smaller of the BS
and CO upper endpoints and dividing by the positive newsvendor lower
endpoint yields each gap certificate. An independent high-precision evaluation
starts from the actual mass $e^{-q}$ and uses
$\E\left[(s-D_L)^+\right]=s\Prb(D_L\le s)-q\Prb(D_L\le s-1)$.
Its values lie inside all rational lower-bound intervals.
For the shorter-window evaluations, use the same interval routine with
$L= p$. The quantile is the integer $ p$ in the nine added cases.
As a separate check, the Poisson call identity gives
$B_{\rm NV}( p, p)=2 p\Prb(\operatorname{Pois}( p)= p)$;
an independent high-precision evaluation of this expression lies in every
added interval. Take the larger of the two certified lower endpoints in
the denominator of each strengthened gap bound. Every displayed gap is
rounded upward. These are evaluations of existing lower comparisons for
specified instances, not a new uniform approximation result.

\paragraph{Dual-sourcing evaluations.}
The DI regular target difference is fixed at
$U=\max\{0,\lfloor\lambda L+\beta(\alpha)\sqrt{\lambda L}\rfloor\}$.
At this $U$, the best integer $S$ is the smallest integer whose CDF for
$N_1+N_2$ is at least $p/(1+p)$. This follows because increasing $S$
by one changes the inventory cost by
$(1+p)\Prb(N_1+N_2\le S)-p$.
The premium cost does not depend on $S$.

For TBS we fix the rate trial in \Cref{sec:implementation} and optimize
only its integer express target. For $0<r<\lambda$, put $\rho=r/\lambda$
and $\sigma=e^{-w}$, where $\rho=(1-e^{-w})/w$.
The time-average inventory tail in \eqref{eq:co-inventory-tail}
gives the positive-part moments. For an integer $A$,
\[
 \E\left[(I_r-A)^+\right]=
 \begin{cases}1/w-A,&A\le0,\\ \sigma^A/w,&A\ge0,\end{cases}
 \qquad
 \E\left[(A-I_r)^+\right]=A-1/w+\E\left[(I_r-A)^+\right].
\]
Condition on $A=N_2-S$. The smallest optimal integer target is the
$p/(1+p)$ quantile of $N_2-I_r$, whose CDF at $S$ is
\[
 \sum_{d=0}^{\infty}\Prb(N_2=d)
 \begin{cases}1,&d\le S,\\ \rho\sigma^{d-S-1},&d>S.\end{cases}
\]
These formulas reduce both target optimizations to integer bisection and
nonnegative probability sums. They do not optimize $U$ or the TBS rate.

The express-demand sums retain $0\le N_2\le80$. If
$\mu=\lambda L_E$ and $\varrho_j=e^{-\mu}\mu^j/j!$, then
\[
 \Prb(N_2>M)\le\frac{\varrho_{M+1}}{1-\mu/(M+2)},\qquad
 \E\left[N_2\mathbf1_{\{N_2>M\}}\right]=\mu\Prb(N_2\ge M),
 \quad M+2>\mu.
\]
Use $\max(1,p)(|S|+U+N_2)$ to bound a DI inventory cost on this tail.
For TBS use $\max(1,p)(|S|+I_r+N_2)$ and independence, with
$\E\left[I_r\right]=1/w$. The resulting omitted cost is below $4.641\times10^{-92}$
in every reported dual-sourcing instance. This is an explicit truncation
bound, separate from roundoff. Its rational check uses
$1/w\le\lceil\sqrt{c(p+1)/(2p)}\rceil$, which follows from
$(1-e^{-w})/w\ge1-w/2$ and $\log(1+1/p)\ge1/(p+1)$.
The degree-$30$ Taylor polynomial lower-bounds $e^2$ and hence
upper-bounds each Poisson mass with mean two. Recomputation at 95-digit working precision reproduces the rounded costs
in \Cref{tab:finite-dual} and agrees with the 20 finite costs plotted in
\Cref{fig:finite-dual} to relative error below $1.4\times10^{-14}$.
The supplement records the chosen integer targets, rates, costs, and checks.

For the limit column in \Cref{tab:finite-dual}, keep the same fixed
normal trial $\beta=\beta(\alpha)$. The optimal limiting DI shift has
$t_{\beta,p}=\Phi^{-1}(p\Phi(\beta)/(1+p))$ and
$s_p=t_{\beta,p}-\beta$ in the objective of \Cref{lem:di-shift}.
Its minimized inventory term is
$((1+p)\phi(t_{\beta,p})-p\phi(\beta))/\Phi(\beta)$.
Thus the displayed reference limit is
\[
 \min\left\{\sqrt\alpha\left[
       (\alpha^{-1}-p)\psi(-\beta)+
       \frac{(1+p)\phi(t_{\beta,p})}{\Phi(\beta)}\right],\ F_{\rm TBS}(p)\right\}.
\]
The truncated Poisson CLT and first-moment convergence used in
\Cref{lem:di-shift} give the DI limit. The CO tail in
\eqref{eq:co-inventory-tail} gives the TBS limit by the same argument
as in \Cref{lem:policy-facts}. Optimizing an
integer shift changes each limiting quantile choice by a vanishing amount
on the diffusion scale. The column is therefore the limit of these
fixed-$U$-rule and fixed-rate comparisons.
It is not the unrestricted optimal cost or a lower bound on that cost.

\bibliographystyle{alpha}
\bibliography{manuscript}

\end{document}